\documentclass[%
 reprint,
 amsmath,amssymb,
 aps,twocolumn
]{revtex4}

\usepackage{natbib}
\usepackage{aas_macros}
\usepackage{graphicx}
\usepackage{dcolumn}
\usepackage{bm}
\usepackage{xcolor}
\usepackage[normalem]{ulem}

\newcommand\ba{\begin{eqnarray}}
\newcommand\ea{\end{eqnarray}}

\usepackage{color}
 
\begin{document}


\title{Analyzing the 21cm forest with Wavelet Scattering Transform: Insight into non-Gaussian features of the 21cm forest}

\author{Hayato Shimabukuro}
 \affiliation{South-Western Institute for Astronomy Research (SWIFAR), Yunnan University, Kunming, Yunnan 650500, People's Republic of China
\\
Key Laboratory of Survey Science of Yunnan Province, Yunnan University, Kunming, Yunnan 650500, People's Republic of China\\
Graduate School of Science, Division of Particle and Astrophysical Science, Nagoya University, Chikusa-Ku, Nagoya, 464-8602, Japan}
  
 \email{shimabukuro@ynu.edu.cn}
 
\author{Yidong Xu}
 \affiliation{%
National Astronomical Observatories, Chinese Academy of Sciences, Beijing 100101, People's Republic of China \\
State Key Laboratory of Radio Astronomy and Technology, Beijing 100101, People's Republic of China
}%
 
 \email{xuyd@nao.cas.cn}

\author{Yue Shao}
 \affiliation{%
Key Laboratory of Cosmology and Astrophysics, College of Sciences, Northeastern
University, Shenyang 110819, China
}%
 
 \email{shaoyue@stumail.neu.edu.cn}

\date{\today}

\begin{abstract}

The 21cm forest—narrow absorption features in the spectra of high-redshift radio sources caused by intervening neutral hydrogen—offers a unique probe of the intergalactic medium and small-scale structures during reionization. While traditional power spectrum methods have been widely used for analyzing the 21cm forest, these techniques are limited in capturing the non-Gaussian nature of the signal. In this work, we introduce the Wavelet Scattering Transform (WST) as a novel diagnostic tool for the 21cm forest, which allows for the extraction of higher-order statistical features that power spectrum methods cannot easily capture. By decomposing simulated brightness temperature spectra into a hierarchy of scattering coefficients, the WST isolates both local intensity fluctuations (first-order coefficients) and scale-scale correlations (second-order coefficients), revealing the complex, multi-scale non-Gaussian interactions inherent in the 21cm forest. This approach enhances the power of 21cm forest in distinguishing between different cosmological models, such as Cold Dark Matter (CDM) and Warm Dark Matter (WDM), as well as scenarios with enhanced X-ray heating. Unlike traditional methods, which focus primarily on Gaussian statistics, the WST captures richer astrophysical and cosmological information. Our analysis shows that WST can significantly improve constraints on key parameters, such as the X-ray heating efficiency and the WDM particle mass, providing deeper insights into the early stages of cosmic structure formation.
\end{abstract}


\maketitle


\section{Introduction}

Over the past few decades, the 21cm hyperfine transition of neutral hydrogen (HI) has become a key observable in the quest to understand the history of the universe, such as the dark ages, cosmic dawn, and the epoch of reionization (EoR). Most efforts have focused on two main approaches to 21cm studies: the global 21cm signal and large-scale interferometric observations targeting the 21cm power spectrum\citep[e.g.][]{fur,2012RPPh...75h6901P,2013PhRvD..87d3002L,2023PASJ...75S...1S}. Both methods aim to probe the thermal and ionization state of the intergalactic medium (IGM) across cosmic time, providing insights into how the first stars, galaxies, and black holes formed in the early universe.

The 21cm global signal corresponds to the sky-averaged brightness temperature of neutral hydrogen as a function of redshift (or observing frequency). Experiments such as EDGES\citep{2018Natur.555...67B}, SARAS\citep{2018ApJ...858...54S,2021arXiv210401756N}, LEDA\citep{2018MNRAS.478.4193P}, and future instruments such as DAPPER\citep{Burns2021},
LuSEE-night\citep{2023arXiv230110345B}, PRATUSH\citep{2023ExA....56..741S,2025arXiv250705654S}, and DSL\citep{2021RSPTA.37990566C} attempt to detect this global signal, which encodes the evolution of the IGM from the dark ages through cosmic dawn. Despite significant challenges arising from strong foreground emission and instrumental systematics, these experiments aim to constrain key epochs, including the formation of the first luminous objects and the heating of the early IGM.

Meanwhile, large-scale interferometric observations of the 21cm power spectrum focus on measuring spatial fluctuations in the IGM's neutral fraction. Facilities such as LOFAR\citep[e.g.][]{2013A&A...556A...2V}, MWA\citep[e.g.][]{2018PASA...35...33W} and HERA\citep[e.g.][]{2017PASP..129d5001D}seek to characterize the three-dimensional distribution of neutral hydrogen. These measurements, while extremely sensitive to foreground contamination and requiring long integration times, have yielded progressively tighter upper limits on the amplitude of the 21cm power spectrum\citep[e.g., see Fig. 19 of ][]{2023PASJ...75S...1S}. 
Furthermore, the SKA will start observation at the end of 2020's and it is expected to constrain the 21cm power spectrum more tightly \citep{2013ExA....36..235M, 2015aska.confE...1K}. Such constraints inform theoretical models of the timing, duration, and topology of reionization, as well as the nature of the first astrophysical heating and ionizing sources.

In addition to these high-redshift efforts, low-redshift ($z\lesssim5$) 21 cm surveys serve as crucial probes of the evolution of large-scale structure and the nature of dark energy—particularly through measurements of baryon acoustic oscillations (BAO)—and also provide insights into the formation and evolution of gas within galaxies. Currently, this field sees the active involvement and planning of various ground-based radio telescopes, including Canadian Hydrogen Intensity Mapping Experiment (CHIME)\citep[e.g.][]{2022ApJS..261...29C,2019clrp.2020....9L,2023ApJ...947...16A}, the Canadian Hydrogen Observatory and Radio-transient Detector (CHORD)\citep[e.g.][]{2019clrp.2020...28V}, upgraded Giant Metrewave Radio Telescope (uGMRT)\citep[e.g.][]{2017CSci..113..707G,2019MNRAS.482.5597A,2020ApJ...900L..30C,2021MNRAS.500..998A}, MeerKAT\citep[e.g.][]{2016mks..confE...1J,2023arXiv230111943P,2023MNRAS.518.6262C,2025arXiv250117564M}, Five-hundred-meter Aperture Spherical Telescope (FAST)\citep[e.g.][]{2011IJMPD..20..989N,2025ApJS..277...25H,2025ApJS..276....6Z}, Baryon Acoustic Oscillations from Integrated Neutral Gas Observatations (BINGO) \citep[e.g.][]{2022A&A...664A..15W,2022A&A...664A..14A,2022MNRAS.510.1495X,2022A&A...666A..83N} and the developing Hydrogen Intensity and Real-time Analysis eXperiment (HIRAX)\citep[e.g.][]{2016SPIE.9906E..5XN}. 


In parallel to these approaches, the so-called ``21cm forest'' has emerged as a powerful and complementary technique for probing the high-redshift universe\citep[e.g.][]{2002ApJ...577...22C,2002ApJ...579....1F,2006MNRAS.370.1867F,2015aska.confE...6C}. By analogy with the Ly-$\alpha$ forest, the 21cm forest consists of narrow absorption lines imprinted by intervening HI along the line of sight to a bright background radio source such as a quasar or radio galaxy. Some theoretical studies demonstrated that, unlike Ly-$\alpha$ absorption (which saturates in a highly neutral IGM), the 21cm transition remains sensitive even when the universe is predominantly neutral. This characteristic makes the 21cm forest particularly valuable for exploring the detailed structure and physical conditions of the IGM during the EoR and beyond, potentially at redshifts $z \gtrsim 6$--7 \citep[e.g.][]{2002ApJ...577...22C,2002ApJ...579....1F,2006MNRAS.370.1867F,2009ApJ...704.1396X,2011MNRAS.410.2025X,2012MNRAS.425.2988M,2015aska.confE...6C,2016MNRAS.455..962S,2021MNRAS.506.5818S}.


A key advantage of the 21cm forest lies in its ability to probe small-scale structures—on the order of tens of kiloparsecs to a few megaparsecs—through high-resolution “pencil-beam” observations along individual line of sight, unlike standard 21cm emission surveys that focus on large-scale fluctuations of hundreds to thousands of comoving megaparsecs. This fine-grained information helps break parameter degeneracies inherent in emission-based measurements and provides unique insights into local thermal and radiative feedback mechanisms in the early IGM. Moreover, by capturing these small-scale density fluctuations, the 21cm forest offers robust probes of cosmological effects such as alternative dark matter scenarios, neutrino mass, and primordial fluctuations\citep[e.g.][]{2014PhRvD..90h3003S,2020PhRvD.101d3516S,2020PhRvD.102b3522S,2021JCAP...04..019K,2023PASJ...75S..33V,2023PhRvD.107l3520S}. 

All three observational modes are powerful techniques for probing the neutral hydrogen IGM, but they capture different aspects of its structure. 
The sky-averaged signal characterize the global evolution of the IGM properties, the 21cm forest provides high-resolution, small-scale insights, whereas 21cm tomography 
surveys offer broader, large-scale statistical information. Together, these complementary approaches open a new window into the physics of the IGM (including reionization, X-ray heating, etc.), the role of mini-halos, and the interplay between galaxy formation and the IGM\citep[e.g.][]{2009ApJ...704.1396X,2011MNRAS.410.2025X,2023JCAP...03..017K,2024JCAP...10..091N}. As future radio telescopes improve in sensitivity and more high-redshift radio sources are discovered, the 21cm forest will become an increasingly powerful tool for testing fundamental physics and refining our understanding of the earliest stages of cosmic structure formation.

Recent theoretical work has also highlighted the importance of the 21cm forest power spectrum\citep{2014MNRAS.441.2476E,2023NatAs...7.1116S,2025MNRAS.tmp...29S,2025arXiv250100769S,2025CmPhy...8..220S}. While the three-dimensional power spectrum from 21cm observations remains the primary tool for exploring large-scale structures such as the IGM, the one-dimensional 21cm forest power spectrum extracted from absorption-line data encodes complementary information about small-scale fluctuations. The introduction of the 21cm forest power spectrum marks a significant step forward from traditional methods based solely on number-counting discrete absorption lines. By capturing the collective statistical properties of 21cm forest along many lines of sight, this approach enables a more comprehensive and robust characterization of small-scale structures in the IGM. 

While power spectrum analyses of the 21cm forest provide a powerful way to statistically describe 21cm absorption lines, they are not necessarily sufficient to capture complex structures such as non-Gaussian features of 21cm absorption lines. In this paper, we propose the Wavelet Scattering Transform (WST) to characterize 21cm forest. The WST first applies a wavelet transform at multiple scales and then performs a nonlinear pooling of certain wavelet coefficients to stably extract higher-order correlation information. A particularly important advantage of WST is its ability to efficiently capture correlations across different scales—so-called scale coupling—which are often missed by power spectrum analyses. The WST has already been introduced in large-scale 21cm map analyses, providing non-Gaussian information on the spatial distribution of 21cm lines\citep{2022MNRAS.513.1719G,2023MNRAS.519.5288G,2023MNRAS.524.4239P,2024ApJ...973...41Z,2024A&A...688A.199P,2024A&A...686A.212H}. Not only in the context of 21cm studies, the WST is also applied to the analysis of the Lyman-$\alpha$ forest\citep{2024PhRvL.132w1002T}. 

Traditional power‐spectrum analyses, which focus on Gaussian properties, can miss critical non‐Gaussian, multi‐scale features of the 21 cm forest. By applying the Wavelet Scattering Transform (WST), we can comprehensively extract higher‐order statistical signatures of both astrophysics and cosmology, thereby advancing our understanding of IGM physics beyond what is accessible with traditional methods. In this study, we demonstrate this approach by investigating the impacts of the Warm Dark Matter (WDM) scenario and X-ray heating on the IGM.



The paper is structured as follows: in Section II, we define the formalism of the 21cm forest, including the relevant theoretical models for the signal, density field, and gas distribution. Section III introduces the methodology of the WST, explaining how the transformation works to decompose the signal into scale-dependent coefficients. Section IV presents the results of applying WST to simulated 21cm forest data under different cosmological models, including Cold Dark Matter (CDM), Warm Dark Matter (WDM), and various X-ray heating scenarios. Section V explores the impact of thermal noise on the results, while Section VI provides a Fisher forecast to estimate parameter constraints. Finally, Section VII offers a summary of the findings, emphasizing the potential of WST as a powerful diagnostic tool for extracting astrophysical and cosmological information from the 21cm forest.

\section{Formalism of the 21cm forest and mock data}


In order to evaluate 21cm forest properties, we first characterize the 21cm optical depth $\tau_{\nu_0}(s,z)$, which quantifies the absorption of 21cm radiation by neutral hydrogen along the line of sight. In the optically thin limit, the optical depth is given by \citep[e.g.][]{2002ApJ...579....1F,2006MNRAS.370.1867F,2023NatAs...7.1116S}
\begin{eqnarray}
\tau_{\nu_0}(s, z) &\approx& 0.0085 [1 + \delta(s, z)] (1 + z)^{3/2} \notag \\
&& \times \left( \frac{x_{\text{HI}}(s, z)}{T_S(s, z)} \right) 
\left( \frac{H(z)/(1+z)}{dv_\parallel/dr_\parallel} \right),
\end{eqnarray}
where $s$ denotes the direction of the line of sight at redshift $z$. $\delta(s, z)$ is the local gas overdensity, $x_{\text{HI}}$ is the neutral fraction of hydrogen, $H(z)$ is the Hubble parameter, and $dv_\parallel/dr_\parallel$ represents the velocity gradient along the line of sight. 

Although the optical depth is a key theoretical parameter that describes the intrinsic absorption strength of the 21cm forest, in practice, the absorption features appear as lines in the spectrum of background radio sources. Since radio observations are based on the Rayleigh-Jeans law, we usually measure brightness temperature rather than optical depth directly in radio astronomy. Consequently, we express the absorption signal in terms of the differential brightness temperature $\delta T_b$, defined by
\begin{equation}
\delta T_b(s, \nu) \approx \frac{T_S(s, z) - T_\gamma(s, \nu_0, z)}{1 + z} \, \tau_{\nu_0}(s, z),
\end{equation}
where $\nu_0=$1420.4 MHz and $T_S$ is the spin temperature of neutral hydrogen, $T_\gamma$ is the background radiation temperature, thereby providing a direct link between the theoretical modeling of the IGM's absorption properties and the observable brightness temperature fluctuations.

We run 10 sets of high-resolution simulations for 10 different environments at redshift 9, each with a different local mean density. The box size of these simulations is 10 Mpc on a side. From each of these simulated boxes, 10 lines of sight (LoS) are randomly selected to calculate the 21cm forest spectra assuming the same flux density of $S_{\rm 150} = 10$ mJy for the background sources. The details of the simulations were described in Ref. \cite{2023NatAs...7.1116S}.  
When calculating the optical depth and brightness temperature, we modeled the dark matter halos based on extended Press-Schechter formalism, the neutral hydrogen profile within dark matter halos, the thermal evolution of the gas as well as infalling gas surrounding halos. Interested readers are encouraged to refer to previous works for further details\citep[e.g.][]{2011MNRAS.410.2025X,2014PhRvD..90h3003S,2023NatAs...7.1116S,2025CmPhy...8..220S}. In practice, realistic 21\,cm forest data must be obtained through forward modeling based on numerical simulations, since nonlinear astrophysical processes and reionization histories cannot be fully captured analytically. The reliability of the statistical inference thus depends critically on the completeness of the models included in the simulation pipeline. While our present work includes variations in dark matter and X-ray heating, further extensions to encompass a broader range of reionization and astrophysical scenarios—such as UV ionization from star formation—will be necessary to ensure fully robust and comprehensive inference in future applications.
The UV ionization from star formation is actually incorporated in the large-scale semi-numerical simulation with 21cmFAST. From this larger box, we select neutral patches of 10 Mpc long to model in detail the small-scale structures and calculate the 21 cm forest signals.

\section{The 1D Wavelet Scattering Transform}
\label{sec:scattering_transform}

The Wavelet Scattering Transform (WST) is a multi-scale method that decomposes a signal into coefficients at different scales, making it particularly effective for analyzing complex signals with multiscale structures and non-Gaussian features\citep[e.g.][]{mallat2012group}. To implement the Wavelet Scattering Transform, we use the Kymatio package \citep{2018arXiv181211214A}, which facilitates efficient computation of scattering coefficients and provides an effective tool for analyzing spectral data.

In this section, we present the methodology for applying the WST to analyze the statistical characteristics of the 21cm forest brightness temperature. The WST is designed to decompose the input brightness temperature of the 21cm forest spectrum, \(\delta T_b\), into a series of recursive convolutions with a family of wavelets of varying scales (and orientation in the general multidimensional case). After the convolution, we take the modulus of the resulting fields and apply low-pass filters to remove rapidly-oscillating components. This process extracts scattering coefficients that describe the statistical characteristics of the 21cm forest spectrum. This process results in scattering coefficients that effectively characterize the features of the spectrum. 



The wavelet decomposition is governed by two quantities, $J$ and $Q$. $J$ defines the maximum value of $j$, which controls the dilation (scale) of each wavelet: larger $j$ produces a more broadly stretched wavelet that captures slowly oscillating, coarse‐scale features, while smaller $j$ yields a more compact wavelet sensitive to rapidly oscillating, fine‐scale variations.  We smooth our mock spectra to 1 kHz resolution, set our base frequency $\lambda_0=1\,$kHz, and then define
\begin{equation}
  \lambda_j = 2^j\,\lambda_0,\quad j=1,\dots,J,
\end{equation}
so that each scale $j$ corresponds to a characteristic oscillation period of order $2^j$ kHz.

Within each octave we subdivide into $Q$ wavelets: for $q=0,\dots,Q-1$, the $q$-th wavelet has center period
\begin{equation}
  \xi_{j,q} = \lambda_0\,2^{\,j-1 + \frac{q}{Q}},
\end{equation}
which spans the full range of coarse‐to‐fine oscillations within that octave.  Increasing $q$ shifts the wavelet toward more rapidly oscillating components at the same overall scale. In practice, we set the maximum scale \(J = 8\), which captures the majority of relevant features in our simulations, and adopt \(Q = 8\) wavelets per octave as a reference configuration.

Wavelet filters are localized in both space and frequency, capturing fluctuations over a local region while focusing on specific frequency bands determined by the scale parameter \(j\). In our implementation we use the complex Morlet wavelet \citep[e.g.][]{2012arXiv1203.1513B,vahidi2023kymatio}

\begin{equation}
\psi_{j,q}(x)
= \frac{1}{\sigma_{j,q}\sqrt{2\pi}}\,
\exp\bigl(i\,2\pi\,\xi_{j,q}\,x\bigr)\,
\exp\!\Bigl(-\frac{x^2}{2\,\sigma_{j,q}^2}\Bigr),
\end{equation}
where in Kymatio the time-domain width \(\sigma_{j,q}\) is automatically chosen by
\begin{equation}
\sigma_{j,q}
= \frac{1}{\xi_{j,q}\,\bigl(2^{1/Q}-2^{-1/Q}\bigr)}.
\end{equation}

Scattering coefficients are computed hierarchically: after each convolution with \(\psi_{j,q}\), we take the complex modulus and then perform a local average over a window of width \(2^j\)—the same scale as the wavelet—to remove any residual rapid oscillations.

\begin{figure}[htbp]
  \centering
  \includegraphics[width=1.0\hsize]{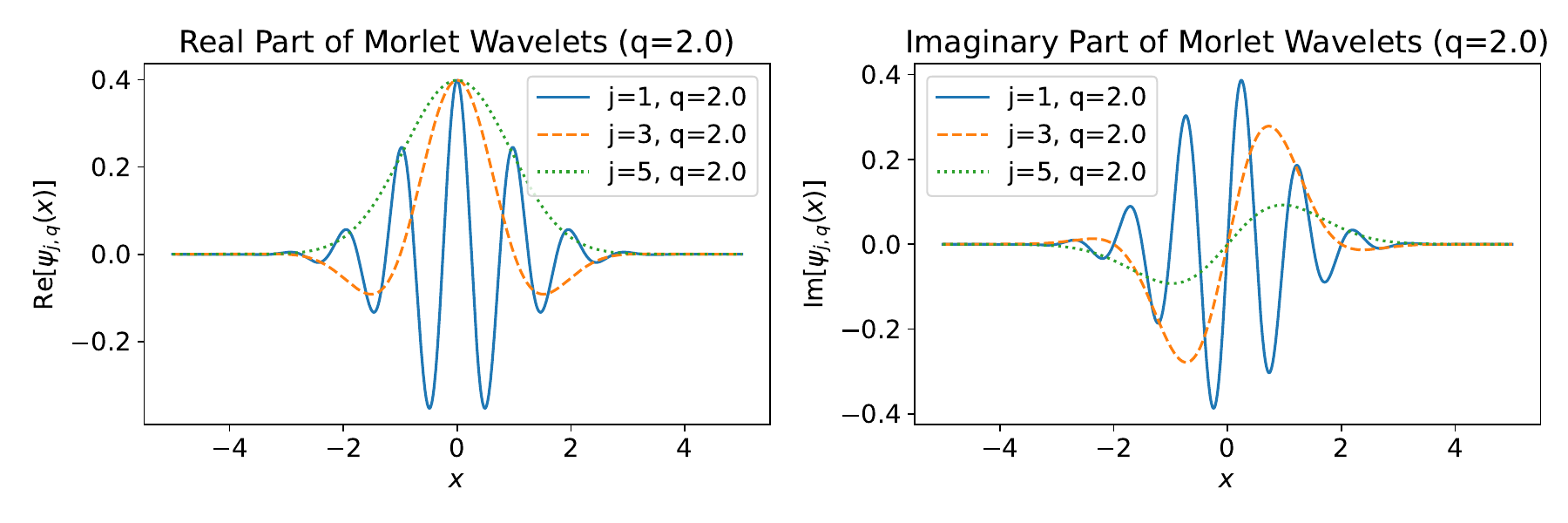}
  \includegraphics[width=1.0\hsize]{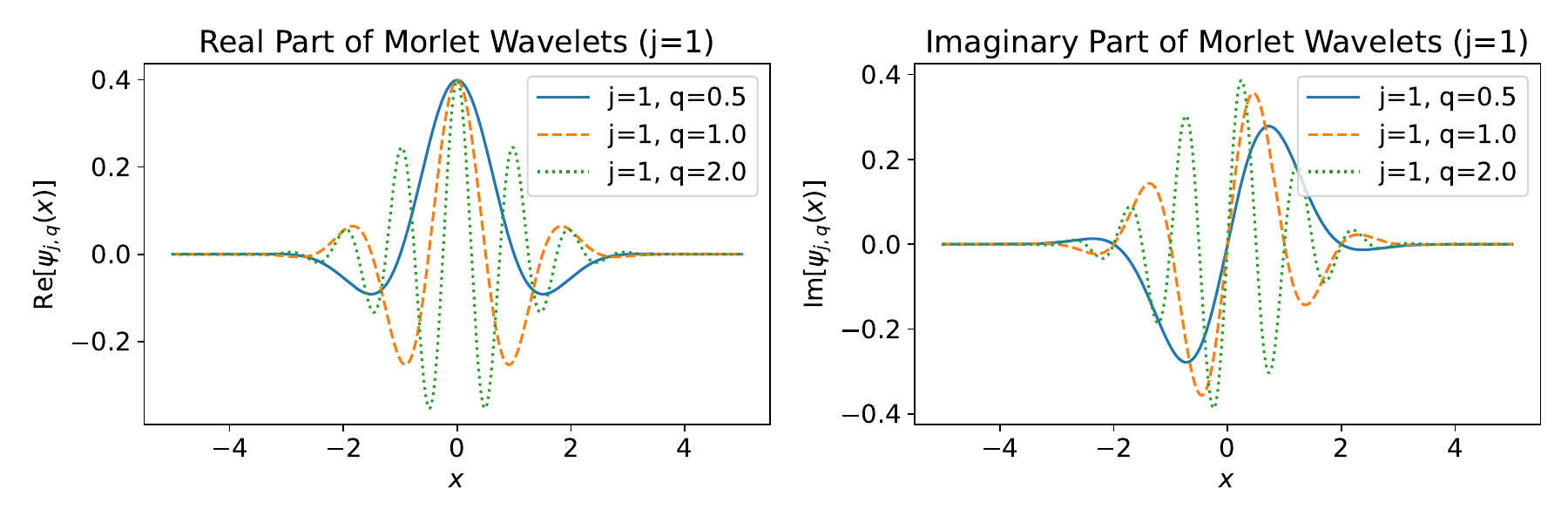}
  
  \caption{(Top): Real (left) and imaginary (right) parts of the Morlet wavelet $\psi_{j,q}(x)$ for fixed scale $q=2$ and $j=1\; (solid),\,3\; (dashed),\,5\; (dotted)$. (Bottom): Real (left) and imaginary (right) parts of the Morlet wavelet $\psi_{j,q}(x)$ for fixed central frequency index $j=1$ and $q=0.5\ (solid),\,1.0\ (dashed),\,2.0\ (dotted)$.}
  \label{fig:morlet}
\end{figure}

In Fig.\ref{fig:morlet}, we show the Morlet wavelet for $j=1,3,5$ with fixed $q=2.0$ and for $q=0.5,1.0,2.0$ with fixed $j=1$. We can see that increasing $j$ causes the Gaussian envelope to broaden, so that the Morlet wavelet becomes sensitive to more slowly oscillating, coarse-scale structure. Conversely, increasing $q$ raises the internal oscillation of the Morlet wavelet, sharpening its sensitivity to rapid, fine-scale variations.

The scattering transform begins by computing the zeroth-order coefficient:
\begin{equation}
  S_0 = \langle \delta T_b \rangle,
\end{equation}
The mean brightness temperature of the 21 cm forest spectrum.  First-order coefficients
\begin{equation}
  S_1(j,q) = \bigl\langle \bigl|\delta T_b * \psi_{j,q}\bigr|\bigr\rangle
\end{equation}
measure the average amplitude of localized oscillations at scale \(j\) and wavelet index \(q\), and are then averaged over \(q\) to yield
\(S_1(j)=\tfrac{1}{Q}\sum_{q}S_1(j,q)\).  Second-order coefficients
\begin{equation}
  S_2(j_1,q_1;\,j_2,q_2)
  = \bigl\langle \bigl|\;|\delta T_b * \psi_{j_1,q_1}|\;\ast \psi_{j_2,q_2}\bigr|\bigr\rangle
\end{equation}
quantify the coupling between features at scales \((j_1,q_1)\) and \((j_2,q_2)\), and are likewise averaged over \((q_1,q_2)\) to form \(S_2(j_1,j_2)\).

For example, in a signal containing both sharp and smooth fluctuations, \(S_1\) captures the sharp, localized variations—rapidly oscillating features—while \(S_2\) reveals how those rapid fluctuations interact with broader-scale structures.  This hierarchical, multi-scale decomposition thus uncovers non-Gaussian signatures, such as those arising from different dark matter free-streaming or X-ray heating histories, which are not fully accessible via traditional power-spectrum analyses.

Furthermore, the WST is robust to small deformations and noise due to its multi-scale wavelet decomposition, which effectively isolates features across various resolutions while filtering out rapidly-oscillating noise. The use of a non-linear modulus operator removes sensitive phase information, ensuring that minor shifts or distortions in the input result in only minimal changes in the output coefficients. Additionally, local averaging smooths the feature representation by aggregating information over small neighborhoods, thereby stabilizing the overall statistical characterization. Moreover, the inherent Lipschitz continuity of the WST guarantees that small perturbations in the input lead to only proportionally limited variations in the scattering coefficients. By separating the signal into components that reflect both local and global features, the WST offers a more comprehensive understanding of the 21cm forest signal.

\section{Results without thermal noise}

In this section, we present the results of applying the Wavelet Scattering Transform (WST) to simulated 21cm forest data. We analyze both the first- and second-order WST coefficients calculated from simulated 21cm forest data. In addition to the Cold Dark Matter (CDM) model without X-ray heating, we also investigate the impact of varying key parameters such as the warm dark matter (WDM) mass and the X-ray heating efficiency. Specifically, we explore models with different warm dark matter masses ($m_{\text{WDM}}$ = 3 keV, 6 keV) and X-ray heating efficiency values (\(f_X = 0.1, 0.3\)) to assess how these parameters influence the 21cm forest signal. We use the calculated mean value of each coefficient averaged over 10 lines of sight (LoS) for all WST analyses.

\subsection{21cm forest brightness temperature spectra}

\begin{figure}
    \centering
    \includegraphics[width=1.0\hsize]{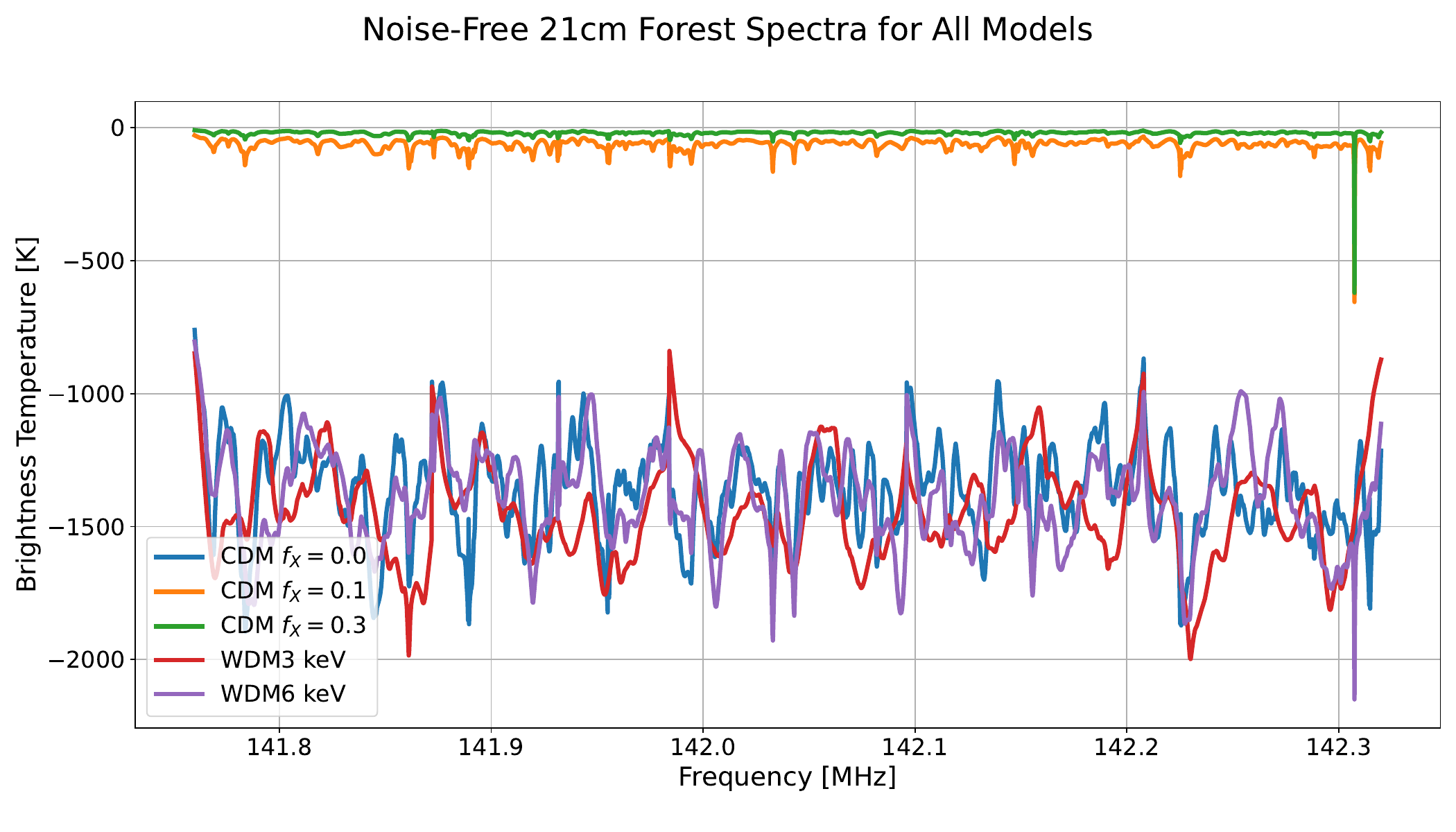}
    \caption{21cm forest brightness temperature spectra in the CDM model for $f_X$=0, 0.1, and 0.3, and in WDM models with particle masses of 3keV and 6keV for $f_X$=0, respectively.}
    \label{fig:spectrum}
\end{figure}

In Fig.~\ref{fig:spectrum}, we first present the brightness temperature spectra of the 21cm forest for different cosmological scenarios, namely the CDM and WDM models, along with cases of varying X-ray heating efficiencies. The spectra shown correspond to a single representative line of sight. In the CDM scenario, multiple deep absorption troughs and sharp variations are prominently visible. In contrast, the WDM scenario exhibits fewer and less pronounced absorption features, resulting in a smoother spectrum due to the suppression of small-scale fluctuations caused by free-streaming of warm dark matter particles. Additionally, the degree of suppression depends on the dark matter particle mass; a lower mass (3 keV) produces stronger suppression effects compared to a higher mass (6 keV). Nonetheless, even the 6 keV WDM model does not exhibit the rich small-scale structure found in the CDM scenario. In models incorporating X-ray heating, the spectrum becomes relatively uniform with significantly fewer prominent troughs. This occurs because X-ray radiation heats the intergalactic medium, increasing the spin temperature and consequently reducing the optical depth, which suppresses the overall 21cm forest absorption features. While strong X-ray heating can still occasionally generate localized sharp absorption features, it predominantly flattens the spectrum over broader frequency ranges.

\subsection{First-Order Coefficients}
\begin{figure}[h!]
\centering
\includegraphics[width=1.0\hsize]{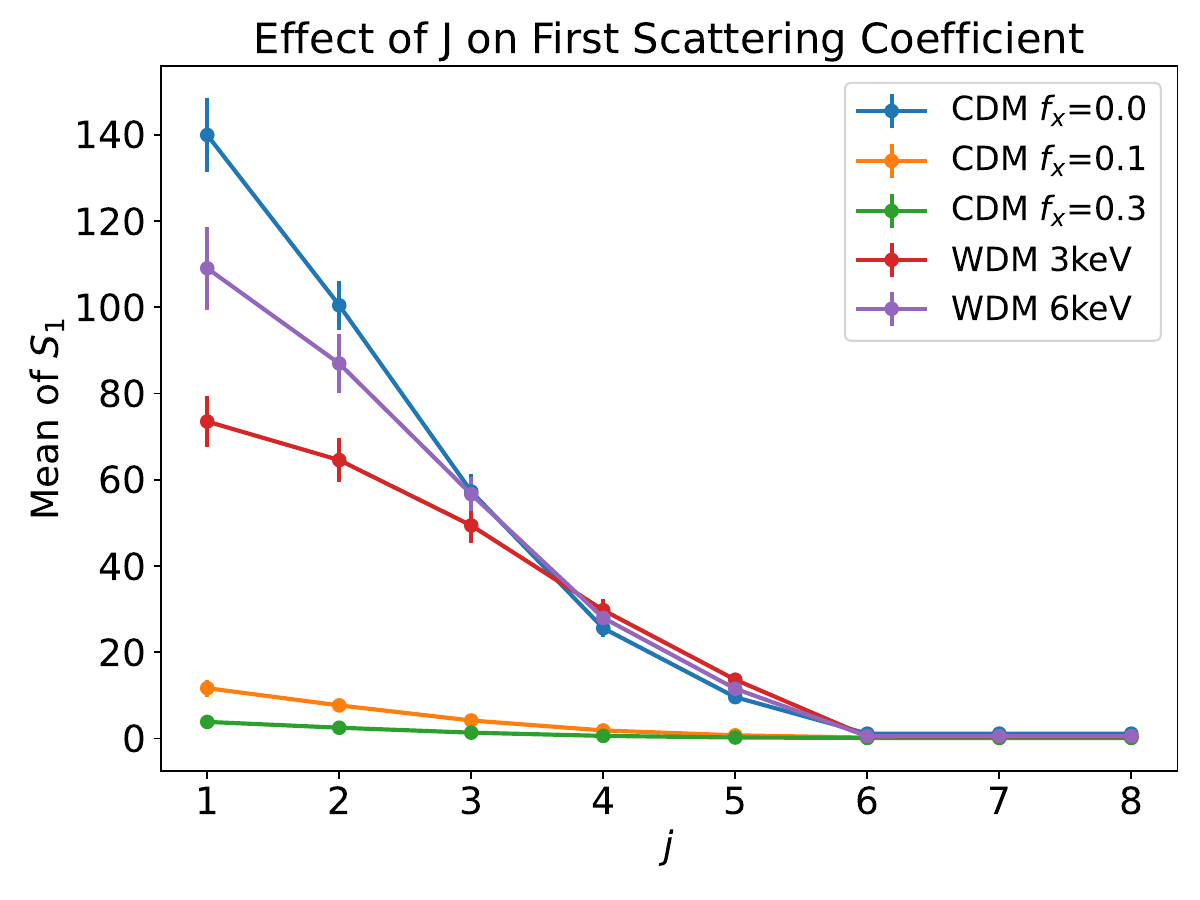}
\caption{First-order WST coefficients \( S_1(j) \) for different models: CDM, WDM, and enhanced X-ray heating. The error bars represent the sample variance calculated over 10 independent lines of sight.
}
\label{fig:wst_first_order}
\end{figure}

The first-order WST coefficients, \( S_1(j) \), represent the intensity distribution of the 21cm forest brightness temperature across different frequency scales. Fig. \ref{fig:wst_first_order} shows the mean first-order scattering coefficients as a function of the scale parameter \(j\) for the different models: CDM, WDM (3 keV and 6 keV), and CDM with enhanced X-ray heating (\(f_X = 0.1, 0.3\)). The error bars represent the sample variance calculated over 10 independent lines of sight. In all cases, the coefficients decrease rapidly as \(j\) increases, indicating that smaller scales (lower \(j\)) capture stronger local variations in the brightness temperature spectra. At low values of \(j\), CDM exhibits the largest coefficients because it retains abundant small-scale structure, leading to deep absorption troughs and pronounced peaks across multiple frequency intervals. By contrast, WDM partially suppresses such fine structures through free-streaming, which reduces the first-order coefficients, though not as dramatically as strong X-ray heating can. Indeed, the CDM model with \(f_X=0.1, 0.3\) shows smaller values at all \(j\), reflecting that the heating process flattens or homogenizes the spectrum, leaving fewer simultaneous fluctuations over multiple frequency ranges. At higher \(j\) (larger scales), all models converge toward low coefficients. This convergence occurs because the intrinsic 21cm forest signal is inherently smooth at larger scales—broad, smooth structures naturally yield less variation. Consequently, model-specific differences become less significant once the wavelet analysis probes these broader structures. Taken together, these trends confirm that small-to-intermediate scales are crucial for discriminating among CDM, WDM, and heated scenarios. The first-order scattering coefficients thus highlight how both the suppression of small-scale power (in WDM) and the smoothing effect of intense X-ray heating diminish the overall amplitude of local spectral variations, whereas standard CDM maintains strong small-scale features and, therefore, higher first-order coefficients at low \(j\).

The first-order coefficients from the WST provide a measure of the amplitude of wavelet responses averaged over different scales, which closely resembles the scale-dependent power captured by the 21cm forest power spectrum\citep[see][]{2023NatAs...7.1116S}. Both approaches quantify how fluctuations vary with scale, thereby revealing the impacts of parameters such as warm dark matter mass—which suppresses small-scale structure (lower $J$ in the first-order coefficient of WST)—and X-ray heating—which generally reduces the overall signal amplitude (while $J$ in the first-order coefficient of WST). However, while the power spectrum is a second-order statistic based on the Fourier transform that offers a global measure of power yet loses local phase information, the WST first-order coefficients are derived using a wavelet basis that retains some degree of local structural detail (Appendix \ref{sec:local}). Despite this, when only first-order coefficients are considered, the WST primarily captures amplitude information akin to the power spectrum without fully incorporating the higher-order nonlinearities and non-Gaussian features present in the data. Furthermore, the second-order coefficients are highly significant because they capture interactions between wavelet responses at different scales, thereby revealing subtle nonlinear and non-Gaussian structures that the first-order analysis may overlook. These higher-order features can provide deeper insights into the complex interplay of astrophysical processes affecting the 21cm forest signal. In the next section, we will delve into the role and interpretation of the second-order coefficients, discussing how they enhance our understanding of the underlying physical processes beyond what is accessible through first-order analysis alone.

We also find that the sample variance for the $f_X=0.1,0.3$ case is small. This can be interpreted as follows. As $f_X$ increases, the enhanced X-ray heating raises the gas temperature and brings the spin temperature closer to or above the background radiation temperature. Consequently, the depth of the 21cm absorption features decreases, effectively ``flattening'' the spectrum. This homogenizes the brightness temperature distribution across different lines of sight, thereby reducing the overall sample variance in the wavelet scattering transform coefficients.


\subsection{Second-Order Wavelet Scattering Coefficients}

To quantify the multi-scale correlations of our simulated 21cm forest brightness temperature, we next focus on the second-order WST scattering coefficient. The second-order scattering coefficients, \(S_{2}(j_1,j_2)\), measure correlations between 21cm forest brightness temperature spectrum at two different scales \(j_1\) and \(j_2\). High values indicate that the signal exhibits strongly coupled fluctuations at both these scales---an important signature of a non-Gaussian, multi-scale structure.

\begin{figure}[h!]
\centering
\includegraphics[width=1.0\hsize]{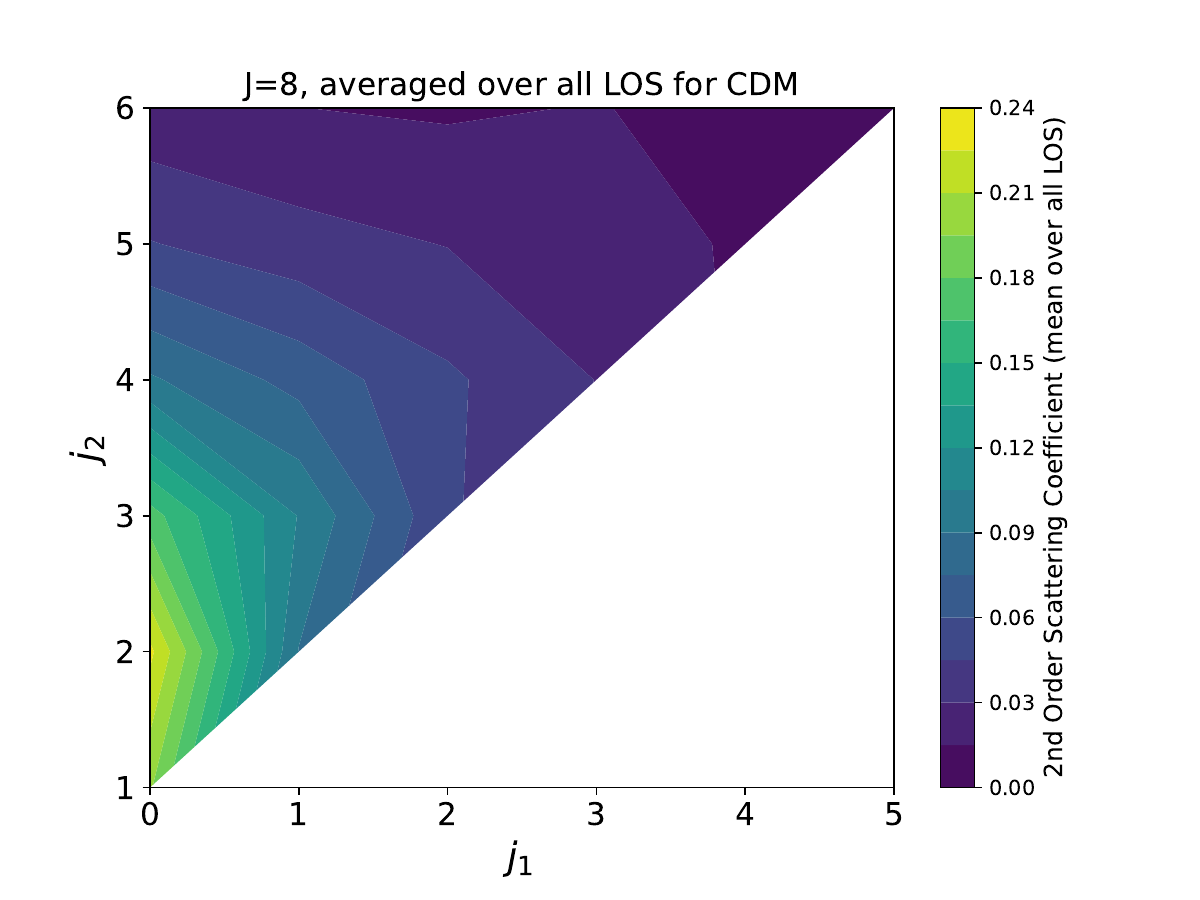}
\includegraphics[width=1.0\hsize]{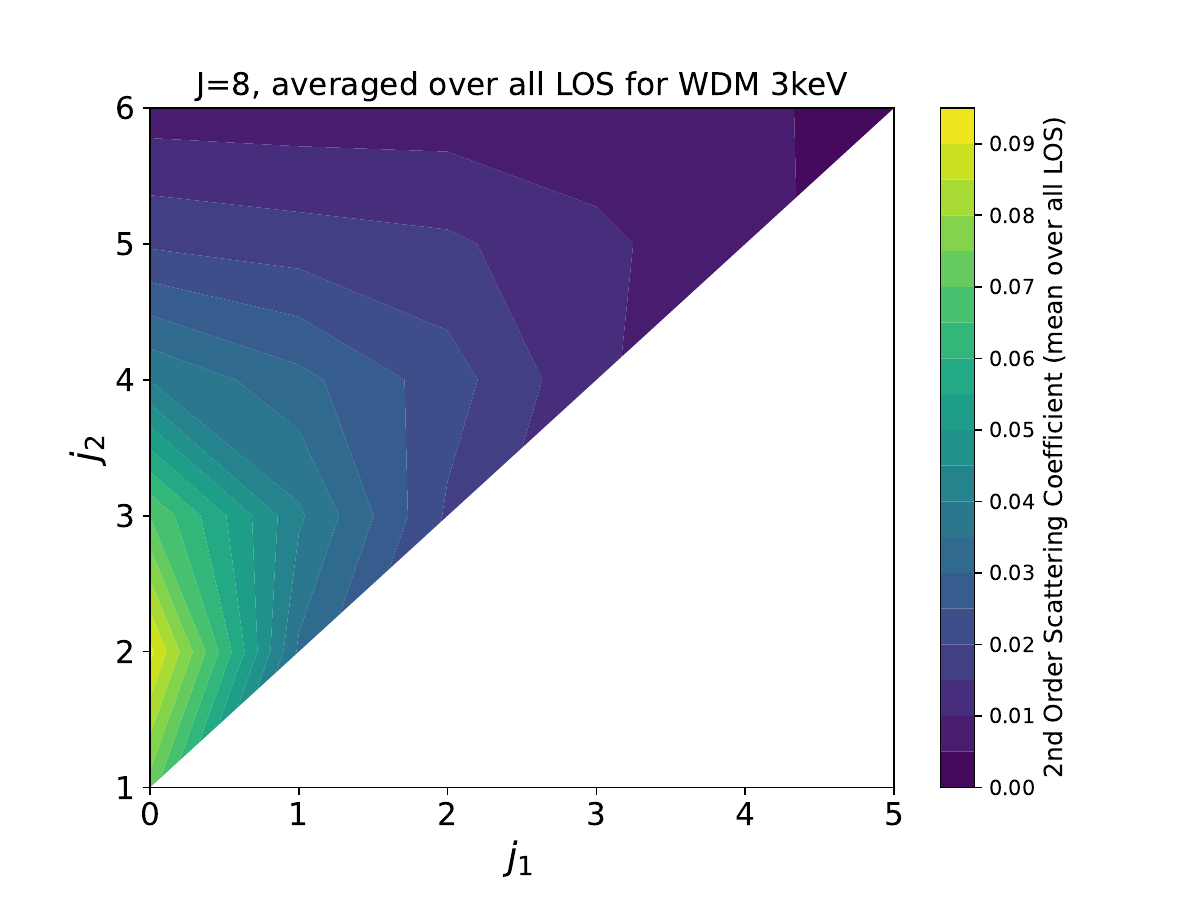}
\includegraphics[width=1.0\hsize]{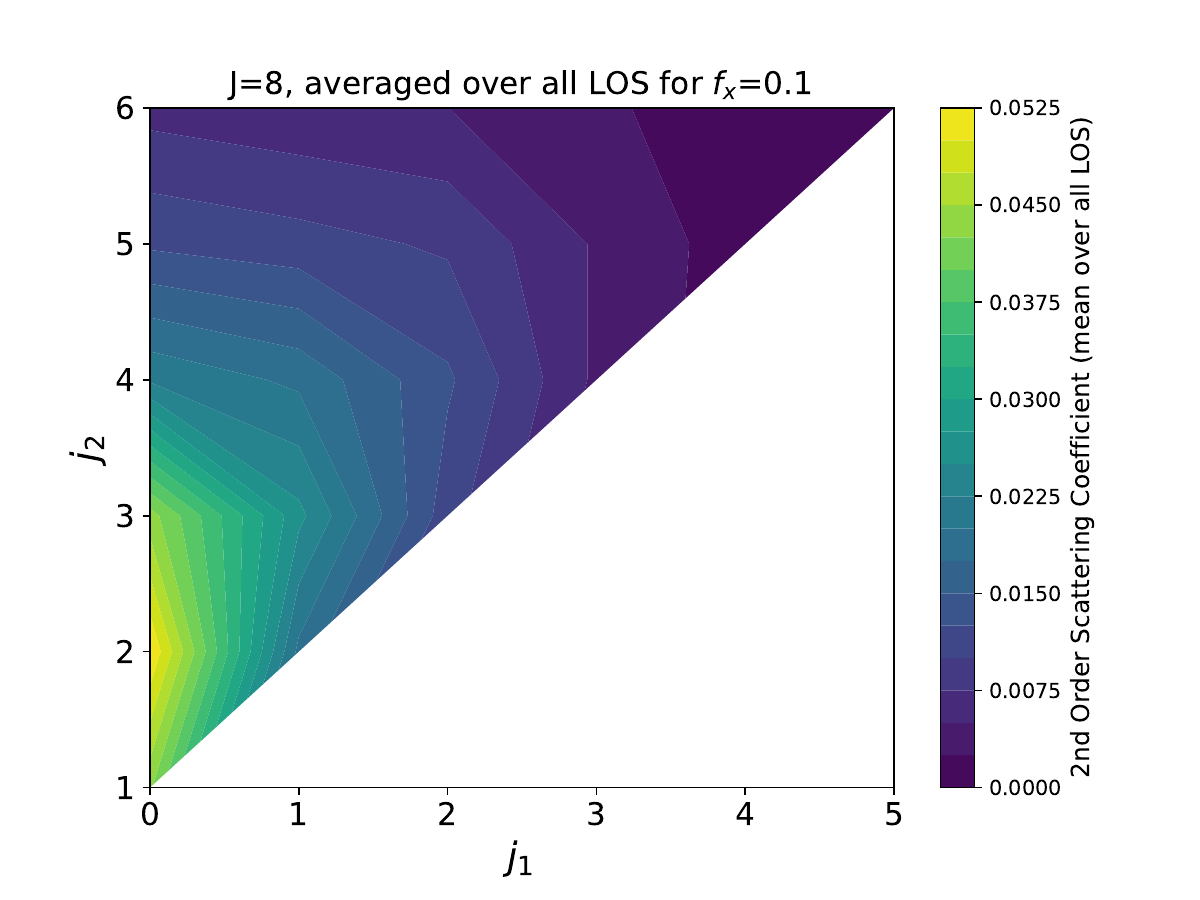}
\caption{Second-order WST coefficients \( S_2(j_1, j_2) \) for the CDM model (top), the 3~keV WDM scenario (middle), and the $f_X=0.1$ CDM scenario (bottom), each averaged over 10 lines of sight.}
\label{fig:wst_second_order}
\end{figure}

Figure~\ref{fig:wst_second_order} presents the second-order wavelet scattering coefficients, \(S_{2}(j_1,j_2)\), for three different scenarios---CDM (top), WDM at 3 keV (middle), and CDM with elevated X-ray heating (\(f_X=0.1\), bottom)---averaged over 10 lines of sight. Each panel plots the average scattering coefficient as a function of the scales \(j_1\) (horizontal axis) and \(j_2\) (vertical axis), with the color scale indicating the amplitude of the second-order coefficients. Note that we only show \(S_{2}(j_1,j_2)\) at \(j_2 > j_1\) because of symmetry.

In the CDM case (top panel), the maximum coefficient reaches \(\sim 0.24\), substantially higher than in the other two scenarios. This high amplitude over a wide range of \((j_1, j_2)\) highlights CDM's strong multi-scale coupling characteristic, where large- and small-scale features tend to overlap or reinforce one another. Such robust non-Gaussian interactions are reduced in the WDM (3 keV) case (middle panel), where the peak value drops to \(\sim 0.09\). Although deep absorption troughs still occur, the free-streaming effect in WDM suppresses small-scale structure, reducing the correlations between different scales and thus lowering the overall amplitude of \(S_{2}(j_1,j_2)\).

In the bottom panel, the CDM model with strong X-ray heating (\(f_X=0.1\)) shows a further reduction, with a maximum on the order of \(\sim 0.05\). Although we still find sharply localized features in the 21cm forest brightness temperature spectrum, the global multi-scale correlation is largely smoothed out by X-ray heating. This effect arises because, while the WDM model primarily suppresses small-scale fluctuations via free-streaming, strong X-ray heating homogenizes the IGM temperature over extended regions. Consequently, the brightness temperature field becomes more uniform, leading to a greater reduction in inter-scale correlations, particularly at larger scales. As a result, large-scale and small-scale fluctuations become less correlated, causing the second-order coefficients to be even smaller than in the WDM case. Although the differences between the WDM and \(f_X=0.1\) models are not overwhelmingly large in the noise-free case, the observed trend is consistent with the underlying physical mechanisms.

These outcomes collectively indicate how different dark matter scenarios and heating mechanisms leave distinct multi-scale non-Gaussian imprints on 21cm forest brightness temperature spectra. The CDM model generates rich structures across all scales, reflected in large \((j_1, j_2)\) coupling in second-order coefficients. Warm Dark Matter suppresses small-scale power, lowering the amplitude of second-order coefficients but not as dramatically as strong X-ray heating can. Meanwhile, strong X-ray heating flattens or homogenizes the 21cm forest spectrum, leaving less overlapping fluctuations and hence the smallest overall \(S_2(j_1, j_2)\) despite occasional strong absorption lines. These characteristics highlight the utility of second-order scattering coefficients in capturing subtle differences in multi-scale structure and non-Gaussianity.

In addition to examining the absolute values of the second-order scattering coefficients, we computed the ratios of \(S_2\) for the 3keV of WDM and \(f_X=0.1\) models relative to the CDM baseline, as shown by

\begin{equation}
\frac{S_2(j_1, j_2)\left(\text { 3 keV mass of WDM or } f_X=0.1 \text{with CDM}\right)}{S_2(j_1, j_2)(\mathrm{CDM})}.
\label{eq:ratio}
\end{equation}

\begin{figure}[h!]
    \centering
    \includegraphics[width=1.0\linewidth]{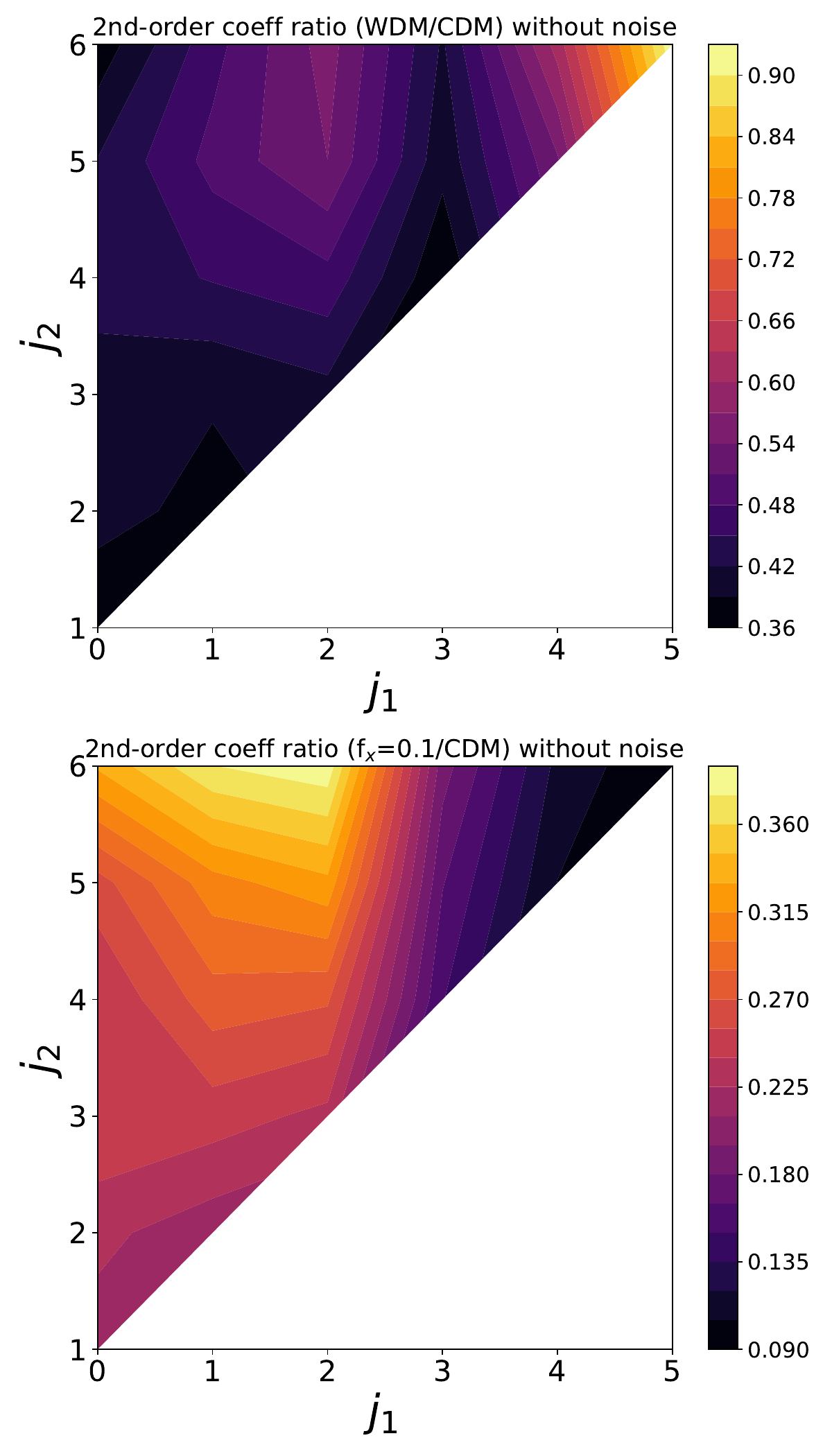}
    \caption{
        Two-dimensional maps of the second-order WST coefficient ratio relative to the CDM model in the noise-free case.
        \textbf{Top:} $S_2(\mathrm{WDM}) / S_2(\mathrm{CDM})$ for a 3~keV warm dark matter scenario.
        \textbf{Bottom:} $S_2(f_X = 0.1) / S_2(\mathrm{CDM})$.
        Both plots highlight how WDM and strong X-ray heating suppress multi-scale correlations compared to CDM.
    }
    \label{fig:ratio_no_noise}
\end{figure}

The results are shown in Fig.~\ref{fig:ratio_no_noise}. In this noise-free analysis, we isolate the pure effects of free-streaming and strong X-ray heating on multi-scale correlations. The top panel presents \(S_2(\mathrm{WDM})/S_2(\mathrm{CDM})\) for a 3~keV warm dark matter model, while the bottom panel shows \(S_2(f_X=0.1)/S_2(\mathrm{CDM})\).

Both ratios lie below unity across most of the \((j_1, j_2)\) plane, confirming that both WDM and strong X-ray heating reduce multi-scale correlations relative to CDM. Notably, in the strong X-ray heating case, the suppression is particularly pronounced at higher \(j_1\) and \(j_2\) values (i.e., on larger scales). This pronounced suppression at large scales is revealed by the second-order WST coefficients, which capture correlations between different scales—a key strength of the WST method. By probing these inter-scale correlations, the second-order coefficients provide detailed insights into the multi-scale structure of the 21cm forest that are difficult to obtain with more conventional techniques. The observed effect arises because strong X-ray heating homogenizes the IGM temperature over extended regions, thereby flattening the brightness temperature spectrum and diminishing the multi-scale correlations more significantly at larger scales. In contrast, the suppression observed in the WDM model is more moderate and mainly localized to smaller scales. Comparing these ratio maps with the absolute \(S_2\) maps in Fig.~\ref{fig:wst_second_order} clearly shows that the regions with the strongest second-order correlations in the CDM model are most significantly attenuated under both the WDM and \(f_X=0.1\) conditions.

Overall, Figs.~\ref{fig:wst_second_order} and \ref{fig:ratio_no_noise} together demonstrate that WDM and strong X-ray heating both weaken the second-order WST coefficients, albeit in somewhat different manners. While WDM reduces the amplitude of small-scale correlations through free-streaming, strong X-ray heating more dramatically flattens the brightness temperature field, especially on larger scales, leading to a lower overall \(S_2\) across a wide range of scales. This combined analysis of absolute \(S_2\) values and their ratios to the CDM baseline underscores the distinct physical signatures each scenario leaves on the 21cm forest.

\section{WST analysis with thermal noise}

Thermal noise is an unavoidable aspect of any realistic 21cm forest observation and can significantly affect both the brightness temperature spectra and the derived WST coefficients. In this section, we explore WST coefficients of 21cm forest spectrum under the condition that includes thermal noise. 

\subsection{Thermal Noise Formalism}

In the context of direct measurement of individual absorption lines, we consider the contribution of thermal noise to brightness temperature measurements. We follow the standard radiometer equation, specialized to interferometric or single-dish imaging, and employ the following key expression for the noise level per frequency bin

\begin{equation}
\delta T^{N} \;\approx\; 
\frac{\lambda^2 \,T_{\mathrm{sys}}}{A_{\mathrm{eff}}\,\Omega\,\sqrt{2\,\delta \nu \,\delta t}}
\,,
\label{eq:thermal-noise}
\end{equation}
where $\lambda = c / \nu$ is the (observed) wavelength, with $c$ being the speed of light and $\nu$ the observed frequency.  $T_{\mathrm{sys}}$ is the system temperature (in Kelvins), which includes receiver noise, sky temperature, etc. $A_{\mathrm{eff}}$ is the effective collecting area of the telescope or interferometer. $\Omega$ is the solid angle of the beam. For a diffraction-limited circular aperture of diameter $D$,  we approximate $\theta \approx 1.22\lambda/D$. From this, we can calculate the solid angle of the beam as $\Omega \approx \pi (\theta/ 2)^2$. $\delta\nu$ is the frequency resolution (bandwidth per bin). $\delta t$ is the total integration time. The factor $\sqrt{2}$ in the denominator arises from assuming thermal noise with two polarization states (or equivalently from the standard radiometer equation that includes $\sqrt{2 \delta\nu \delta t}$ in the noise term). 

In our calculations, we adopt the following representative values as a reference, motivated by SKA1-LOW design specifications\citep{2019arXiv191212699B}:

\begin{itemize}
  \item $A_{\mathrm{eff}} / T_{\mathrm{sys}} \,\approx\, 800\,\mathrm{m^2/K}.$ 
  \item We assume a maximum baseline or dish diameter of $D = 65\,\mathrm{km}$, which is a rough estimate for the core or extended configuration for SKA1-LOW.
  \item The integration time $\delta t$ is set to $100\,\mathrm{hours}$ ($= 3600 \times 100\,\mathrm{s}$).
  \item The frequency resolution $\delta \nu$ is taken to be $1\,\mathrm{kHz}$.
  \item Over each frequency bin $\nu_i$, we compute $\lambda_i = c/\nu_i$, and insert it into Eq.~(\ref{eq:thermal-noise}).
\end{itemize}

Hence, for each frequency bin $i$, the standard deviation of the thermal noise is computed as
\begin{equation}
\Delta T^N_i \;=\; \frac{\left(\frac{c}{\nu_i}\right)^2 \, T_{\mathrm{sys}}}{A_{\mathrm{eff}} \pi \Bigl(\frac{1.22\,\frac{c}{\nu_i}}{2\,D}\Bigr)^2 \sqrt{2\,\delta\nu \delta t}}.
\end{equation}

This noise is then treated as a Gaussian random variable $N_i \sim \mathcal{N}(0,\,(\Delta T^N_i)^2)$, and is added to the original 21cm brightness temperature signal $T_{\mathrm{21}}(\nu_i)$ to obtain the noisy measurement:
\begin{equation}
T_{\mathrm{obs}}(\nu_i) \;=\; T_{\mathrm{21}}(\nu_i) \;+\; N_i.
\label{eq:noisy_spec}
\end{equation}
These noisy measurements are then used as input for the 
WST analysis.

\subsection{Noisy spectra and wavelet scattering transform coefficients}

\begin{figure}
    \centering
    \includegraphics[width=1.0\hsize]{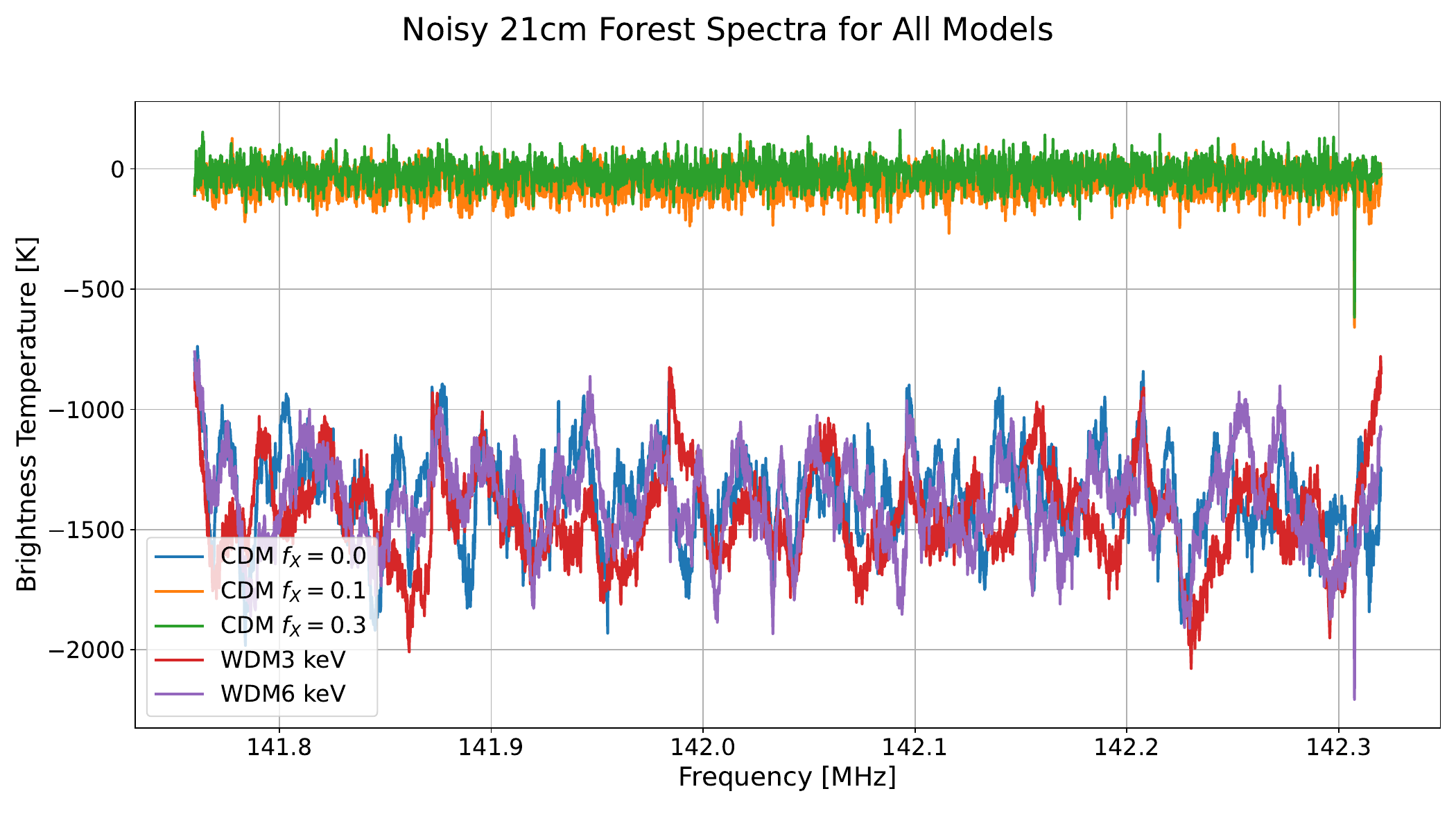}
    \caption{21cm forest brightness temperature spectrum including thermal noise in the CDM model for $f_X$=0, 0.1, and 0.3, and in WDM models with particle masses of 3 keV and 6 keV for $f_X$=0, respectively.}
\label{fig:spectrum_w_noise}
\end{figure}

In Fig.\ref{fig:spectrum_w_noise}, we plot a noisy 21cm forest brightness temperature spectrum for one line of sight. When introducing thermal noise, random fluctuations overlay the intrinsic noiseless spectral shapes, reducing the signal-to-noise ratio (SNR). As a result, previously distinct absorption troughs and peaks become blurred, making small-scale features difficult to precisely identify; noise partially obscures the fundamental differences among CDM, WDM, and heated CDM—especially those on small scales that are lost in random fluctuations—and, because thermal noise is modeled as white noise affecting all frequencies roughly equally, the overall spectrum appears more homogeneous.

When random thermal noise, as described by Eq.~\ref{eq:noisy_spec}, is added to the 21cm forest signal, the observed spectrum becomes the sum of the original signal and the noise. The first-order WST coefficients, \(S^{(1)}_j\), are calculated through a nonlinear operation—taking the modulus of the wavelet transform followed by averaging—so they do not simply add linearly. However, for a signal \(x\) and independent noise \(N\), we can approximate
\begin{equation}
S^{(1)}_j(x+N) \approx S^{(1)}_j(x) + S^{(1)}_j(N) + \text{(cross terms)},
\end{equation}
where the cross terms vanish if the noise and the signal are uncorrelated.

\begin{figure}[htbp]
\centering
\includegraphics[width=1.0\hsize]{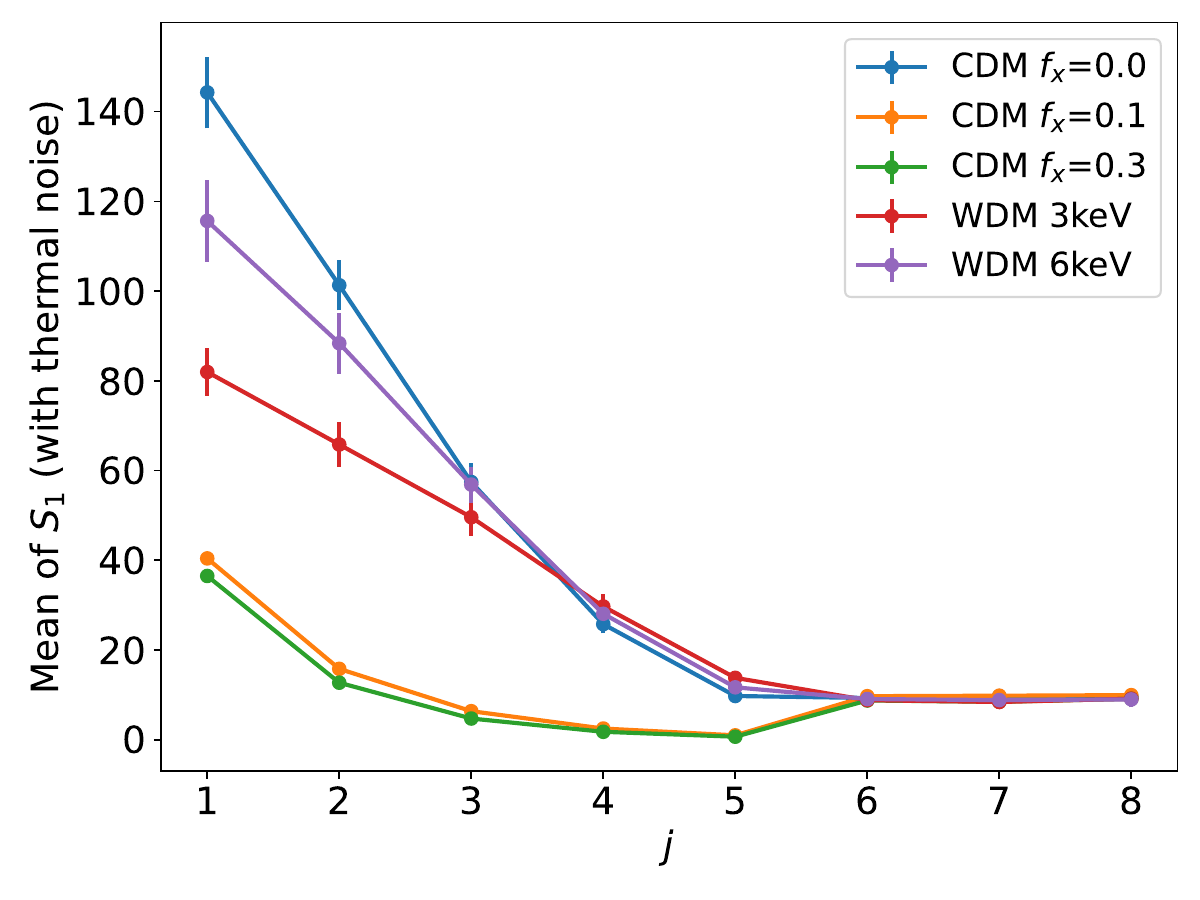}
\caption{Same with Fig.\ref{fig:wst_first_order}. But, these first-order WST coefficients are calculated from 21cm forest brightness temperature with thermal noise.
}
\label{fig:wst_first_order_w_noise}
\end{figure}

In practice, the coefficients \(S^{(1)}_j(N)\) for thermal noise are nearly constant across all scales \(j\), introducing an approximately constant offset to the overall first-order coefficients. This behavior is illustrated in Fig.~\ref{fig:wst_first_order_w_noise}, which shows that the presence of thermal noise results in a nearly scale-independent upward shift in the first-order WST coefficients compared to the case without noise.

We can therefore decompose the total first‐order coefficient into signal and noise contributions,

\begin{equation}
      S^{(1)}_j(x') = S_{\rm forest}(j) + S_{\rm noise}(j),
\end{equation}
where \(S_{\rm noise}(j)\) arises from white thermal noise that has been smoothed by a 1kHz moving average. This smoothing is equivalent to a low‐pass filter whose amplitude response exhibits a local minimum at intermediate scales (\(J\approx3\)–5). As both the decaying forest signal \(S_{\rm forest}(j)\) and the smoothed noise term \(S_{\rm noise}(j)\) dip in this region, their sum produces the pronounced valley in \(S^{(1)}(j)\). At larger scales (\(J\ge6\)), the slowly-oscillating tail of the noise spectrum dominates, causing \(S^{(1)}(j)\) to rise again and converge toward noise‐dominated values. This behavior contrasts with Fourier power‐spectrum analyses, where truly white noise contributes equally at all wavenumbers; in the WST, each Morlet wavelet filters a specific band, leading to scale‐dependent noise bias.

The inherent robustness of the WST to thermal noise arises from the localization of the wavelet transform in frequency and the scattering procedure (see Appendix \ref{sec:robustness}). Specifically, by taking the absolute value of the wavelet coefficients and then averaging, the process effectively smooths out the random fluctuations introduced by the noise. However, because the small-scale features of the 21cm forest signal—namely, the slight variations between absorption troughs and peaks—are relatively weak (especially in models with enhanced X-ray heating that smooth out these details), the addition of thermal noise further masks these minor differences, making it more difficult to distinguish between models compared to the noiseless case. Nevertheless, when sample variance is taken into account, the WST coefficients still keep sufficient discriminatory potential to differentiate among the models.


\begin{figure}
    \centering
    \includegraphics[width=1.0\linewidth]{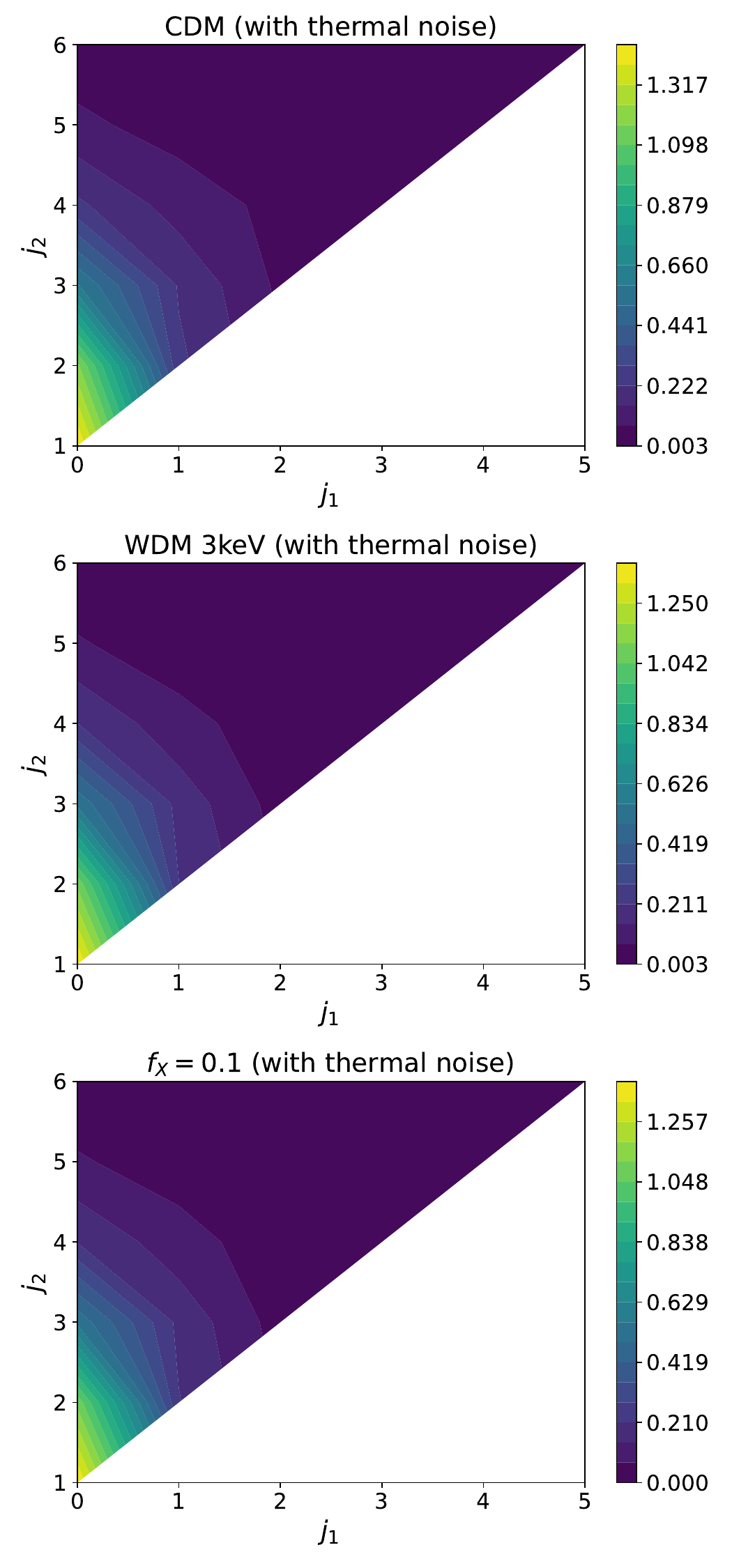}
    \caption{Same as Fig.~\ref{fig:wst_second_order} but including thermal noise.}
    \label{fig:2nd_w_noise}
\end{figure}

We next examine the second-order WST coefficients computed from the simulated 21cm forest brightness temperature spectra, including thermal noise. Fig.~\ref{fig:2nd_w_noise} shows the second-order WST coefficients \(S_2(j_1,j_2)\) averaged over 10 lines of sight for the CDM (top), WDM 3keV (middle), and strong X-ray heating (\(f_X=0.1\); bottom) models. When compared to the noise-free results shown previously (Fig.~\ref{fig:wst_second_order}), thermal noise tends to increase the apparent amplitude of \(S_2\) in certain regions, especially where the intrinsic 21cm signal is weaker. This occurs due to the non-linear modulus operation of the WST, which introduces a positive bias from the noise contribution, complicating the direct interpretation of absolute values.

We further note that second-order coefficients involve a second convolution and modulus step, making them more sensitive to any residual noise carried through from the first-order stage.  To enhance their robustness under realistic thermal noise, we expect to subtract an analytically estimated noise bias for each \((j_1,j_2)\) pair or apply a small soft-threshold to the first-order modulus outputs prior to computing the second-order coefficients.  These simple denoising strategies can substantially improve the stability of \(S_2\) and thus better isolate the non-Gaussian, multi-scale correlations induced by dark matter free-streaming and X-ray heating.

\begin{figure}[htbp]
\centering
\includegraphics[width=1.0\hsize]{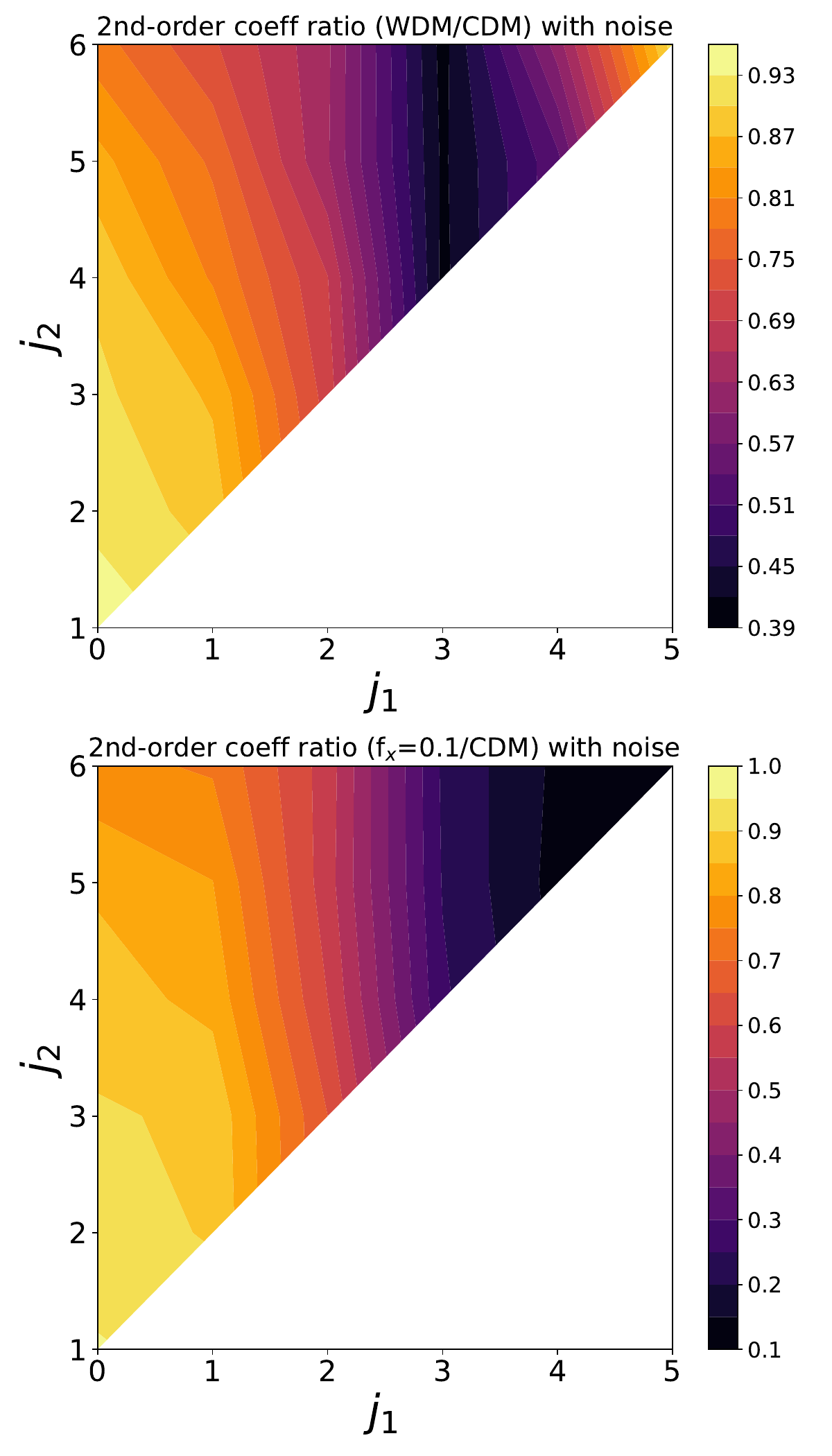}
\caption{\textbf{Top:} the ratio of second-order WST coefficient between 3 keV mass of WDM case and CDM case. \textbf{Bottom:} the ratio of second-order WST coefficient between \(f_X=0.1\) case and CDM case.}
\label{fig:ratio}
\end{figure}

To better understand how physical processes such as free-streaming and X-ray heating influence the multi-scale correlations in the presence of thermal noise, we also present the ratios of the second-order WST coefficients relative to the CDM baseline in Fig.~\ref{fig:ratio}, defined as equation (\ref{eq:ratio}).

Across most of the \((j_1,j_2)\) plane, these ratios remain clearly below unity for both the WDM and \(f_X=0.1\) models. For the WDM scenario, in the noise-free case (Fig.~\ref{fig:ratio_no_noise}), the WDM scenario exhibits substantial suppression at smaller scales (\(j_1,j_2 \lesssim 3\)), reflecting the free-streaming effect. However, once thermal noise is included, this small-scale suppression appears less pronounced, and the ratio in small-scale regions moves closer to unity (ranging from about 0.39 to 0.93, top panel of Fig.~\ref{fig:ratio}). This apparent reduction in suppression at smaller scales occurs because thermal noise disproportionately increases the amplitude of the originally weaker small-scale signals in the WDM scenario, thereby artificially elevating the \(S_2\) values. Conversely, at larger scales (\(j_1,j_2 \gtrsim 3\)), the noise effect is relatively uniform, preserving the stronger intrinsic suppression observed in the WDM model relative to CDM. For the \(f_X=0.1\) scenario, the intrinsic effect of strong X-ray heating is to uniformly suppress brightness temperature fluctuations across all scales, resulting in a homogenized IGM temperature distribution. In the noise-free case (bottom panel of Fig.~\ref{fig:ratio_no_noise}), this suppression is particularly pronounced at large scales, causing the second-order WST coefficients—and thus the ratios relative to CDM—to be especially small in these regions. However, when thermal noise is introduced, the small-scale regions experience a relatively larger apparent increase in amplitude due to the non-linear modulus operation within the WST analysis. As a consequence, the ratios at smaller scales shift closer to unity, making the suppression appear weaker compared to the noise-free scenario. Meanwhile, at larger scales, where the intrinsic signal is even weaker, thermal noise has a relatively smaller effect on the ratio, preserving the strong suppression originally observed. Thus, despite X-ray heating acting uniformly across scales, the scale-dependent impact of thermal noise results in an apparent reduction of suppression at smaller scales and emphasizes the suppression at larger scales relative to CDM.

Thus, despite the presence of thermal noise, the second-order WST coefficients and their ratios to the CDM baseline remain effective in distinguishing between scenarios. The differences observed in the ratio maps clearly reflect the distinct multi-scale correlation structures induced by free-streaming in the WDM model and temperature homogenization in the X-ray heating model. These results demonstrate that the second-order WST coefficients provide robust insights into the underlying physical processes shaping the 21cm forest, even under realistic observational conditions.

\subsection{Fisher forecast}

We next perform a Fisher forecast to estimate parameter constraints from the Wavelet Scattering Transform applied to the 21cm forest brightness temperature spectrum. Specifically, we use both first- and second-order WST coefficients, \(\mathbf{S}\), whose dependence on the model parameters \(\theta_i\) (As parameters, we choose \(f_X\) and \(m_\text{WDM}\) ) is described by the derivatives \(\partial \mathbf{S} / \partial \theta_i\). Assuming a Gaussian likelihood in the WST coefficients, the Fisher matrix elements \(F_{ij}\) are given by
\begin{equation}
\mathbf{F}_{i j} \equiv-\left\langle\frac{\partial^2 \ln (\mathcal{L})}{\partial \theta_i \partial \theta_j}\right\rangle=\sum \frac{\partial^2}{\partial \theta_i \partial \theta_j} \mathbf{S}^{\boldsymbol{\top}} \cdot \mathbf{\Sigma}^{-1} \cdot \mathbf{S},
\end{equation}
where \(\Sigma\) is the total covariance matrix. 

To account for cosmic variance, we generate ten independent realizations of the 21cm forest signal and compute the covariance of the WST coefficients across these realizations. We include thermal noise at levels consistent with SKA sensitivity, adding it in quadrature to the intrinsic scatter from the cosmic variance. The resulting covariance matrix, \(\Sigma\), thus incorporates cosmic variance and thermal noise. We numerically evaluate the partial derivatives \(\partial \mathbf{S} / \partial \theta_i\) around the fiducial parameters, and subsequently invert the Fisher matrix \(F_{ij}\) to obtain forecasted uncertainties. The diagonal elements of the inverse of the Fisher matrix represent the standard deviation (1$\sigma$ error) for parameter \(\theta_i\) given by:
\begin{equation}
\sigma_i=\sqrt{\left(F^{-1}\right)_{i i}}.
\end{equation}

\begin{figure}[htbp]
\centering
\includegraphics[width=1.0\hsize]{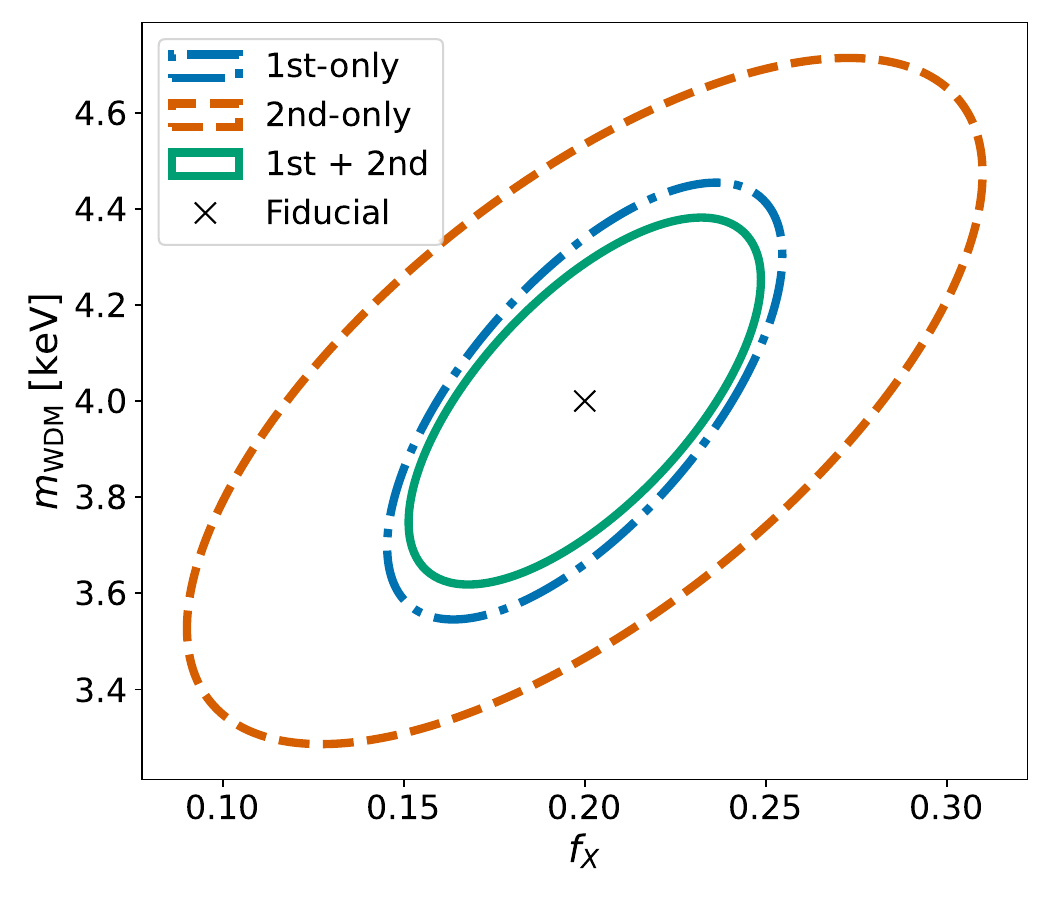}

\caption{Fisher forecast contours for the parameters \(f_X\)and \(m_{\text{WDM}}\), showing the 95 \% confidence level regions for different combinations of WST coefficients. The blue dashed contours represent the constraint from the first-order WST coefficients only, the orange dashed-dotted contours represent the second-order WST coefficients only, and the green solid contours represent the combined constraint from both the first- and second-order coefficients. The fiducial values are marked by the cross symbol.}
\label{fig:fisher_tot}
\end{figure}

Fig.\ref{fig:fisher_tot} shows the 95 $\%$ confidence ellipses for the parameters $f_X$ and $m_{\mathrm{WDM}}$.  The blue dashed curve (1st-only) and the orange dashed curve (2nd-only) each represent analyses based on 1st and 2nd WST coefficients, respectively, while the green solid curve (1st + 2nd) shows the result of combining both WST coefficients. The ``Fiducial'' mark represents the fiducial value of parameters (\(f_X, m_{\text{WDM}}\))=(0.2, 4.0 \text{keV}). Note that although the \(S_1\) values appear very similar for models with higher \(f_X\), strong X-ray heating significantly smooths out temperature fluctuations in the IGM. This smoothing results in a nearly uniform 21cm brightness temperature spectrum, leading to small variations across different lines of sight and, consequently, small error bars on the \(S_1\) measurements. Because these error bars are small, even subtle differences in the \(S_1\) values become statistically meaningful, allowing us to tightly constrain \(f_X\) using only the first-order statistic.


We can see that the error ellipse shrinks when the 1st and 2nd WST coefficients are combined, implying a tighter constraint on both parameters. For instance, the standard deviations of $f_X$ decrease from approximately $\sigma(f_X)_\mathrm{1st} = 0.0273$ (1st-only) and $\sigma(f_X)_\mathrm{2nd} = 0.0460$ (2nd-only) to $\sigma(f_X)_\mathrm{tot} = 0.0229$ in the combined analysis. Similarly, the uncertainties in $m_{\mathrm{WDM}}$ improve from $\sigma(m_{\mathrm{WDM}})_\mathrm{1st} \simeq 0.2273~\mathrm{keV}$ and $\sigma(m_{\mathrm{WDM}})_\mathrm{2nd} \simeq 0.2731~\mathrm{keV}$ to $\sigma(m_{\mathrm{WDM}})_\mathrm{tot} \simeq 0.1747~\mathrm{keV}$. 

We can interpret the inclination of the ellipse as follows. The Fisher contour can be interpreted as an approximate isocontour for maintaining a certain level of the WST coefficients. If \(f_X\) becomes larger, the amplitude of the WST coefficients becomes smaller (See Figs. \ref{fig:wst_first_order} and \ref{fig:wst_second_order}). However, simultaneously increasing \(m_{\text{WDM}}\) can offset the effect of raising \(f_X\) effects, thus preserving a particular first-order coefficient level. Consequently, (\(f_X\), \(m_{\text{WDM}}\)) exhibits a positive correlation, manifested as an upward-right inclination of the Fisher ellipse.



\section{Summary \& Discussion}

The 21cm forest, a series of absorption features in radio spectra caused by neutral hydrogen along cosmological sightlines, provides valuable insights into the intergalactic medium (IGM) and the underlying dark matter structure. Traditional power spectrum analyses, which focus on second-order statistical moments, are limited in their ability to characterize the non-Gaussian and multi-scale nature of these signals. To address this limitation, this study applies the Wavelet Scattering Transform (WST) to simulated 21cm forest spectra, enabling the extraction of higher-order statistical information beyond what the power spectrum can capture.

In our analysis, the first-order WST coefficients demonstrate a clear distinction between different cosmological models. For the Cold Dark Matter (CDM) scenario, the first-order coefficients exhibit the strongest variations, as CDM preserves small-scale structures in the 21cm forest. In contrast, the Warm Dark Matter (WDM) scenario suppresses small-scale fluctuations due to the free-streaming effect, leading to smaller first-order coefficients. Similarly, strong X-ray heating flattens the spectrum, resulting in a reduction in the first-order coefficients. The second-order WST coefficients, which capture multi-scale correlations, further emphasize the differences between models. In the CDM scenario, the second-order coefficients are largest, reflecting the strong non-Gaussian interactions between large- and small-scale features. In the WDM scenario, these correlations are weaker, with a notable decrease in the amplitude of the second-order coefficients. The X-ray heating scenario also results in reduced multi-scale correlations, as the heating smooths out the small-scale variations, leading to smaller second-order coefficients.

To quantify the precision of our parameter constraints, we performed a Fisher forecast using both first- and second-order WST coefficients. Our results indicate that combining both coefficients provides tighter constraints on key parameters such as the X-ray heating efficiency and the WDM particle mass. This improvement highlights the power of the WST in providing more accurate parameter estimates compared to traditional methods.

These findings show the potential of WST as a powerful diagnostic tool for extracting astrophysical and cosmological information from the 21cm forest. By providing a more comprehensive statistical characterization of the 21cm absorption spectrum, WST offers new avenues for probing the nature of dark matter and the thermal history of the IGM, further advancing our understanding of the high-redshift universe.

Here, we continue the discussion. When we compare the Fisher forecast using the first-order WST coefficients with the 21cm forest power spectrum discussed in \citet{2023NatAs...7.1116S}, we find that the use of the first-order WST coefficients provides more stringent restrictions on the parameters. Quantitatively, the power spectrum yields \(\sigma(f_X)=0.0447\) and \(\sigma(m_{\rm WDM})=0.7784\ \mathrm{keV}\), whereas the first-order WST forecast achieves \(\sigma(f_X)_{\rm 1st}=0.0273\) and \(\sigma(m_{\rm WDM})_{\rm 1st}=0.2273\ \mathrm{keV}\). The primary reason that the Fisher analysis based on first-order WST coefficients provides tighter parameter constraints than the standard 21cm forest power spectrum is that WST captures multi-scale and non-Gaussian features that the power spectrum alone cannot. While the power spectrum is limited to second-order statistics and thus less sensitive to higher-order correlations or local structures, WST decomposes the signal into multiple frequency scales and applies nonlinear operations, thereby encoding richer information, including some sensitivity to higher-order statistics. As a result, the WST coefficients tend to correlate more strongly with underlying model parameters and help break degeneracies that remain when only second-order information is used. Moreover, WST is inherently stable to small perturbations and noise, yielding more robust features for parameter estimation. Consequently, Fisher matrices constructed from WST coefficients yield significantly tighter constraints on cosmological and astrophysical parameters compared to those derived from the power spectrum alone.

Although second‑order WST coefficients encode non‑Gaussian information, the constraints obtained from them alone are markedly weaker than those derived from first‑order statistics. However powerful the Fisher‐matrix approach may be for Gaussian statistics, it fundamentally relies on assuming a multivariate Gaussian likelihood for the data summaries—an assumption that fails when applied to non‑Gaussian quantities such as the second‑order WST coefficients. In particular, the Fisher matrix formalism relies on approximating the data likelihood as a multivariate Gaussian in the chosen summary statistics \(\mathbf{d}\) for theoretical parameters $\boldsymbol\theta$:
\[
\mathcal{L}(\mathbf{d}\mid\boldsymbol\theta)\propto
\exp\Bigl[-\tfrac12\bigl(\mathbf{d}-\bar{\mathbf{d}}(\boldsymbol\theta)\bigr)^T
\mathbf{C}^{-1}
\bigl(\mathbf{d}-\bar{\mathbf{d}}(\boldsymbol\theta)\bigr)\Bigr],
\]
where \(\mathbf{C}\) is the covariance matrix.  This Gaussian approximation discards all higher‑order moments—skewness, kurtosis, and heavy tails—that are intrinsic to the true distribution of the second‑order WST coefficients \(S_2(j_1,j_2)\). As a result, it underestimates variance from heavy tails and nonlinear noise propagation through the modulus, and throws away non‑Gaussian mode couplings that carry real information. This mismatch leads the Fisher errors on \(S_2\) alone to appear artificially large and the resulting parameter bounds weaker than they could be. To overcome these limitations and fully exploit the non‑Gaussian information content of the WST, we must instead turn to likelihood‑free (or simulation‑based) inference methods—such as Approximate Bayesian Computation or neural density estimators—which make no Gaussian approximation and can therefore capture the true distribution of the scattering coefficients\citep[e.g.][]{2022ApJ...926..151Z,2022ApJ...933..236Z,2025CmPhy...8..220S}. But we leave it as a future work.


Second-order WST coefficients are actually very powerful for studying the non-Gaussian nature of the 21cm signal. Compared to the first-order WST coefficients, they are able to capture more complicated scale interactions, which can be very important to understanding the detailed physics of reionization. Previous works have emphasized that non-Gaussian statistics often contain unique information, and second-order WST can provide new insights not fully seen by the power spectrum or first-order WST terms \citep[e.g.][]{2016MNRAS.458.3003S,2017MNRAS.468.1542S,2017MNRAS.472.2436W,2018MNRAS.476.4007M,2020MNRAS.492..653H,2024ApJ...974..141D}. One of the reasons why the second-order WST is attractive is that it naturally captures mode couplings that arise from nonlinear processes. Although some studies show that non-Gaussian statistics can increase error bars if the covariance is properly accounted for \citep{2019MNRAS.487.4951S}, this also suggests that these statistics are sensitive to more complex structures in the 21cm field. In that sense, second-order WST coefficients are not just repeating first-order information; instead, they reveal additional correlations that can improve our understanding of reionization if used carefully \citep{2023mla..confE..30S}.

Furthermore, even when non-Gaussian statistics do not always provide large improvements in parameter constraints, they can still complement existing statistics by breaking certain degeneracies. For example, \citet{2024ApJ...974..141D} shows that Minkowski functionals offer extra information compared to the power spectrum alone. Although the parameter improvement was around 30\%, it still indicates that non-Gaussian methods can add value. Similarly, second-order WST coefficients might not always yield dramatic gains, but they often deliver independent information that can be combined with other statistics.

It is also the case that observational challenges, such as foreground contamination, may affect second-order WST more strongly \citep{2022MNRAS.510.3838W,2024A&A...688A.199P}. Nevertheless, if we can develop better ways to handle observational limitations, the non-Gaussian nature captured by second-order WST can become more significant. Recent works confirm that scattering transform methods can outperform simpler approaches for certain parameter regimes \citep{2024A&A...688A.199P}. This implies that, with careful application and robust noise treatment, second-order WST coefficients have the potential to greatly enhance our analysis of the 21cm reionization signal.

While second-order WST coefficients do sometimes correlate with first-order coefficients and can be sensitive to observational limitations, they also contain valuable non-Gaussian information that is otherwise missed. Therefore, it remains an important avenue to investigate how best to use second-order WST coefficients alongside other statistics. With more advanced analysis techniques and improved data, second-order WST could significantly boost our ability to extract cosmological and astrophysical insights from the 21cm signal.


Recent discoveries of radio-loud quasars and blazars at high redshifts (with some sources now reaching $z \gtrsim 6$) have dramatically expanded the opportunities to detect the 21cm forest---absorption features arising from intervening neutral hydrogen in the IGM\citep[e.g.,][]{2020A&A...635L...7B,2021ApJ...909...80B,2022A&A...664A..39K,2022A&A...668A..27G,2024ApJ...977L..46B,2024NatAs.tmp..293B}. These bright radio sources serve as critical background illuminators against which the 21cm absorption can be measured. Although current observations with facilities such as the GMRT and LOFAR have not yet yielded definitive detections of 21cm absorption features, the growing number of high-$z$ radio sources paves the way for future, more sensitive surveys that will be carried out with the SKA, ultimately probing the IGM and the reionization process more directly.

A key metric in assessing the detectability of the 21cm forest is the minimum background flux density, $S_{\min}$, needed to achieve a targeted signal-to-noise ratio. In the expression for $S_{\min}$, we often encounter the term $F_{21,\text{th}} = e^{-\tau_{21}}$, which describes the transmitted fraction of the continuum flux in the presence of an optical depth $\tau_{21}$. Consequently, the factor $(1 - F_{21,\text{th}})$ represents the depth of the 21cm absorption line. The minimum background flux density can be estimated by \citep{2021MNRAS.506.5818S,2023MNRAS.519.3027S}
\begin{eqnarray}
S_{\min} &=& 38.5\mathrm{mJy}
\left(\frac{0.01}{1 - F_{21,\text{th}}}\right)
\left(\frac{\mathrm{S/N}}{5}\right)^{1/2}\\
&& \times\left(\frac{1\,\mathrm{kHz}}{\Delta \nu}\right)^{1/2}\;
\left(\frac{1000\mathrm{h}}{t_{\mathrm{int}}}\right)^{1/2}\;
\left(\frac{600\mathrm{m^2K^{-1}}}{A_{\mathrm{eff}}/T_{\mathrm{sys}}}\right)^{1/2},
\label{eq:Smin}
\end{eqnarray}
where $T_{\mathrm{sys}}$ is the system temperature, $\Delta \nu$ is the channel bandwidth, $A_{\mathrm{eff}}$ is the effective telescope area, and $t_{\mathrm{int}}$ is the integration time. For SKA1-low sensitivity ($A_{\mathrm{eff}}/T_{\mathrm{sys}} \simeq 600\mathrm{m^2\,K^{-1}}$:see \citep{2019arXiv191212699B}), an integration time of $t_{\mathrm{int}} = 1000$h, and $\mathrm{S/N} = 5$, we obtain $S_{\min} \approx 38.5$mJy. Reducing the integration time to 100h increases $S_{\min}$ to $\sim 121.6$mJy. For SKA2 sensitivity ($5500\mathrm{m^2\,K^{-1}}$:see \citep{2019arXiv191212699B}), even if $t_{\mathrm{int}} = 100$h with $\mathrm{S/N} = 5$, $S_{\min} \approx 12.7$mJy.

In practice, known high-redshift sources exhibit a range of observed flux densities. For instance, the quasar PSO\,J172$+18$ has a 3$\sigma$ upper limit of $S_{147.5\,\mathrm{MHz}} < 8.5$mJy \citep{2021ApJ...909...80B}, whereas the bright radio-loud blazar PSO\,J0309$+27$ at $z = 6.1$ has $S_{147.5\,\mathrm{MHz}} = 64.2 \pm 6.2$mJy \citep{2020A&A...635L...7B}. Both objects thus remain viable targets for attempts to detect 21cm absorption from neutral hydrogen along their lines of sight, although the feasibility depends strongly on the combination of integration time, spectral resolution, and telescope sensitivity as reflected in Eq.~\ref{eq:Smin}. As next-generation radio facilities achieve lower $S_{\min}$ thresholds, detecting the 21cm forest from diffuse IGM gas at $z \gtrsim 6$ will become increasingly feasible, shedding new light on the 21cm forest study.

Recent studies have shown that the detectability of the 21cm forest depends critically on the spectral resolution of the observational data. In our work, we preprocess the 21cm forest spectral data by applying a smoothing step to match a target resolution of 1\,kHz. Although the raw data are sampled at an effective resolution of about 0.22\,kHz, smoothing the data to 1\,kHz effectively filters out rapidly-oscillating noise while preserving the signal’s intrinsic local features. In the case of the first-order coefficients \(S_1\), our results indicate that smoothing to 2\,kHz (i.e., applying more aggressive averaging) slightly attenuates fine-scale rapidly-oscillating fluctuations, resulting in modestly lower \(S_1\) values in the rapidly-oscillating (small-scale) regime compared to 1\,kHz smoothing. Nevertheless, the overall trends across different scale parameters \(j\) remain robust. Similarly, the second-order scattering coefficients \(S_2\), which capture inter-scale correlations and thus provide insight into the signal’s non-linear, multi-scale structure, show only minor quantitative differences between the 1\,kHz and 2\,kHz smoothing settings. While the more aggressive smoothing tends to further diminish fine-scale local interactions, the overall pattern of multi-scale correlations is well preserved regardless of the smoothing resolution applied. These findings demonstrate that, although the absolute values of both \(S_1\) and \(S_2\) may vary modestly with different smoothing resolutions, the essential multi-scale structural information extracted by the WST remains stable. Therefore, by carefully matching the smoothing resolution to the effective observational resolution, our methodology reliably filters out rapidly-oscillating noise while preserving the local features of the 21 cm forest spectrum, as captured by both the first-order and second-order scattering coefficients.

From a theoretical standpoint, previous work indicates that the necessary high-redshift radio sources for 21cm forest observations may indeed exist\citep{2025ApJ...978..145N}. Their analyses of high-redshift quasar luminosity functions suggest that, under the assumption of a roughly constant radio-loud fraction of $\sim10\%$,  $\sim 20$ of radio-loud quasars could persist even at redshifts $z>9$. If sufficiently luminous, these quasars would provide viable background sources for detecting 21cm absorption features. However, uncertainties remain. If the radio-loud fraction evolves with redshift and decreases at higher $z$, the abundance of these sources may be significantly reduced, complicating individual 21cm line detections. While challenges exist in confirming the exact abundance of high-redshift radio sources, previous work supports the possibility that a non-negligible population is present. These results provide a promising basis for future 21cm forest studies, especially when combined with multi-wavelength survey strategies.

Finally, we also should mention the effects of radio frequency interference (RFI). In this study, we have focused on idealized 21cm forest spectra with only thermal noise and have not considered the RFI. However, in real observations, RFI is unavoidable and leads to the masking of certain frequency channels, resulting in gaps in the data. This poses a significant challenge for the analysis of 21cm forest signals, especially for methods like the WST that rely on local convolutions. The presence of masked channels can introduce biases or increased uncertainty in the estimated scattering coefficients, particularly near the edges of masked regions. Nevertheless, because the WST is intrinsically a local, multi-scale technique, we expect that it is possible to mitigate RFI effects by excluding masked regions from the convolution, down-weighting or expanding the uncertainty for coefficients computed near gaps, or applying gap-filling and interpolation techniques prior to the analysis. Developing and validating robust WST-based inference pipelines that account for realistic RFI masking will be an important direction for future work.

\section*{Acknowledgements}

HS appreciates Tsutomu.T. Takeuch's for insightful comments. This work is supported by the National SKA Program of China (No.2020SKA0110401), NSFC (Grant No.~12103044), and Yunnan Provincial Key Laboratory of Survey Science with project No. 202449CE340002.
Y. Xu acknowledges the support from the National Key R\&D Program of China No. 2022YFF0504300.

\appendix

\section{Comparison of Fourier and Wavelet Scattering Transform Representations}
\label{sec:local}

Note that the 21cm forest brightness temperature spectrum, \(\delta T_b(\nu)\), is analyzed as a one-dimensional signal in the frequency domain (i.e., as a function of observing frequency) rather than in the time domain. In the following Appendices, we describe the general properties of the scattering transform using time-stream data only for pedagogical illustration; our main results use frequency-domain spectra.

In Fourier analysis, a signal \(x(t)\) is decomposed into globally supported sinusoids via the transform
\[
\hat{x}(k) = \int_{-\infty}^{\infty} x(t)\, e^{-i k t}\, dt,
\]
and the power spectrum is defined as
\[
P(k) = |\hat{x}(k)|^2.
\]
This function describes how the signal’s energy is distributed over wavenumber \(k\).  While \(P(k)\) itself retains scale-dependent peaks and troughs—such as the acoustic peaks of the CMB or BAO wiggles—collapsing it to a single scalar
\[
\int_{-\infty}^{\infty} P(k)\, dk
\]
eliminates all scale-specific information, yielding only the total variance.  Thus, although \(P(k)\) contains rich cosmological information when viewed as a function of \(k\), reducing it to its integral discards the “where” and “which scale” details entirely.

In contrast, the Wavelet Scattering Transform (WST) employs localized wavelets.  For a given mother wavelet \(\psi(t)\), the scaled and shifted wavelet is defined as
\[
\psi_\lambda(t) = \frac{1}{\sqrt{\lambda}}\,\psi\!\bigl(\tfrac{t}{\lambda}\bigr),
\]
where \(\lambda\) denotes the scale.  The continuous wavelet transform of \(x(t)\) is
\[
W_x(\lambda,t) = (x * \psi_\lambda)(t)
= \int_{-\infty}^{\infty} x(u)\,\psi_\lambda(t-u)\,du.
\]
Because \(\psi_\lambda(t)\) is localized in time (or space) with support \(O(\lambda)\), the coefficients \(W_x(\lambda,t)\) capture local features at scale \(\lambda\) and position \(t\).  The first-order scattering coefficient
\[
S_1(\lambda) = \int \bigl|W_x(\lambda,t)\bigr|\,dt
\]
thereby preserves not only the amplitude at each scale (as in the Fourier spectrum) but also the localization of those features, although the modulus operation removes raw phase information as similar as Fourier transformation.  Higher-order coefficients, for example
\[
S_2(\lambda_1,\lambda_2)
= \int \bigl|\;|W_x(\lambda_1,t)| * \psi_{\lambda_2}(t)\bigr|\,dt,
\]
capture scale–scale couplings (i.e. non-Gaussian interactions) that a single-scale statistic like \(P(k)\) cannot detect.

Hence, by using localized wavelets and a nonlinear modulus‐averaging cascade, the WST retains crucial “where-and-which-scale” information, enabling it to distinguish signals with identical global power but differing in their local, multiscale structure—something the Fourier power spectrum alone cannot achieve.

\vspace{+0.1in}

\section{Robustness of the Wavelet Scattering Transform to Thermal Noise}
\label{sec:robustness}

We consider an observed signal 
\[
y(t) = x(t) + n(t),
\]
where \(x(t)\) 
is in analogy to 
the true 21cm forest signal and \(n(t)\) denotes additive thermal noise (assumed to have zero mean). The wavelet transform of \(y(t)\) at scale \(\lambda\) and time \(t\) is defined as
\[
W_y(\lambda, t) = \int_{-\infty}^{\infty} y(u) \, \psi_\lambda(t-u)\, du.
\]
This expression can be decomposed into
\[
W_y(\lambda,t) = W_x(\lambda,t) + W_n(\lambda,t),
\]
where \(W_x(\lambda,t)\) and \(W_n(\lambda,t)\) are the wavelet transforms of the true signal \(x(t)\) and the noise \(n(t)\), respectively.

The first-order scattering coefficients are obtained by taking the modulus of the wavelet coefficients and then averaging over time:
\[
U_y(\lambda,t) = |W_y(\lambda,t)|, \quad\text{and}\quad S_1(\lambda) = \int |W_y(\lambda,t)|\, dt.
\]

A key property in this process is the Lipschitz continuity of the modulus operator, which implies
\[
\bigl||W_y(\lambda,t)| - |W_x(\lambda,t)|\bigr| \le |W_y(\lambda,t) - W_x(\lambda,t)| = |W_n(\lambda,t)|.
\]
Since \(n(t)\) represents thermal noise, \(W_n(\lambda,t)\) typically exhibits random fluctuations that are localized in time due to the compact support of the wavelet \(\psi_\lambda(t)\). The subsequent averaging over \(t\) in the computation of \(S_1(\lambda)\) further reduces these random fluctuations.

Thus, even though thermal noise perturbs the wavelet coefficients \(W_y(\lambda,t)\) locally, its overall impact on the first-order scattering coefficients \(S_1(\lambda)\) is attenuated by both the inherent localization of the wavelet transform and the averaging process. This combination of factors renders the first-order WST coefficients relatively robust to additive thermal noise, thereby preserving the essential features of the 21cm forest signal even in the presence of noise.



\bibliography{reference}

\begin{thebibliography}{91}
\expandafter\ifx\csname natexlab\endcsname\relax\def\natexlab#1{#1}\fi
\expandafter\ifx\csname bibnamefont\endcsname\relax
  \def\bibnamefont#1{#1}\fi
\expandafter\ifx\csname bibfnamefont\endcsname\relax
  \def\bibfnamefont#1{#1}\fi
\expandafter\ifx\csname citenamefont\endcsname\relax
  \def\citenamefont#1{#1}\fi
\expandafter\ifx\csname url\endcsname\relax
  \def\url#1{\texttt{#1}}\fi
\expandafter\ifx\csname urlprefix\endcsname\relax\def\urlprefix{URL }\fi
\providecommand{\bibinfo}[2]{#2}
\providecommand{\eprint}[2][]{\url{#2}}

\bibitem[{\citenamefont{{Furlanetto} et~al.}(2006)\citenamefont{{Furlanetto}, {Oh}, and {Briggs}}}]{fur}
\bibinfo{author}{\bibfnamefont{S.~R.} \bibnamefont{{Furlanetto}}}, \bibinfo{author}{\bibfnamefont{S.~P.} \bibnamefont{{Oh}}}, \bibnamefont{and} \bibinfo{author}{\bibfnamefont{F.~H.} \bibnamefont{{Briggs}}}, \bibinfo{journal}{\physrep} \textbf{\bibinfo{volume}{433}}, \bibinfo{pages}{181} (\bibinfo{year}{2006}), \eprint{astro-ph/0608032}.

\bibitem[{\citenamefont{{Pritchard} and {Loeb}}(2012)}]{2012RPPh...75h6901P}
\bibinfo{author}{\bibfnamefont{J.~R.} \bibnamefont{{Pritchard}}} \bibnamefont{and} \bibinfo{author}{\bibfnamefont{A.}~\bibnamefont{{Loeb}}}, \bibinfo{journal}{Reports on Progress in Physics} \textbf{\bibinfo{volume}{75}}, \bibinfo{eid}{086901} (\bibinfo{year}{2012}), \eprint{1109.6012}.

\bibitem[{\citenamefont{{Liu} et~al.}(2013)\citenamefont{{Liu}, {Pritchard}, {Tegmark}, and {Loeb}}}]{2013PhRvD..87d3002L}
\bibinfo{author}{\bibfnamefont{A.}~\bibnamefont{{Liu}}}, \bibinfo{author}{\bibfnamefont{J.~R.} \bibnamefont{{Pritchard}}}, \bibinfo{author}{\bibfnamefont{M.}~\bibnamefont{{Tegmark}}}, \bibnamefont{and} \bibinfo{author}{\bibfnamefont{A.}~\bibnamefont{{Loeb}}}, \bibinfo{journal}{\prd} \textbf{\bibinfo{volume}{87}}, \bibinfo{eid}{043002} (\bibinfo{year}{2013}), \eprint{1211.3743}.

\bibitem[{\citenamefont{{Shimabukuro} et~al.}(2023{\natexlab{a}})\citenamefont{{Shimabukuro}, {Hasegawa}, {Kuchinomachi}, {Yajima}, and {Yoshiura}}}]{2023PASJ...75S...1S}
\bibinfo{author}{\bibfnamefont{H.}~\bibnamefont{{Shimabukuro}}}, \bibinfo{author}{\bibfnamefont{K.}~\bibnamefont{{Hasegawa}}}, \bibinfo{author}{\bibfnamefont{A.}~\bibnamefont{{Kuchinomachi}}}, \bibinfo{author}{\bibfnamefont{H.}~\bibnamefont{{Yajima}}}, \bibnamefont{and} \bibinfo{author}{\bibfnamefont{S.}~\bibnamefont{{Yoshiura}}}, \bibinfo{journal}{\pasj} \textbf{\bibinfo{volume}{75}}, \bibinfo{pages}{S1} (\bibinfo{year}{2023}{\natexlab{a}}), \eprint{2303.07594}.

\bibitem[{\citenamefont{{Bowman} et~al.}(2018)\citenamefont{{Bowman}, {Rogers}, {Monsalve}, {Mozdzen}, and {Mahesh}}}]{2018Natur.555...67B}
\bibinfo{author}{\bibfnamefont{J.~D.} \bibnamefont{{Bowman}}}, \bibinfo{author}{\bibfnamefont{A.~E.~E.} \bibnamefont{{Rogers}}}, \bibinfo{author}{\bibfnamefont{R.~A.} \bibnamefont{{Monsalve}}}, \bibinfo{author}{\bibfnamefont{T.~J.} \bibnamefont{{Mozdzen}}}, \bibnamefont{and} \bibinfo{author}{\bibfnamefont{N.}~\bibnamefont{{Mahesh}}}, \bibinfo{journal}{\nat} \textbf{\bibinfo{volume}{555}}, \bibinfo{pages}{67} (\bibinfo{year}{2018}), \eprint{1810.05912}.

\bibitem[{\citenamefont{{Singh} et~al.}(2018)\citenamefont{{Singh}, {Subrahmanyan}, {Udaya Shankar}, {Sathyanarayana Rao}, {Fialkov}, {Cohen}, {Barkana}, {Girish}, {Raghunathan}, {Somashekar} et~al.}}]{2018ApJ...858...54S}
\bibinfo{author}{\bibfnamefont{S.}~\bibnamefont{{Singh}}}, \bibinfo{author}{\bibfnamefont{R.}~\bibnamefont{{Subrahmanyan}}}, \bibinfo{author}{\bibfnamefont{N.}~\bibnamefont{{Udaya Shankar}}}, \bibinfo{author}{\bibfnamefont{M.}~\bibnamefont{{Sathyanarayana Rao}}}, \bibinfo{author}{\bibfnamefont{A.}~\bibnamefont{{Fialkov}}}, \bibinfo{author}{\bibfnamefont{A.}~\bibnamefont{{Cohen}}}, \bibinfo{author}{\bibfnamefont{R.}~\bibnamefont{{Barkana}}}, \bibinfo{author}{\bibfnamefont{B.~S.} \bibnamefont{{Girish}}}, \bibinfo{author}{\bibfnamefont{A.}~\bibnamefont{{Raghunathan}}}, \bibinfo{author}{\bibfnamefont{R.}~\bibnamefont{{Somashekar}}}, \bibnamefont{et~al.}, \bibinfo{journal}{\apj} \textbf{\bibinfo{volume}{858}}, \bibinfo{eid}{54} (\bibinfo{year}{2018}), \eprint{1711.11281}.

\bibitem[{\citenamefont{{Nambissan T.} et~al.}(2021)\citenamefont{{Nambissan T.}, {Subrahmanyan}, {Somashekar}, {Udaya Shankar}, {Singh}, {Raghunathan}, {Girish}, {Srivani}, and {Sathyanarayana Rao}}}]{2021arXiv210401756N}
\bibinfo{author}{\bibfnamefont{J.}~\bibnamefont{{Nambissan T.}}}, \bibinfo{author}{\bibfnamefont{R.}~\bibnamefont{{Subrahmanyan}}}, \bibinfo{author}{\bibfnamefont{R.}~\bibnamefont{{Somashekar}}}, \bibinfo{author}{\bibfnamefont{N.}~\bibnamefont{{Udaya Shankar}}}, \bibinfo{author}{\bibfnamefont{S.}~\bibnamefont{{Singh}}}, \bibinfo{author}{\bibfnamefont{A.}~\bibnamefont{{Raghunathan}}}, \bibinfo{author}{\bibfnamefont{B.~S.} \bibnamefont{{Girish}}}, \bibinfo{author}{\bibfnamefont{K.~S.} \bibnamefont{{Srivani}}}, \bibnamefont{and} \bibinfo{author}{\bibfnamefont{M.}~\bibnamefont{{Sathyanarayana Rao}}}, \bibinfo{journal}{arXiv e-prints} \bibinfo{eid}{arXiv:2104.01756} (\bibinfo{year}{2021}), \eprint{2104.01756}.

\bibitem[{\citenamefont{{Price} et~al.}(2018)\citenamefont{{Price}, {Greenhill}, {Fialkov}, {Bernardi}, {Garsden}, {Barsdell}, {Kocz}, {Anderson}, {Bourke}, {Craig} et~al.}}]{2018MNRAS.478.4193P}
\bibinfo{author}{\bibfnamefont{D.~C.} \bibnamefont{{Price}}}, \bibinfo{author}{\bibfnamefont{L.~J.} \bibnamefont{{Greenhill}}}, \bibinfo{author}{\bibfnamefont{A.}~\bibnamefont{{Fialkov}}}, \bibinfo{author}{\bibfnamefont{G.}~\bibnamefont{{Bernardi}}}, \bibinfo{author}{\bibfnamefont{H.}~\bibnamefont{{Garsden}}}, \bibinfo{author}{\bibfnamefont{B.~R.} \bibnamefont{{Barsdell}}}, \bibinfo{author}{\bibfnamefont{J.}~\bibnamefont{{Kocz}}}, \bibinfo{author}{\bibfnamefont{M.~M.} \bibnamefont{{Anderson}}}, \bibinfo{author}{\bibfnamefont{S.~A.} \bibnamefont{{Bourke}}}, \bibinfo{author}{\bibfnamefont{J.}~\bibnamefont{{Craig}}}, \bibnamefont{et~al.}, \bibinfo{journal}{\mnras} \textbf{\bibinfo{volume}{478}}, \bibinfo{pages}{4193} (\bibinfo{year}{2018}), \eprint{1709.09313}.

\bibitem[{\citenamefont{{Burns}}(2021)}]{Burns2021}
\bibinfo{author}{\bibfnamefont{J.~O.} \bibnamefont{{Burns}}}, \bibinfo{journal}{Philosophical Transactions of the Royal Society of London Series A} \textbf{\bibinfo{volume}{379}}, \bibinfo{eid}{20190564} (\bibinfo{year}{2021}), \eprint{2003.06881}.

\bibitem[{\citenamefont{{Bale} et~al.}(2023)\citenamefont{{Bale}, {Bassett}, {Burns}, {Dorigo Jones}, {Goetz}, {Hellum-Bye}, {Hermann}, {Hibbard}, {Maksimovic}, {McLean} et~al.}}]{2023arXiv230110345B}
\bibinfo{author}{\bibfnamefont{S.~D.} \bibnamefont{{Bale}}}, \bibinfo{author}{\bibfnamefont{N.}~\bibnamefont{{Bassett}}}, \bibinfo{author}{\bibfnamefont{J.~O.} \bibnamefont{{Burns}}}, \bibinfo{author}{\bibfnamefont{J.}~\bibnamefont{{Dorigo Jones}}}, \bibinfo{author}{\bibfnamefont{K.}~\bibnamefont{{Goetz}}}, \bibinfo{author}{\bibfnamefont{C.}~\bibnamefont{{Hellum-Bye}}}, \bibinfo{author}{\bibfnamefont{S.}~\bibnamefont{{Hermann}}}, \bibinfo{author}{\bibfnamefont{J.}~\bibnamefont{{Hibbard}}}, \bibinfo{author}{\bibfnamefont{M.}~\bibnamefont{{Maksimovic}}}, \bibinfo{author}{\bibfnamefont{R.}~\bibnamefont{{McLean}}}, \bibnamefont{et~al.}, \bibinfo{journal}{arXiv e-prints} \bibinfo{eid}{arXiv:2301.10345} (\bibinfo{year}{2023}), \eprint{2301.10345}.

\bibitem[{\citenamefont{{Sathyanarayana Rao} et~al.}(2023)\citenamefont{{Sathyanarayana Rao}, {Singh}, {K.~S.}, {B.~S.}, {Sathish}, {Somashekar}, {Agaram}, {Kavitha}, {Vishwapriya}, {Anand} et~al.}}]{2023ExA....56..741S}
\bibinfo{author}{\bibfnamefont{M.}~\bibnamefont{{Sathyanarayana Rao}}}, \bibinfo{author}{\bibfnamefont{S.}~\bibnamefont{{Singh}}}, \bibinfo{author}{\bibfnamefont{S.}~\bibnamefont{{K.~S.}}}, \bibinfo{author}{\bibfnamefont{G.}~\bibnamefont{{B.~S.}}}, \bibinfo{author}{\bibfnamefont{K.}~\bibnamefont{{Sathish}}}, \bibinfo{author}{\bibfnamefont{R.}~\bibnamefont{{Somashekar}}}, \bibinfo{author}{\bibfnamefont{R.}~\bibnamefont{{Agaram}}}, \bibinfo{author}{\bibfnamefont{K.}~\bibnamefont{{Kavitha}}}, \bibinfo{author}{\bibfnamefont{G.}~\bibnamefont{{Vishwapriya}}}, \bibinfo{author}{\bibfnamefont{A.}~\bibnamefont{{Anand}}}, \bibnamefont{et~al.}, \bibinfo{journal}{Experimental Astronomy} \textbf{\bibinfo{volume}{56}}, \bibinfo{pages}{741} (\bibinfo{year}{2023}).

\bibitem[{\citenamefont{{Sathyanarayana Rao} et~al.}(2025)\citenamefont{{Sathyanarayana Rao}, {Singh}, {S}, {S}, {Satish}, {R}, {Agaram}, {Kalyanasundaram}, {Vishwapriya}, {Anand} et~al.}}]{2025arXiv250705654S}
\bibinfo{author}{\bibfnamefont{M.}~\bibnamefont{{Sathyanarayana Rao}}}, \bibinfo{author}{\bibfnamefont{S.}~\bibnamefont{{Singh}}}, \bibinfo{author}{\bibfnamefont{S.~K.} \bibnamefont{{S}}}, \bibinfo{author}{\bibfnamefont{G.~B.} \bibnamefont{{S}}}, \bibinfo{author}{\bibfnamefont{K.}~\bibnamefont{{Satish}}}, \bibinfo{author}{\bibfnamefont{S.}~\bibnamefont{{R}}}, \bibinfo{author}{\bibfnamefont{R.}~\bibnamefont{{Agaram}}}, \bibinfo{author}{\bibfnamefont{K.}~\bibnamefont{{Kalyanasundaram}}}, \bibinfo{author}{\bibfnamefont{G.}~\bibnamefont{{Vishwapriya}}}, \bibinfo{author}{\bibfnamefont{A.}~\bibnamefont{{Anand}}}, \bibnamefont{et~al.}, \bibinfo{journal}{arXiv e-prints} \bibinfo{eid}{arXiv:2507.05654} (\bibinfo{year}{2025}), \eprint{2507.05654}.

\bibitem[{\citenamefont{{Chen} et~al.}(2021)\citenamefont{{Chen}, {Yan}, {Deng}, {Wu}, {Wu}, {Xu}, and {Zhou}}}]{2021RSPTA.37990566C}
\bibinfo{author}{\bibfnamefont{X.}~\bibnamefont{{Chen}}}, \bibinfo{author}{\bibfnamefont{J.}~\bibnamefont{{Yan}}}, \bibinfo{author}{\bibfnamefont{L.}~\bibnamefont{{Deng}}}, \bibinfo{author}{\bibfnamefont{F.}~\bibnamefont{{Wu}}}, \bibinfo{author}{\bibfnamefont{L.}~\bibnamefont{{Wu}}}, \bibinfo{author}{\bibfnamefont{Y.}~\bibnamefont{{Xu}}}, \bibnamefont{and} \bibinfo{author}{\bibfnamefont{L.}~\bibnamefont{{Zhou}}}, \bibinfo{journal}{Philosophical Transactions of the Royal Society of London Series A} \textbf{\bibinfo{volume}{379}}, \bibinfo{eid}{20190566} (\bibinfo{year}{2021}), \eprint{2007.15794}.

\bibitem[{\citenamefont{{van Haarlem} et~al.}(2013)\citenamefont{{van Haarlem}, {Wise}, {Gunst}, {Heald}, {McKean}, {Hessels}, {de Bruyn}, {Nijboer}, {Swinbank}, {Fallows} et~al.}}]{2013A&A...556A...2V}
\bibinfo{author}{\bibfnamefont{M.~P.} \bibnamefont{{van Haarlem}}}, \bibinfo{author}{\bibfnamefont{M.~W.} \bibnamefont{{Wise}}}, \bibinfo{author}{\bibfnamefont{A.~W.} \bibnamefont{{Gunst}}}, \bibinfo{author}{\bibfnamefont{G.}~\bibnamefont{{Heald}}}, \bibinfo{author}{\bibfnamefont{J.~P.} \bibnamefont{{McKean}}}, \bibinfo{author}{\bibfnamefont{J.~W.~T.} \bibnamefont{{Hessels}}}, \bibinfo{author}{\bibfnamefont{A.~G.} \bibnamefont{{de Bruyn}}}, \bibinfo{author}{\bibfnamefont{R.}~\bibnamefont{{Nijboer}}}, \bibinfo{author}{\bibfnamefont{J.}~\bibnamefont{{Swinbank}}}, \bibinfo{author}{\bibfnamefont{R.}~\bibnamefont{{Fallows}}}, \bibnamefont{et~al.}, \bibinfo{journal}{\aap} \textbf{\bibinfo{volume}{556}}, \bibinfo{eid}{A2} (\bibinfo{year}{2013}), \eprint{1305.3550}.

\bibitem[{\citenamefont{{Wayth} et~al.}(2018)\citenamefont{{Wayth}, {Tingay}, {Trott}, {Emrich}, {Johnston-Hollitt}, {McKinley}, {Gaensler}, {Beardsley}, {Booler}, {Crosse} et~al.}}]{2018PASA...35...33W}
\bibinfo{author}{\bibfnamefont{R.~B.} \bibnamefont{{Wayth}}}, \bibinfo{author}{\bibfnamefont{S.~J.} \bibnamefont{{Tingay}}}, \bibinfo{author}{\bibfnamefont{C.~M.} \bibnamefont{{Trott}}}, \bibinfo{author}{\bibfnamefont{D.}~\bibnamefont{{Emrich}}}, \bibinfo{author}{\bibfnamefont{M.}~\bibnamefont{{Johnston-Hollitt}}}, \bibinfo{author}{\bibfnamefont{B.}~\bibnamefont{{McKinley}}}, \bibinfo{author}{\bibfnamefont{B.~M.} \bibnamefont{{Gaensler}}}, \bibinfo{author}{\bibfnamefont{A.~P.} \bibnamefont{{Beardsley}}}, \bibinfo{author}{\bibfnamefont{T.}~\bibnamefont{{Booler}}}, \bibinfo{author}{\bibfnamefont{B.}~\bibnamefont{{Crosse}}}, \bibnamefont{et~al.}, \bibinfo{journal}{\pasa} \textbf{\bibinfo{volume}{35}}, \bibinfo{pages}{33} (\bibinfo{year}{2018}), \eprint{1809.06466}.

\bibitem[{\citenamefont{{DeBoer} et~al.}(2017)\citenamefont{{DeBoer}, {Parsons}, {Aguirre}, {Alexander}, {Ali}, {Beardsley}, {Bernardi}, {Bowman}, {Bradley}, {Carilli} et~al.}}]{2017PASP..129d5001D}
\bibinfo{author}{\bibfnamefont{D.~R.} \bibnamefont{{DeBoer}}}, \bibinfo{author}{\bibfnamefont{A.~R.} \bibnamefont{{Parsons}}}, \bibinfo{author}{\bibfnamefont{J.~E.} \bibnamefont{{Aguirre}}}, \bibinfo{author}{\bibfnamefont{P.}~\bibnamefont{{Alexander}}}, \bibinfo{author}{\bibfnamefont{Z.~S.} \bibnamefont{{Ali}}}, \bibinfo{author}{\bibfnamefont{A.~P.} \bibnamefont{{Beardsley}}}, \bibinfo{author}{\bibfnamefont{G.}~\bibnamefont{{Bernardi}}}, \bibinfo{author}{\bibfnamefont{J.~D.} \bibnamefont{{Bowman}}}, \bibinfo{author}{\bibfnamefont{R.~F.} \bibnamefont{{Bradley}}}, \bibinfo{author}{\bibfnamefont{C.~L.} \bibnamefont{{Carilli}}}, \bibnamefont{et~al.}, \bibinfo{journal}{\pasp} \textbf{\bibinfo{volume}{129}}, \bibinfo{pages}{045001} (\bibinfo{year}{2017}), \eprint{1606.07473}.

\bibitem[{\citenamefont{{Mellema} et~al.}(2013)\citenamefont{{Mellema}, {Koopmans}, {Abdalla}, {Bernardi}, {Ciardi}, {Daiboo}, {de Bruyn}, {Datta}, {Falcke}, {Ferrara} et~al.}}]{2013ExA....36..235M}
\bibinfo{author}{\bibfnamefont{G.}~\bibnamefont{{Mellema}}}, \bibinfo{author}{\bibfnamefont{L.~V.~E.} \bibnamefont{{Koopmans}}}, \bibinfo{author}{\bibfnamefont{F.~A.} \bibnamefont{{Abdalla}}}, \bibinfo{author}{\bibfnamefont{G.}~\bibnamefont{{Bernardi}}}, \bibinfo{author}{\bibfnamefont{B.}~\bibnamefont{{Ciardi}}}, \bibinfo{author}{\bibfnamefont{S.}~\bibnamefont{{Daiboo}}}, \bibinfo{author}{\bibfnamefont{A.~G.} \bibnamefont{{de Bruyn}}}, \bibinfo{author}{\bibfnamefont{K.~K.} \bibnamefont{{Datta}}}, \bibinfo{author}{\bibfnamefont{H.}~\bibnamefont{{Falcke}}}, \bibinfo{author}{\bibfnamefont{A.}~\bibnamefont{{Ferrara}}}, \bibnamefont{et~al.}, \bibinfo{journal}{Experimental Astronomy} \textbf{\bibinfo{volume}{36}}, \bibinfo{pages}{235} (\bibinfo{year}{2013}), \eprint{1210.0197}.

\bibitem[{\citenamefont{{Koopmans} et~al.}(2015)\citenamefont{{Koopmans}, {Pritchard}, {Mellema}, {Aguirre}, {Ahn}, {Barkana}, {van Bemmel}, {Bernardi}, {Bonaldi}, {Briggs} et~al.}}]{2015aska.confE...1K}
\bibinfo{author}{\bibfnamefont{L.}~\bibnamefont{{Koopmans}}}, \bibinfo{author}{\bibfnamefont{J.}~\bibnamefont{{Pritchard}}}, \bibinfo{author}{\bibfnamefont{G.}~\bibnamefont{{Mellema}}}, \bibinfo{author}{\bibfnamefont{J.}~\bibnamefont{{Aguirre}}}, \bibinfo{author}{\bibfnamefont{K.}~\bibnamefont{{Ahn}}}, \bibinfo{author}{\bibfnamefont{R.}~\bibnamefont{{Barkana}}}, \bibinfo{author}{\bibfnamefont{I.}~\bibnamefont{{van Bemmel}}}, \bibinfo{author}{\bibfnamefont{G.}~\bibnamefont{{Bernardi}}}, \bibinfo{author}{\bibfnamefont{A.}~\bibnamefont{{Bonaldi}}}, \bibinfo{author}{\bibfnamefont{F.}~\bibnamefont{{Briggs}}}, \bibnamefont{et~al.}, \bibinfo{journal}{Advancing Astrophysics with the Square Kilometre Array (AASKA14)} \bibinfo{eid}{1} (\bibinfo{year}{2015}), \eprint{1505.07568}.

\bibitem[{\citenamefont{{CHIME Collaboration} et~al.}(2022)\citenamefont{{CHIME Collaboration}, {Amiri}, {Bandura}, {Boskovic}, {Chen}, {Cliche}, {Deng}, {Denman}, {Dobbs}, {Fandino} et~al.}}]{2022ApJS..261...29C}
\bibinfo{author}{\bibnamefont{{CHIME Collaboration}}}, \bibinfo{author}{\bibfnamefont{M.}~\bibnamefont{{Amiri}}}, \bibinfo{author}{\bibfnamefont{K.}~\bibnamefont{{Bandura}}}, \bibinfo{author}{\bibfnamefont{A.}~\bibnamefont{{Boskovic}}}, \bibinfo{author}{\bibfnamefont{T.}~\bibnamefont{{Chen}}}, \bibinfo{author}{\bibfnamefont{J.-F.} \bibnamefont{{Cliche}}}, \bibinfo{author}{\bibfnamefont{M.}~\bibnamefont{{Deng}}}, \bibinfo{author}{\bibfnamefont{N.}~\bibnamefont{{Denman}}}, \bibinfo{author}{\bibfnamefont{M.}~\bibnamefont{{Dobbs}}}, \bibinfo{author}{\bibfnamefont{M.}~\bibnamefont{{Fandino}}}, \bibnamefont{et~al.}, \bibinfo{journal}{\apjs} \textbf{\bibinfo{volume}{261}}, \bibinfo{eid}{29} (\bibinfo{year}{2022}), \eprint{2201.07869}.

\bibitem[{\citenamefont{{Liu} et~al.}(2019)\citenamefont{{Liu}, {Foreman}, {Padmanabhan}, {Chiang}, {Siegel}, {Wulf}, {Sievers}, {Dobbs}, and {Vanderlinde}}}]{2019clrp.2020....9L}
\bibinfo{author}{\bibfnamefont{A.}~\bibnamefont{{Liu}}}, \bibinfo{author}{\bibfnamefont{S.}~\bibnamefont{{Foreman}}}, \bibinfo{author}{\bibfnamefont{H.}~\bibnamefont{{Padmanabhan}}}, \bibinfo{author}{\bibfnamefont{H.~C.} \bibnamefont{{Chiang}}}, \bibinfo{author}{\bibfnamefont{S.}~\bibnamefont{{Siegel}}}, \bibinfo{author}{\bibfnamefont{D.}~\bibnamefont{{Wulf}}}, \bibinfo{author}{\bibfnamefont{J.}~\bibnamefont{{Sievers}}}, \bibinfo{author}{\bibfnamefont{M.}~\bibnamefont{{Dobbs}}}, \bibnamefont{and} \bibinfo{author}{\bibfnamefont{K.}~\bibnamefont{{Vanderlinde}}}, in \emph{\bibinfo{booktitle}{Canadian Long Range Plan for Astronomy and Astrophysics White Papers}} (\bibinfo{year}{2019}), vol. \bibinfo{volume}{2020}, p.~\bibinfo{pages}{9}, \eprint{1910.02889}.

\bibitem[{\citenamefont{{Amiri} et~al.}(2023)\citenamefont{{Amiri}, {Bandura}, {Chen}, {Deng}, {Dobbs}, {Fandino}, {Foreman}, {Halpern}, {Hill}, {Hinshaw} et~al.}}]{2023ApJ...947...16A}
\bibinfo{author}{\bibfnamefont{M.}~\bibnamefont{{Amiri}}}, \bibinfo{author}{\bibfnamefont{K.}~\bibnamefont{{Bandura}}}, \bibinfo{author}{\bibfnamefont{T.}~\bibnamefont{{Chen}}}, \bibinfo{author}{\bibfnamefont{M.}~\bibnamefont{{Deng}}}, \bibinfo{author}{\bibfnamefont{M.}~\bibnamefont{{Dobbs}}}, \bibinfo{author}{\bibfnamefont{M.}~\bibnamefont{{Fandino}}}, \bibinfo{author}{\bibfnamefont{S.}~\bibnamefont{{Foreman}}}, \bibinfo{author}{\bibfnamefont{M.}~\bibnamefont{{Halpern}}}, \bibinfo{author}{\bibfnamefont{A.~S.} \bibnamefont{{Hill}}}, \bibinfo{author}{\bibfnamefont{G.}~\bibnamefont{{Hinshaw}}}, \bibnamefont{et~al.}, \bibinfo{journal}{\apj} \textbf{\bibinfo{volume}{947}}, \bibinfo{eid}{16} (\bibinfo{year}{2023}), \eprint{2202.01242}.

\bibitem[{\citenamefont{{Vanderlinde} et~al.}(2019)\citenamefont{{Vanderlinde}, {Liu}, {Gaensler}, {Bond}, {Hinshaw}, {Ng}, {Chiang}, {Stairs}, {Brown}, {Sievers} et~al.}}]{2019clrp.2020...28V}
\bibinfo{author}{\bibfnamefont{K.}~\bibnamefont{{Vanderlinde}}}, \bibinfo{author}{\bibfnamefont{A.}~\bibnamefont{{Liu}}}, \bibinfo{author}{\bibfnamefont{B.}~\bibnamefont{{Gaensler}}}, \bibinfo{author}{\bibfnamefont{D.}~\bibnamefont{{Bond}}}, \bibinfo{author}{\bibfnamefont{G.}~\bibnamefont{{Hinshaw}}}, \bibinfo{author}{\bibfnamefont{C.}~\bibnamefont{{Ng}}}, \bibinfo{author}{\bibfnamefont{C.}~\bibnamefont{{Chiang}}}, \bibinfo{author}{\bibfnamefont{I.}~\bibnamefont{{Stairs}}}, \bibinfo{author}{\bibfnamefont{J.-A.} \bibnamefont{{Brown}}}, \bibinfo{author}{\bibfnamefont{J.}~\bibnamefont{{Sievers}}}, \bibnamefont{et~al.}, in \emph{\bibinfo{booktitle}{Canadian Long Range Plan for Astronomy and Astrophysics White Papers}} (\bibinfo{year}{2019}), vol. \bibinfo{volume}{2020}, p.~\bibinfo{pages}{28}, \eprint{1911.01777}.

\bibitem[{\citenamefont{{Gupta} et~al.}(2017)\citenamefont{{Gupta}, {Ajithkumar}, {Kale}, {Nayak}, {Sabhapathy}, {Sureshkumar}, {Swami}, {Chengalur}, {Ghosh}, {Ishwara-Chandra} et~al.}}]{2017CSci..113..707G}
\bibinfo{author}{\bibfnamefont{Y.}~\bibnamefont{{Gupta}}}, \bibinfo{author}{\bibfnamefont{B.}~\bibnamefont{{Ajithkumar}}}, \bibinfo{author}{\bibfnamefont{H.~S.} \bibnamefont{{Kale}}}, \bibinfo{author}{\bibfnamefont{S.}~\bibnamefont{{Nayak}}}, \bibinfo{author}{\bibfnamefont{S.}~\bibnamefont{{Sabhapathy}}}, \bibinfo{author}{\bibfnamefont{S.}~\bibnamefont{{Sureshkumar}}}, \bibinfo{author}{\bibfnamefont{R.~V.} \bibnamefont{{Swami}}}, \bibinfo{author}{\bibfnamefont{J.~N.} \bibnamefont{{Chengalur}}}, \bibinfo{author}{\bibfnamefont{S.~K.} \bibnamefont{{Ghosh}}}, \bibinfo{author}{\bibfnamefont{C.~H.} \bibnamefont{{Ishwara-Chandra}}}, \bibnamefont{et~al.}, \bibinfo{journal}{Current Science} \textbf{\bibinfo{volume}{113}}, \bibinfo{pages}{707} (\bibinfo{year}{2017}).

\bibitem[{\citenamefont{{Aditya}}(2019)}]{2019MNRAS.482.5597A}
\bibinfo{author}{\bibfnamefont{J.~N.~H.~S.} \bibnamefont{{Aditya}}}, \bibinfo{journal}{\mnras} \textbf{\bibinfo{volume}{482}}, \bibinfo{pages}{5597} (\bibinfo{year}{2019}), \eprint{1811.03048}.

\bibitem[{\citenamefont{{Chowdhury} et~al.}(2020)\citenamefont{{Chowdhury}, {Kanekar}, and {Chengalur}}}]{2020ApJ...900L..30C}
\bibinfo{author}{\bibfnamefont{A.}~\bibnamefont{{Chowdhury}}}, \bibinfo{author}{\bibfnamefont{N.}~\bibnamefont{{Kanekar}}}, \bibnamefont{and} \bibinfo{author}{\bibfnamefont{J.~N.} \bibnamefont{{Chengalur}}}, \bibinfo{journal}{\apjl} \textbf{\bibinfo{volume}{900}}, \bibinfo{eid}{L30} (\bibinfo{year}{2020}), \eprint{2008.11403}.

\bibitem[{\citenamefont{{Aditya} et~al.}(2021)\citenamefont{{Aditya}, {Jorgenson}, {Joshi}, {Singh}, {An}, and {Chandola}}}]{2021MNRAS.500..998A}
\bibinfo{author}{\bibfnamefont{J.~N.~H.~S.} \bibnamefont{{Aditya}}}, \bibinfo{author}{\bibfnamefont{R.}~\bibnamefont{{Jorgenson}}}, \bibinfo{author}{\bibfnamefont{V.}~\bibnamefont{{Joshi}}}, \bibinfo{author}{\bibfnamefont{V.}~\bibnamefont{{Singh}}}, \bibinfo{author}{\bibfnamefont{T.}~\bibnamefont{{An}}}, \bibnamefont{and} \bibinfo{author}{\bibfnamefont{Y.}~\bibnamefont{{Chandola}}}, \bibinfo{journal}{\mnras} \textbf{\bibinfo{volume}{500}}, \bibinfo{pages}{998} (\bibinfo{year}{2021}), \eprint{2010.10565}.

\bibitem[{\citenamefont{{Jonas} and {MeerKAT Team}}(2016)}]{2016mks..confE...1J}
\bibinfo{author}{\bibfnamefont{J.}~\bibnamefont{{Jonas}}} \bibnamefont{and} \bibinfo{author}{\bibnamefont{{MeerKAT Team}}}, in \emph{\bibinfo{booktitle}{MeerKAT Science: On the Pathway to the SKA}} (\bibinfo{year}{2016}), p.~\bibinfo{pages}{1}.

\bibitem[{\citenamefont{{Paul} et~al.}(2023)\citenamefont{{Paul}, {Santos}, {Chen}, and {Wolz}}}]{2023arXiv230111943P}
\bibinfo{author}{\bibfnamefont{S.}~\bibnamefont{{Paul}}}, \bibinfo{author}{\bibfnamefont{M.~G.} \bibnamefont{{Santos}}}, \bibinfo{author}{\bibfnamefont{Z.}~\bibnamefont{{Chen}}}, \bibnamefont{and} \bibinfo{author}{\bibfnamefont{L.}~\bibnamefont{{Wolz}}}, \bibinfo{journal}{arXiv e-prints} \bibinfo{eid}{arXiv:2301.11943} (\bibinfo{year}{2023}), \eprint{2301.11943}.

\bibitem[{\citenamefont{{Cunnington} et~al.}(2023)\citenamefont{{Cunnington}, {Li}, {Santos}, {Wang}, {Carucci}, {Irfan}, {Pourtsidou}, {Spinelli}, {Wolz}, {Soares} et~al.}}]{2023MNRAS.518.6262C}
\bibinfo{author}{\bibfnamefont{S.}~\bibnamefont{{Cunnington}}}, \bibinfo{author}{\bibfnamefont{Y.}~\bibnamefont{{Li}}}, \bibinfo{author}{\bibfnamefont{M.~G.} \bibnamefont{{Santos}}}, \bibinfo{author}{\bibfnamefont{J.}~\bibnamefont{{Wang}}}, \bibinfo{author}{\bibfnamefont{I.~P.} \bibnamefont{{Carucci}}}, \bibinfo{author}{\bibfnamefont{M.~O.} \bibnamefont{{Irfan}}}, \bibinfo{author}{\bibfnamefont{A.}~\bibnamefont{{Pourtsidou}}}, \bibinfo{author}{\bibfnamefont{M.}~\bibnamefont{{Spinelli}}}, \bibinfo{author}{\bibfnamefont{L.}~\bibnamefont{{Wolz}}}, \bibinfo{author}{\bibfnamefont{P.~S.} \bibnamefont{{Soares}}}, \bibnamefont{et~al.}, \bibinfo{journal}{\mnras} \textbf{\bibinfo{volume}{518}}, \bibinfo{pages}{6262} (\bibinfo{year}{2023}), \eprint{2206.01579}.

\bibitem[{\citenamefont{{Mazumder} et~al.}(2025)\citenamefont{{Mazumder}, {Wolz}, {Chen}, {Paul}, {Santos}, {Jarvis}, {Townsend}, {Sekhar}, and {Taylor}}}]{2025arXiv250117564M}
\bibinfo{author}{\bibfnamefont{A.}~\bibnamefont{{Mazumder}}}, \bibinfo{author}{\bibfnamefont{L.}~\bibnamefont{{Wolz}}}, \bibinfo{author}{\bibfnamefont{Z.}~\bibnamefont{{Chen}}}, \bibinfo{author}{\bibfnamefont{S.}~\bibnamefont{{Paul}}}, \bibinfo{author}{\bibfnamefont{M.}~\bibnamefont{{Santos}}}, \bibinfo{author}{\bibfnamefont{M.}~\bibnamefont{{Jarvis}}}, \bibinfo{author}{\bibfnamefont{J.}~\bibnamefont{{Townsend}}}, \bibinfo{author}{\bibfnamefont{S.}~\bibnamefont{{Sekhar}}}, \bibnamefont{and} \bibinfo{author}{\bibfnamefont{R.}~\bibnamefont{{Taylor}}}, \bibinfo{journal}{arXiv e-prints} \bibinfo{eid}{arXiv:2501.17564} (\bibinfo{year}{2025}), \eprint{2501.17564}.

\bibitem[{\citenamefont{{Nan} et~al.}(2011)\citenamefont{{Nan}, {Li}, {Jin}, {Wang}, {Zhu}, {Zhu}, {Zhang}, {Yue}, and {Qian}}}]{2011IJMPD..20..989N}
\bibinfo{author}{\bibfnamefont{R.}~\bibnamefont{{Nan}}}, \bibinfo{author}{\bibfnamefont{D.}~\bibnamefont{{Li}}}, \bibinfo{author}{\bibfnamefont{C.}~\bibnamefont{{Jin}}}, \bibinfo{author}{\bibfnamefont{Q.}~\bibnamefont{{Wang}}}, \bibinfo{author}{\bibfnamefont{L.}~\bibnamefont{{Zhu}}}, \bibinfo{author}{\bibfnamefont{W.}~\bibnamefont{{Zhu}}}, \bibinfo{author}{\bibfnamefont{H.}~\bibnamefont{{Zhang}}}, \bibinfo{author}{\bibfnamefont{Y.}~\bibnamefont{{Yue}}}, \bibnamefont{and} \bibinfo{author}{\bibfnamefont{L.}~\bibnamefont{{Qian}}}, \bibinfo{journal}{International Journal of Modern Physics D} \textbf{\bibinfo{volume}{20}}, \bibinfo{pages}{989} (\bibinfo{year}{2011}), \eprint{1105.3794}.

\bibitem[{\citenamefont{{Hu} et~al.}(2025)\citenamefont{{Hu}, {Wang}, {Li}, {Yang}, {Xu}, {Wu}, {Pen}, {Wang}, {Jing}, {Xu} et~al.}}]{2025ApJS..277...25H}
\bibinfo{author}{\bibfnamefont{W.}~\bibnamefont{{Hu}}}, \bibinfo{author}{\bibfnamefont{Y.}~\bibnamefont{{Wang}}}, \bibinfo{author}{\bibfnamefont{Y.}~\bibnamefont{{Li}}}, \bibinfo{author}{\bibfnamefont{W.}~\bibnamefont{{Yang}}}, \bibinfo{author}{\bibfnamefont{Y.}~\bibnamefont{{Xu}}}, \bibinfo{author}{\bibfnamefont{F.}~\bibnamefont{{Wu}}}, \bibinfo{author}{\bibfnamefont{U.-L.} \bibnamefont{{Pen}}}, \bibinfo{author}{\bibfnamefont{J.}~\bibnamefont{{Wang}}}, \bibinfo{author}{\bibfnamefont{Y.}~\bibnamefont{{Jing}}}, \bibinfo{author}{\bibfnamefont{C.}~\bibnamefont{{Xu}}}, \bibnamefont{et~al.}, \bibinfo{journal}{\apjs} \textbf{\bibinfo{volume}{277}}, \bibinfo{eid}{25} (\bibinfo{year}{2025}), \eprint{2407.14411}.

\bibitem[{\citenamefont{{Zhang} et~al.}(2025)\citenamefont{{Zhang}, {Zhu}, {Jiang}, {Cheng}, {Xu}, {Yu}, {Liu}, and {Zhang}}}]{2025ApJS..276....6Z}
\bibinfo{author}{\bibfnamefont{C.-P.} \bibnamefont{{Zhang}}}, \bibinfo{author}{\bibfnamefont{M.}~\bibnamefont{{Zhu}}}, \bibinfo{author}{\bibfnamefont{P.}~\bibnamefont{{Jiang}}}, \bibinfo{author}{\bibfnamefont{C.}~\bibnamefont{{Cheng}}}, \bibinfo{author}{\bibfnamefont{J.-L.} \bibnamefont{{Xu}}}, \bibinfo{author}{\bibfnamefont{N.-P.} \bibnamefont{{Yu}}}, \bibinfo{author}{\bibfnamefont{X.-L.} \bibnamefont{{Liu}}}, \bibnamefont{and} \bibinfo{author}{\bibfnamefont{B.}~\bibnamefont{{Zhang}}}, \bibinfo{journal}{\apjs} \textbf{\bibinfo{volume}{276}}, \bibinfo{eid}{6} (\bibinfo{year}{2025}), \eprint{2407.15467}.

\bibitem[{\citenamefont{{Wuensche} et~al.}(2022)\citenamefont{{Wuensche}, {Villela}, {Abdalla}, {Liccardo}, {Vieira}, {Browne}, {Peel}, {Radcliffe}, {Abdalla}, {Marins} et~al.}}]{2022A&A...664A..15W}
\bibinfo{author}{\bibfnamefont{C.~A.} \bibnamefont{{Wuensche}}}, \bibinfo{author}{\bibfnamefont{T.}~\bibnamefont{{Villela}}}, \bibinfo{author}{\bibfnamefont{E.}~\bibnamefont{{Abdalla}}}, \bibinfo{author}{\bibfnamefont{V.}~\bibnamefont{{Liccardo}}}, \bibinfo{author}{\bibfnamefont{F.}~\bibnamefont{{Vieira}}}, \bibinfo{author}{\bibfnamefont{I.}~\bibnamefont{{Browne}}}, \bibinfo{author}{\bibfnamefont{M.~W.} \bibnamefont{{Peel}}}, \bibinfo{author}{\bibfnamefont{C.}~\bibnamefont{{Radcliffe}}}, \bibinfo{author}{\bibfnamefont{F.~B.} \bibnamefont{{Abdalla}}}, \bibinfo{author}{\bibfnamefont{A.}~\bibnamefont{{Marins}}}, \bibnamefont{et~al.}, \bibinfo{journal}{\aap} \textbf{\bibinfo{volume}{664}}, \bibinfo{eid}{A15} (\bibinfo{year}{2022}), \eprint{2107.01634}.

\bibitem[{\citenamefont{{Abdalla} et~al.}(2022)\citenamefont{{Abdalla}, {Ferreira}, {Landim}, {Costa}, {Fornazier}, {Abdalla}, {Barosi}, {Brito}, {Queiroz}, {Villela} et~al.}}]{2022A&A...664A..14A}
\bibinfo{author}{\bibfnamefont{E.}~\bibnamefont{{Abdalla}}}, \bibinfo{author}{\bibfnamefont{E.~G.~M.} \bibnamefont{{Ferreira}}}, \bibinfo{author}{\bibfnamefont{R.~G.} \bibnamefont{{Landim}}}, \bibinfo{author}{\bibfnamefont{A.~A.} \bibnamefont{{Costa}}}, \bibinfo{author}{\bibfnamefont{K.~S.~F.} \bibnamefont{{Fornazier}}}, \bibinfo{author}{\bibfnamefont{F.~B.} \bibnamefont{{Abdalla}}}, \bibinfo{author}{\bibfnamefont{L.}~\bibnamefont{{Barosi}}}, \bibinfo{author}{\bibfnamefont{F.~A.} \bibnamefont{{Brito}}}, \bibinfo{author}{\bibfnamefont{A.~R.} \bibnamefont{{Queiroz}}}, \bibinfo{author}{\bibfnamefont{T.}~\bibnamefont{{Villela}}}, \bibnamefont{et~al.}, \bibinfo{journal}{\aap} \textbf{\bibinfo{volume}{664}}, \bibinfo{eid}{A14} (\bibinfo{year}{2022}), \eprint{2107.01633}.

\bibitem[{\citenamefont{{Xiao} et~al.}(2022)\citenamefont{{Xiao}, {Costa}, and {Wang}}}]{2022MNRAS.510.1495X}
\bibinfo{author}{\bibfnamefont{L.}~\bibnamefont{{Xiao}}}, \bibinfo{author}{\bibfnamefont{A.~A.} \bibnamefont{{Costa}}}, \bibnamefont{and} \bibinfo{author}{\bibfnamefont{B.}~\bibnamefont{{Wang}}}, \bibinfo{journal}{\mnras} \textbf{\bibinfo{volume}{510}}, \bibinfo{pages}{1495} (\bibinfo{year}{2022}), \eprint{2103.01796}.

\bibitem[{\citenamefont{{Novaes} et~al.}(2022)\citenamefont{{Novaes}, {Zhang}, {de Mericia}, {Abdalla}, {Liccardo}, {Wuensche}, {Delabrouille}, {Remazeilles}, {Santos}, {Landim} et~al.}}]{2022A&A...666A..83N}
\bibinfo{author}{\bibfnamefont{C.~P.} \bibnamefont{{Novaes}}}, \bibinfo{author}{\bibfnamefont{J.}~\bibnamefont{{Zhang}}}, \bibinfo{author}{\bibfnamefont{E.~J.} \bibnamefont{{de Mericia}}}, \bibinfo{author}{\bibfnamefont{F.~B.} \bibnamefont{{Abdalla}}}, \bibinfo{author}{\bibfnamefont{V.}~\bibnamefont{{Liccardo}}}, \bibinfo{author}{\bibfnamefont{C.~A.} \bibnamefont{{Wuensche}}}, \bibinfo{author}{\bibfnamefont{J.}~\bibnamefont{{Delabrouille}}}, \bibinfo{author}{\bibfnamefont{M.}~\bibnamefont{{Remazeilles}}}, \bibinfo{author}{\bibfnamefont{L.}~\bibnamefont{{Santos}}}, \bibinfo{author}{\bibfnamefont{R.~G.} \bibnamefont{{Landim}}}, \bibnamefont{et~al.}, \bibinfo{journal}{\aap} \textbf{\bibinfo{volume}{666}}, \bibinfo{eid}{A83} (\bibinfo{year}{2022}), \eprint{2207.12125}.

\bibitem[{\citenamefont{{Newburgh} et~al.}(2016)\citenamefont{{Newburgh}, {Bandura}, {Bucher}, {Chang}, {Chiang}, {Cliche}, {Dav{\'e}}, {Dobbs}, {Clarkson}, {Ganga} et~al.}}]{2016SPIE.9906E..5XN}
\bibinfo{author}{\bibfnamefont{L.~B.} \bibnamefont{{Newburgh}}}, \bibinfo{author}{\bibfnamefont{K.}~\bibnamefont{{Bandura}}}, \bibinfo{author}{\bibfnamefont{M.~A.} \bibnamefont{{Bucher}}}, \bibinfo{author}{\bibfnamefont{T.~C.} \bibnamefont{{Chang}}}, \bibinfo{author}{\bibfnamefont{H.~C.} \bibnamefont{{Chiang}}}, \bibinfo{author}{\bibfnamefont{J.~F.} \bibnamefont{{Cliche}}}, \bibinfo{author}{\bibfnamefont{R.}~\bibnamefont{{Dav{\'e}}}}, \bibinfo{author}{\bibfnamefont{M.}~\bibnamefont{{Dobbs}}}, \bibinfo{author}{\bibfnamefont{C.}~\bibnamefont{{Clarkson}}}, \bibinfo{author}{\bibfnamefont{K.~M.} \bibnamefont{{Ganga}}}, \bibnamefont{et~al.}, in \emph{\bibinfo{booktitle}{Ground-based and Airborne Telescopes VI}}, edited by \bibinfo{editor}{\bibfnamefont{H.~J.} \bibnamefont{{Hall}}}, \bibinfo{editor}{\bibfnamefont{R.}~\bibnamefont{{Gilmozzi}}}, \bibnamefont{and} \bibinfo{editor}{\bibfnamefont{H.~K.} \bibnamefont{{Marshall}}} (\bibinfo{year}{2016}), vol. \bibinfo{volume}{9906} of \emph{\bibinfo{series}{Society of
  Photo-Optical Instrumentation Engineers (SPIE) Conference Series}}, p. \bibinfo{pages}{99065X}, \eprint{1607.02059}.

\bibitem[{\citenamefont{{Carilli} et~al.}(2002)\citenamefont{{Carilli}, {Gnedin}, and {Owen}}}]{2002ApJ...577...22C}
\bibinfo{author}{\bibfnamefont{C.~L.} \bibnamefont{{Carilli}}}, \bibinfo{author}{\bibfnamefont{N.~Y.} \bibnamefont{{Gnedin}}}, \bibnamefont{and} \bibinfo{author}{\bibfnamefont{F.}~\bibnamefont{{Owen}}}, \bibinfo{journal}{ApJ} \textbf{\bibinfo{volume}{577}}, \bibinfo{pages}{22} (\bibinfo{year}{2002}), \eprint{astro-ph/0205169}.

\bibitem[{\citenamefont{{Furlanetto} and {Loeb}}(2002)}]{2002ApJ...579....1F}
\bibinfo{author}{\bibfnamefont{S.~R.} \bibnamefont{{Furlanetto}}} \bibnamefont{and} \bibinfo{author}{\bibfnamefont{A.}~\bibnamefont{{Loeb}}}, \bibinfo{journal}{\apj} \textbf{\bibinfo{volume}{579}}, \bibinfo{pages}{1} (\bibinfo{year}{2002}), \eprint{astro-ph/0206308}.

\bibitem[{\citenamefont{{Furlanetto}}(2006)}]{2006MNRAS.370.1867F}
\bibinfo{author}{\bibfnamefont{S.~R.} \bibnamefont{{Furlanetto}}}, \bibinfo{journal}{\mnras} \textbf{\bibinfo{volume}{370}}, \bibinfo{pages}{1867} (\bibinfo{year}{2006}), \eprint{astro-ph/0604223}.

\bibitem[{\citenamefont{{Ciardi} et~al.}(2015)\citenamefont{{Ciardi}, {Inoue}, {Mack}, {Xu}, and {Bernardi}}}]{2015aska.confE...6C}
\bibinfo{author}{\bibfnamefont{B.}~\bibnamefont{{Ciardi}}}, \bibinfo{author}{\bibfnamefont{S.}~\bibnamefont{{Inoue}}}, \bibinfo{author}{\bibfnamefont{K.}~\bibnamefont{{Mack}}}, \bibinfo{author}{\bibfnamefont{Y.}~\bibnamefont{{Xu}}}, \bibnamefont{and} \bibinfo{author}{\bibfnamefont{G.}~\bibnamefont{{Bernardi}}}, in \emph{\bibinfo{booktitle}{Advancing Astrophysics with the Square Kilometre Array (AASKA14)}} (\bibinfo{year}{2015}), p.~\bibinfo{pages}{6}, \eprint{1501.04425}.

\bibitem[{\citenamefont{{Xu} et~al.}(2009)\citenamefont{{Xu}, {Chen}, {Fan}, {Trac}, and {Cen}}}]{2009ApJ...704.1396X}
\bibinfo{author}{\bibfnamefont{Y.}~\bibnamefont{{Xu}}}, \bibinfo{author}{\bibfnamefont{X.}~\bibnamefont{{Chen}}}, \bibinfo{author}{\bibfnamefont{Z.}~\bibnamefont{{Fan}}}, \bibinfo{author}{\bibfnamefont{H.}~\bibnamefont{{Trac}}}, \bibnamefont{and} \bibinfo{author}{\bibfnamefont{R.}~\bibnamefont{{Cen}}}, \bibinfo{journal}{\apj} \textbf{\bibinfo{volume}{704}}, \bibinfo{pages}{1396} (\bibinfo{year}{2009}), \eprint{0904.4254}.

\bibitem[{\citenamefont{{Xu} et~al.}(2011)\citenamefont{{Xu}, {Ferrara}, and {Chen}}}]{2011MNRAS.410.2025X}
\bibinfo{author}{\bibfnamefont{Y.}~\bibnamefont{{Xu}}}, \bibinfo{author}{\bibfnamefont{A.}~\bibnamefont{{Ferrara}}}, \bibnamefont{and} \bibinfo{author}{\bibfnamefont{X.}~\bibnamefont{{Chen}}}, \bibinfo{journal}{\mnras} \textbf{\bibinfo{volume}{410}}, \bibinfo{pages}{2025} (\bibinfo{year}{2011}), \eprint{1009.1149}.

\bibitem[{\citenamefont{{Mack} and {Wyithe}}(2012)}]{2012MNRAS.425.2988M}
\bibinfo{author}{\bibfnamefont{K.~J.} \bibnamefont{{Mack}}} \bibnamefont{and} \bibinfo{author}{\bibfnamefont{J.~S.~B.} \bibnamefont{{Wyithe}}}, \bibinfo{journal}{\mnras} \textbf{\bibinfo{volume}{425}}, \bibinfo{pages}{2988} (\bibinfo{year}{2012}), \eprint{1101.5431}.

\bibitem[{\citenamefont{{Semelin}}(2016)}]{2016MNRAS.455..962S}
\bibinfo{author}{\bibfnamefont{B.}~\bibnamefont{{Semelin}}}, \bibinfo{journal}{\mnras} \textbf{\bibinfo{volume}{455}}, \bibinfo{pages}{962} (\bibinfo{year}{2016}), \eprint{1510.02296}.

\bibitem[{\citenamefont{{{\v{S}}oltinsk{\'y}} et~al.}(2021)\citenamefont{{{\v{S}}oltinsk{\'y}}, {Bolton}, {Hatch}, {Haehnelt}, {Keating}, {Kulkarni}, {Puchwein}, {Chardin}, and {Aubert}}}]{2021MNRAS.506.5818S}
\bibinfo{author}{\bibfnamefont{T.}~\bibnamefont{{{\v{S}}oltinsk{\'y}}}}, \bibinfo{author}{\bibfnamefont{J.~S.} \bibnamefont{{Bolton}}}, \bibinfo{author}{\bibfnamefont{N.}~\bibnamefont{{Hatch}}}, \bibinfo{author}{\bibfnamefont{M.~G.} \bibnamefont{{Haehnelt}}}, \bibinfo{author}{\bibfnamefont{L.~C.} \bibnamefont{{Keating}}}, \bibinfo{author}{\bibfnamefont{G.}~\bibnamefont{{Kulkarni}}}, \bibinfo{author}{\bibfnamefont{E.}~\bibnamefont{{Puchwein}}}, \bibinfo{author}{\bibfnamefont{J.}~\bibnamefont{{Chardin}}}, \bibnamefont{and} \bibinfo{author}{\bibfnamefont{D.}~\bibnamefont{{Aubert}}}, \bibinfo{journal}{\mnras} \textbf{\bibinfo{volume}{506}}, \bibinfo{pages}{5818} (\bibinfo{year}{2021}), \eprint{2105.02250}.

\bibitem[{\citenamefont{{Shimabukuro} et~al.}(2014)\citenamefont{{Shimabukuro}, {Ichiki}, {Inoue}, and {Yokoyama}}}]{2014PhRvD..90h3003S}
\bibinfo{author}{\bibfnamefont{H.}~\bibnamefont{{Shimabukuro}}}, \bibinfo{author}{\bibfnamefont{K.}~\bibnamefont{{Ichiki}}}, \bibinfo{author}{\bibfnamefont{S.}~\bibnamefont{{Inoue}}}, \bibnamefont{and} \bibinfo{author}{\bibfnamefont{S.}~\bibnamefont{{Yokoyama}}}, \bibinfo{journal}{\prd} \textbf{\bibinfo{volume}{90}}, \bibinfo{eid}{083003} (\bibinfo{year}{2014}), \eprint{1403.1605}.

\bibitem[{\citenamefont{{Shimabukuro} et~al.}(2020{\natexlab{a}})\citenamefont{{Shimabukuro}, {Ichiki}, and {Kadota}}}]{2020PhRvD.101d3516S}
\bibinfo{author}{\bibfnamefont{H.}~\bibnamefont{{Shimabukuro}}}, \bibinfo{author}{\bibfnamefont{K.}~\bibnamefont{{Ichiki}}}, \bibnamefont{and} \bibinfo{author}{\bibfnamefont{K.}~\bibnamefont{{Kadota}}}, \bibinfo{journal}{\prd} \textbf{\bibinfo{volume}{101}}, \bibinfo{eid}{043516} (\bibinfo{year}{2020}{\natexlab{a}}), \eprint{1910.06011}.

\bibitem[{\citenamefont{{Shimabukuro} et~al.}(2020{\natexlab{b}})\citenamefont{{Shimabukuro}, {Ichiki}, and {Kadota}}}]{2020PhRvD.102b3522S}
\bibinfo{author}{\bibfnamefont{H.}~\bibnamefont{{Shimabukuro}}}, \bibinfo{author}{\bibfnamefont{K.}~\bibnamefont{{Ichiki}}}, \bibnamefont{and} \bibinfo{author}{\bibfnamefont{K.}~\bibnamefont{{Kadota}}}, \bibinfo{journal}{\prd} \textbf{\bibinfo{volume}{102}}, \bibinfo{eid}{023522} (\bibinfo{year}{2020}{\natexlab{b}}), \eprint{2005.05589}.

\bibitem[{\citenamefont{{Kawasaki} et~al.}(2021)\citenamefont{{Kawasaki}, {Nakano}, {Nakatsuka}, and {Sonomoto}}}]{2021JCAP...04..019K}
\bibinfo{author}{\bibfnamefont{M.}~\bibnamefont{{Kawasaki}}}, \bibinfo{author}{\bibfnamefont{W.}~\bibnamefont{{Nakano}}}, \bibinfo{author}{\bibfnamefont{H.}~\bibnamefont{{Nakatsuka}}}, \bibnamefont{and} \bibinfo{author}{\bibfnamefont{E.}~\bibnamefont{{Sonomoto}}}, \bibinfo{journal}{\jcap} \textbf{\bibinfo{volume}{2021}}, \bibinfo{eid}{019} (\bibinfo{year}{2021}), \eprint{2010.13504}.

\bibitem[{\citenamefont{{Villanueva-Domingo} and {Ichiki}}(2023)}]{2023PASJ...75S..33V}
\bibinfo{author}{\bibfnamefont{P.}~\bibnamefont{{Villanueva-Domingo}}} \bibnamefont{and} \bibinfo{author}{\bibfnamefont{K.}~\bibnamefont{{Ichiki}}}, \bibinfo{journal}{\pasj} \textbf{\bibinfo{volume}{75}}, \bibinfo{pages}{S33} (\bibinfo{year}{2023}), \eprint{2104.10695}.

\bibitem[{\citenamefont{{Shimabukuro} et~al.}(2023{\natexlab{b}})\citenamefont{{Shimabukuro}, {Ichiki}, and {Kadota}}}]{2023PhRvD.107l3520S}
\bibinfo{author}{\bibfnamefont{H.}~\bibnamefont{{Shimabukuro}}}, \bibinfo{author}{\bibfnamefont{K.}~\bibnamefont{{Ichiki}}}, \bibnamefont{and} \bibinfo{author}{\bibfnamefont{K.}~\bibnamefont{{Kadota}}}, \bibinfo{journal}{\prd} \textbf{\bibinfo{volume}{107}}, \bibinfo{eid}{123520} (\bibinfo{year}{2023}{\natexlab{b}}), \eprint{2212.08409}.

\bibitem[{\citenamefont{{Kadota} et~al.}(2023)\citenamefont{{Kadota}, {Villanueva-Domingo}, {Ichiki}, {Hasegawa}, and {Naruse}}}]{2023JCAP...03..017K}
\bibinfo{author}{\bibfnamefont{K.}~\bibnamefont{{Kadota}}}, \bibinfo{author}{\bibfnamefont{P.}~\bibnamefont{{Villanueva-Domingo}}}, \bibinfo{author}{\bibfnamefont{K.}~\bibnamefont{{Ichiki}}}, \bibinfo{author}{\bibfnamefont{K.}~\bibnamefont{{Hasegawa}}}, \bibnamefont{and} \bibinfo{author}{\bibfnamefont{G.}~\bibnamefont{{Naruse}}}, \bibinfo{journal}{\jcap} \textbf{\bibinfo{volume}{2023}}, \bibinfo{eid}{017} (\bibinfo{year}{2023}), \eprint{2209.01305}.

\bibitem[{\citenamefont{{Naruse} et~al.}(2024)\citenamefont{{Naruse}, {Hasegawa}, {Kadota}, {Tashiro}, and {Ichiki}}}]{2024JCAP...10..091N}
\bibinfo{author}{\bibfnamefont{G.}~\bibnamefont{{Naruse}}}, \bibinfo{author}{\bibfnamefont{K.}~\bibnamefont{{Hasegawa}}}, \bibinfo{author}{\bibfnamefont{K.}~\bibnamefont{{Kadota}}}, \bibinfo{author}{\bibfnamefont{H.}~\bibnamefont{{Tashiro}}}, \bibnamefont{and} \bibinfo{author}{\bibfnamefont{K.}~\bibnamefont{{Ichiki}}}, \bibinfo{journal}{\jcap} \textbf{\bibinfo{volume}{2024}}, \bibinfo{eid}{091} (\bibinfo{year}{2024}), \eprint{2404.01034}.

\bibitem[{\citenamefont{{Ewall-Wice} et~al.}(2014)\citenamefont{{Ewall-Wice}, {Dillon}, {Mesinger}, and {Hewitt}}}]{2014MNRAS.441.2476E}
\bibinfo{author}{\bibfnamefont{A.}~\bibnamefont{{Ewall-Wice}}}, \bibinfo{author}{\bibfnamefont{J.~S.} \bibnamefont{{Dillon}}}, \bibinfo{author}{\bibfnamefont{A.}~\bibnamefont{{Mesinger}}}, \bibnamefont{and} \bibinfo{author}{\bibfnamefont{J.}~\bibnamefont{{Hewitt}}}, \bibinfo{journal}{\mnras} \textbf{\bibinfo{volume}{441}}, \bibinfo{pages}{2476} (\bibinfo{year}{2014}), \eprint{1310.7936}.

\bibitem[{\citenamefont{{Shao} et~al.}(2023)\citenamefont{{Shao}, {Xu}, {Wang}, {Yang}, {Li}, {Zhang}, and {Chen}}}]{2023NatAs...7.1116S}
\bibinfo{author}{\bibfnamefont{Y.}~\bibnamefont{{Shao}}}, \bibinfo{author}{\bibfnamefont{Y.}~\bibnamefont{{Xu}}}, \bibinfo{author}{\bibfnamefont{Y.}~\bibnamefont{{Wang}}}, \bibinfo{author}{\bibfnamefont{W.}~\bibnamefont{{Yang}}}, \bibinfo{author}{\bibfnamefont{R.}~\bibnamefont{{Li}}}, \bibinfo{author}{\bibfnamefont{X.}~\bibnamefont{{Zhang}}}, \bibnamefont{and} \bibinfo{author}{\bibfnamefont{X.}~\bibnamefont{{Chen}}}, \bibinfo{journal}{Nature Astronomy} \textbf{\bibinfo{volume}{7}}, \bibinfo{pages}{1116} (\bibinfo{year}{2023}), \eprint{2307.04130}.

\bibitem[{\citenamefont{{{\v{S}}oltinsk{\'y}} et~al.}(2025)\citenamefont{{{\v{S}}oltinsk{\'y}}, {Kulkarni}, {Tendulkar}, and {Bolton}}}]{2025MNRAS.tmp...29S}
\bibinfo{author}{\bibfnamefont{T.}~\bibnamefont{{{\v{S}}oltinsk{\'y}}}}, \bibinfo{author}{\bibfnamefont{G.}~\bibnamefont{{Kulkarni}}}, \bibinfo{author}{\bibfnamefont{S.~P.} \bibnamefont{{Tendulkar}}}, \bibnamefont{and} \bibinfo{author}{\bibfnamefont{J.~S.} \bibnamefont{{Bolton}}}, \bibinfo{journal}{\mnras}  (\bibinfo{year}{2025}), \eprint{2412.06879}.

\bibitem[{\citenamefont{{Shao} et~al.}(2025)\citenamefont{{Shao}, {Du}, {Li}, and {Zhang}}}]{2025arXiv250100769S}
\bibinfo{author}{\bibfnamefont{Y.}~\bibnamefont{{Shao}}}, \bibinfo{author}{\bibfnamefont{G.-H.} \bibnamefont{{Du}}}, \bibinfo{author}{\bibfnamefont{T.-N.} \bibnamefont{{Li}}}, \bibnamefont{and} \bibinfo{author}{\bibfnamefont{X.}~\bibnamefont{{Zhang}}}, \bibinfo{journal}{Physics Letters B} \textbf{\bibinfo{volume}{862}}, \bibinfo{eid}{139342} (\bibinfo{year}{2025}), \eprint{2501.00769}.

\bibitem[{\citenamefont{{Sun} et~al.}(2025)\citenamefont{{Sun}, {Shao}, {Li}, {Xu}, {Wang}, and {Zhang}}}]{2025CmPhy...8..220S}
\bibinfo{author}{\bibfnamefont{T.-Y.} \bibnamefont{{Sun}}}, \bibinfo{author}{\bibfnamefont{Y.}~\bibnamefont{{Shao}}}, \bibinfo{author}{\bibfnamefont{Y.}~\bibnamefont{{Li}}}, \bibinfo{author}{\bibfnamefont{Y.}~\bibnamefont{{Xu}}}, \bibinfo{author}{\bibfnamefont{H.}~\bibnamefont{{Wang}}}, \bibnamefont{and} \bibinfo{author}{\bibfnamefont{X.}~\bibnamefont{{Zhang}}}, \bibinfo{journal}{Communications Physics} \textbf{\bibinfo{volume}{8}}, \bibinfo{eid}{220} (\bibinfo{year}{2025}), \eprint{2407.14298}.

\bibitem[{\citenamefont{{Greig} et~al.}(2022)\citenamefont{{Greig}, {Ting}, and {Kaurov}}}]{2022MNRAS.513.1719G}
\bibinfo{author}{\bibfnamefont{B.}~\bibnamefont{{Greig}}}, \bibinfo{author}{\bibfnamefont{Y.-S.} \bibnamefont{{Ting}}}, \bibnamefont{and} \bibinfo{author}{\bibfnamefont{A.~A.} \bibnamefont{{Kaurov}}}, \bibinfo{journal}{\mnras} \textbf{\bibinfo{volume}{513}}, \bibinfo{pages}{1719} (\bibinfo{year}{2022}), \eprint{2204.02544}.

\bibitem[{\citenamefont{{Greig} et~al.}(2023)\citenamefont{{Greig}, {Ting}, and {Kaurov}}}]{2023MNRAS.519.5288G}
\bibinfo{author}{\bibfnamefont{B.}~\bibnamefont{{Greig}}}, \bibinfo{author}{\bibfnamefont{Y.-S.} \bibnamefont{{Ting}}}, \bibnamefont{and} \bibinfo{author}{\bibfnamefont{A.~A.} \bibnamefont{{Kaurov}}}, \bibinfo{journal}{\mnras} \textbf{\bibinfo{volume}{519}}, \bibinfo{pages}{5288} (\bibinfo{year}{2023}), \eprint{2207.09082}.

\bibitem[{\citenamefont{{Prelogovi{\'c}} and {Mesinger}}(2023)}]{2023MNRAS.524.4239P}
\bibinfo{author}{\bibfnamefont{D.}~\bibnamefont{{Prelogovi{\'c}}}} \bibnamefont{and} \bibinfo{author}{\bibfnamefont{A.}~\bibnamefont{{Mesinger}}}, \bibinfo{journal}{\mnras} \textbf{\bibinfo{volume}{524}}, \bibinfo{pages}{4239} (\bibinfo{year}{2023}), \eprint{2305.03074}.

\bibitem[{\citenamefont{{Zhao} et~al.}(2024)\citenamefont{{Zhao}, {Mao}, {Zuo}, and {Wandelt}}}]{2024ApJ...973...41Z}
\bibinfo{author}{\bibfnamefont{X.}~\bibnamefont{{Zhao}}}, \bibinfo{author}{\bibfnamefont{Y.}~\bibnamefont{{Mao}}}, \bibinfo{author}{\bibfnamefont{S.}~\bibnamefont{{Zuo}}}, \bibnamefont{and} \bibinfo{author}{\bibfnamefont{B.~D.} \bibnamefont{{Wandelt}}}, \bibinfo{journal}{\apj} \textbf{\bibinfo{volume}{973}}, \bibinfo{eid}{41} (\bibinfo{year}{2024}), \eprint{2310.17602}.

\bibitem[{\citenamefont{{Prelogovi{\'c}} and {Mesinger}}(2024)}]{2024A&A...688A.199P}
\bibinfo{author}{\bibfnamefont{D.}~\bibnamefont{{Prelogovi{\'c}}}} \bibnamefont{and} \bibinfo{author}{\bibfnamefont{A.}~\bibnamefont{{Mesinger}}}, \bibinfo{journal}{\aap} \textbf{\bibinfo{volume}{688}}, \bibinfo{eid}{A199} (\bibinfo{year}{2024}), \eprint{2401.12277}.

\bibitem[{\citenamefont{{Hothi} et~al.}(2024)\citenamefont{{Hothi}, {Allys}, {Semelin}, and {Boulanger}}}]{2024A&A...686A.212H}
\bibinfo{author}{\bibfnamefont{I.}~\bibnamefont{{Hothi}}}, \bibinfo{author}{\bibfnamefont{E.}~\bibnamefont{{Allys}}}, \bibinfo{author}{\bibfnamefont{B.}~\bibnamefont{{Semelin}}}, \bibnamefont{and} \bibinfo{author}{\bibfnamefont{F.}~\bibnamefont{{Boulanger}}}, \bibinfo{journal}{\aap} \textbf{\bibinfo{volume}{686}}, \bibinfo{eid}{A212} (\bibinfo{year}{2024}), \eprint{2311.00036}.

\bibitem[{\citenamefont{{Tohfa} et~al.}(2024)\citenamefont{{Tohfa}, {Bird}, {Ho}, {Qezlou}, and {Fernandez}}}]{2024PhRvL.132w1002T}
\bibinfo{author}{\bibfnamefont{H.~M.} \bibnamefont{{Tohfa}}}, \bibinfo{author}{\bibfnamefont{S.}~\bibnamefont{{Bird}}}, \bibinfo{author}{\bibfnamefont{M.-F.} \bibnamefont{{Ho}}}, \bibinfo{author}{\bibfnamefont{M.}~\bibnamefont{{Qezlou}}}, \bibnamefont{and} \bibinfo{author}{\bibfnamefont{M.}~\bibnamefont{{Fernandez}}}, \bibinfo{journal}{\prl} \textbf{\bibinfo{volume}{132}}, \bibinfo{eid}{231002} (\bibinfo{year}{2024}), \eprint{2310.06010}.

\bibitem[{\citenamefont{Mallat}(2012)}]{mallat2012group}
\bibinfo{author}{\bibfnamefont{S.}~\bibnamefont{Mallat}}, \bibinfo{journal}{Communications on Pure and Applied Mathematics} \textbf{\bibinfo{volume}{65}}, \bibinfo{pages}{1331} (\bibinfo{year}{2012}).

\bibitem[{\citenamefont{{Andreux} et~al.}(2018)\citenamefont{{Andreux}, {Angles}, {Exarchakis}, {Leonarduzzi}, {Rochette}, {Thiry}, {Zarka}, {Mallat}, {and{\'e}n}, {Belilovsky} et~al.}}]{2018arXiv181211214A}
\bibinfo{author}{\bibfnamefont{M.}~\bibnamefont{{Andreux}}}, \bibinfo{author}{\bibfnamefont{T.}~\bibnamefont{{Angles}}}, \bibinfo{author}{\bibfnamefont{G.}~\bibnamefont{{Exarchakis}}}, \bibinfo{author}{\bibfnamefont{R.}~\bibnamefont{{Leonarduzzi}}}, \bibinfo{author}{\bibfnamefont{G.}~\bibnamefont{{Rochette}}}, \bibinfo{author}{\bibfnamefont{L.}~\bibnamefont{{Thiry}}}, \bibinfo{author}{\bibfnamefont{J.}~\bibnamefont{{Zarka}}}, \bibinfo{author}{\bibfnamefont{S.}~\bibnamefont{{Mallat}}}, \bibinfo{author}{\bibfnamefont{J.}~\bibnamefont{{and{\'e}n}}}, \bibinfo{author}{\bibfnamefont{E.}~\bibnamefont{{Belilovsky}}}, \bibnamefont{et~al.}, \bibinfo{journal}{arXiv e-prints} \bibinfo{eid}{arXiv:1812.11214} (\bibinfo{year}{2018}), \eprint{1812.11214}.

\bibitem[{\citenamefont{{Bruna} and {Mallat}}(2012)}]{2012arXiv1203.1513B}
\bibinfo{author}{\bibfnamefont{J.}~\bibnamefont{{Bruna}}} \bibnamefont{and} \bibinfo{author}{\bibfnamefont{S.}~\bibnamefont{{Mallat}}}, \bibinfo{journal}{arXiv e-prints} \bibinfo{eid}{arXiv:1203.1513} (\bibinfo{year}{2012}), \eprint{1203.1513}.

\bibitem[{\citenamefont{Vahidi et~al.}(2023)\citenamefont{Vahidi, Mitcheltree, and Lostanlen}}]{vahidi2023kymatio}
\bibinfo{author}{\bibfnamefont{C.}~\bibnamefont{Vahidi}}, \bibinfo{author}{\bibfnamefont{C.}~\bibnamefont{Mitcheltree}}, \bibnamefont{and} \bibinfo{author}{\bibfnamefont{V.}~\bibnamefont{Lostanlen}}, \emph{\bibinfo{title}{Kymatio: Deep Learning meets Wavelet Theory for Music Signal Processing}} (\bibinfo{publisher}{ISMIR}, \bibinfo{year}{2023}), \urlprefix\url{https://kymatio.github.io/ismir23-tutorial}.

\bibitem[{\citenamefont{{Braun} et~al.}(2019)\citenamefont{{Braun}, {Bonaldi}, {Bourke}, {Keane}, and {Wagg}}}]{2019arXiv191212699B}
\bibinfo{author}{\bibfnamefont{R.}~\bibnamefont{{Braun}}}, \bibinfo{author}{\bibfnamefont{A.}~\bibnamefont{{Bonaldi}}}, \bibinfo{author}{\bibfnamefont{T.}~\bibnamefont{{Bourke}}}, \bibinfo{author}{\bibfnamefont{E.}~\bibnamefont{{Keane}}}, \bibnamefont{and} \bibinfo{author}{\bibfnamefont{J.}~\bibnamefont{{Wagg}}}, \bibinfo{journal}{arXiv e-prints} \bibinfo{eid}{arXiv:1912.12699} (\bibinfo{year}{2019}), \eprint{1912.12699}.

\bibitem[{\citenamefont{{Zhao} et~al.}(2022{\natexlab{a}})\citenamefont{{Zhao}, {Mao}, {Cheng}, and {Wandelt}}}]{2022ApJ...926..151Z}
\bibinfo{author}{\bibfnamefont{X.}~\bibnamefont{{Zhao}}}, \bibinfo{author}{\bibfnamefont{Y.}~\bibnamefont{{Mao}}}, \bibinfo{author}{\bibfnamefont{C.}~\bibnamefont{{Cheng}}}, \bibnamefont{and} \bibinfo{author}{\bibfnamefont{B.~D.} \bibnamefont{{Wandelt}}}, \bibinfo{journal}{\apj} \textbf{\bibinfo{volume}{926}}, \bibinfo{eid}{151} (\bibinfo{year}{2022}{\natexlab{a}}), \eprint{2105.03344}.

\bibitem[{\citenamefont{{Zhao} et~al.}(2022{\natexlab{b}})\citenamefont{{Zhao}, {Mao}, and {Wandelt}}}]{2022ApJ...933..236Z}
\bibinfo{author}{\bibfnamefont{X.}~\bibnamefont{{Zhao}}}, \bibinfo{author}{\bibfnamefont{Y.}~\bibnamefont{{Mao}}}, \bibnamefont{and} \bibinfo{author}{\bibfnamefont{B.~D.} \bibnamefont{{Wandelt}}}, \bibinfo{journal}{\apj} \textbf{\bibinfo{volume}{933}}, \bibinfo{eid}{236} (\bibinfo{year}{2022}{\natexlab{b}}), \eprint{2203.15734}.

\bibitem[{\citenamefont{{Shimabukuro} et~al.}(2016)\citenamefont{{Shimabukuro}, {Yoshiura}, {Takahashi}, {Yokoyama}, and {Ichiki}}}]{2016MNRAS.458.3003S}
\bibinfo{author}{\bibfnamefont{H.}~\bibnamefont{{Shimabukuro}}}, \bibinfo{author}{\bibfnamefont{S.}~\bibnamefont{{Yoshiura}}}, \bibinfo{author}{\bibfnamefont{K.}~\bibnamefont{{Takahashi}}}, \bibinfo{author}{\bibfnamefont{S.}~\bibnamefont{{Yokoyama}}}, \bibnamefont{and} \bibinfo{author}{\bibfnamefont{K.}~\bibnamefont{{Ichiki}}}, \bibinfo{journal}{\mnras} \textbf{\bibinfo{volume}{458}}, \bibinfo{pages}{3003} (\bibinfo{year}{2016}), \eprint{1507.01335}.

\bibitem[{\citenamefont{{Shimabukuro} et~al.}(2017)\citenamefont{{Shimabukuro}, {Yoshiura}, {Takahashi}, {Yokoyama}, and {Ichiki}}}]{2017MNRAS.468.1542S}
\bibinfo{author}{\bibfnamefont{H.}~\bibnamefont{{Shimabukuro}}}, \bibinfo{author}{\bibfnamefont{S.}~\bibnamefont{{Yoshiura}}}, \bibinfo{author}{\bibfnamefont{K.}~\bibnamefont{{Takahashi}}}, \bibinfo{author}{\bibfnamefont{S.}~\bibnamefont{{Yokoyama}}}, \bibnamefont{and} \bibinfo{author}{\bibfnamefont{K.}~\bibnamefont{{Ichiki}}}, \bibinfo{journal}{\mnras} \textbf{\bibinfo{volume}{468}}, \bibinfo{pages}{1542} (\bibinfo{year}{2017}), \eprint{1608.00372}.

\bibitem[{\citenamefont{{Watkinson} et~al.}(2017)\citenamefont{{Watkinson}, {Majumdar}, {Pritchard}, and {Mondal}}}]{2017MNRAS.472.2436W}
\bibinfo{author}{\bibfnamefont{C.~A.} \bibnamefont{{Watkinson}}}, \bibinfo{author}{\bibfnamefont{S.}~\bibnamefont{{Majumdar}}}, \bibinfo{author}{\bibfnamefont{J.~R.} \bibnamefont{{Pritchard}}}, \bibnamefont{and} \bibinfo{author}{\bibfnamefont{R.}~\bibnamefont{{Mondal}}}, \bibinfo{journal}{\mnras} \textbf{\bibinfo{volume}{472}}, \bibinfo{pages}{2436} (\bibinfo{year}{2017}), \eprint{1705.06284}.

\bibitem[{\citenamefont{{Majumdar} et~al.}(2018)\citenamefont{{Majumdar}, {Pritchard}, {Mondal}, {Watkinson}, {Bharadwaj}, and {Mellema}}}]{2018MNRAS.476.4007M}
\bibinfo{author}{\bibfnamefont{S.}~\bibnamefont{{Majumdar}}}, \bibinfo{author}{\bibfnamefont{J.~R.} \bibnamefont{{Pritchard}}}, \bibinfo{author}{\bibfnamefont{R.}~\bibnamefont{{Mondal}}}, \bibinfo{author}{\bibfnamefont{C.~A.} \bibnamefont{{Watkinson}}}, \bibinfo{author}{\bibfnamefont{S.}~\bibnamefont{{Bharadwaj}}}, \bibnamefont{and} \bibinfo{author}{\bibfnamefont{G.}~\bibnamefont{{Mellema}}}, \bibinfo{journal}{\mnras} \textbf{\bibinfo{volume}{476}}, \bibinfo{pages}{4007} (\bibinfo{year}{2018}), \eprint{1708.08458}.

\bibitem[{\citenamefont{{Hutter} et~al.}(2020)\citenamefont{{Hutter}, {Watkinson}, {Seiler}, {Dayal}, {Sinha}, and {Croton}}}]{2020MNRAS.492..653H}
\bibinfo{author}{\bibfnamefont{A.}~\bibnamefont{{Hutter}}}, \bibinfo{author}{\bibfnamefont{C.~A.} \bibnamefont{{Watkinson}}}, \bibinfo{author}{\bibfnamefont{J.}~\bibnamefont{{Seiler}}}, \bibinfo{author}{\bibfnamefont{P.}~\bibnamefont{{Dayal}}}, \bibinfo{author}{\bibfnamefont{M.}~\bibnamefont{{Sinha}}}, \bibnamefont{and} \bibinfo{author}{\bibfnamefont{D.~J.} \bibnamefont{{Croton}}}, \bibinfo{journal}{\mnras} \textbf{\bibinfo{volume}{492}}, \bibinfo{pages}{653} (\bibinfo{year}{2020}), \eprint{1907.04342}.

\bibitem[{\citenamefont{{Diao} et~al.}(2024)\citenamefont{{Diao}, {Chen}, {Chen}, and {Mao}}}]{2024ApJ...974..141D}
\bibinfo{author}{\bibfnamefont{K.}~\bibnamefont{{Diao}}}, \bibinfo{author}{\bibfnamefont{Z.}~\bibnamefont{{Chen}}}, \bibinfo{author}{\bibfnamefont{X.}~\bibnamefont{{Chen}}}, \bibnamefont{and} \bibinfo{author}{\bibfnamefont{Y.}~\bibnamefont{{Mao}}}, \bibinfo{journal}{\apj} \textbf{\bibinfo{volume}{974}}, \bibinfo{eid}{141} (\bibinfo{year}{2024}), \eprint{2406.20058}.

\bibitem[{\citenamefont{{Shaw} et~al.}(2019)\citenamefont{{Shaw}, {Bharadwaj}, and {Mondal}}}]{2019MNRAS.487.4951S}
\bibinfo{author}{\bibfnamefont{A.~K.} \bibnamefont{{Shaw}}}, \bibinfo{author}{\bibfnamefont{S.}~\bibnamefont{{Bharadwaj}}}, \bibnamefont{and} \bibinfo{author}{\bibfnamefont{R.}~\bibnamefont{{Mondal}}}, \bibinfo{journal}{\mnras} \textbf{\bibinfo{volume}{487}}, \bibinfo{pages}{4951} (\bibinfo{year}{2019}), \eprint{1902.08706}.

\bibitem[{\citenamefont{{Sui} et~al.}(2023)\citenamefont{{Sui}, {Zhao}, {Jing}, and {Mao}}}]{2023mla..confE..30S}
\bibinfo{author}{\bibfnamefont{C.}~\bibnamefont{{Sui}}}, \bibinfo{author}{\bibfnamefont{X.}~\bibnamefont{{Zhao}}}, \bibinfo{author}{\bibfnamefont{T.}~\bibnamefont{{Jing}}}, \bibnamefont{and} \bibinfo{author}{\bibfnamefont{Y.}~\bibnamefont{{Mao}}}, in \emph{\bibinfo{booktitle}{Machine Learning for Astrophysics}} (\bibinfo{year}{2023}), p.~\bibinfo{pages}{30}, \eprint{2307.04994}.

\bibitem[{\citenamefont{{Watkinson} et~al.}(2022)\citenamefont{{Watkinson}, {Greig}, and {Mesinger}}}]{2022MNRAS.510.3838W}
\bibinfo{author}{\bibfnamefont{C.~A.} \bibnamefont{{Watkinson}}}, \bibinfo{author}{\bibfnamefont{B.}~\bibnamefont{{Greig}}}, \bibnamefont{and} \bibinfo{author}{\bibfnamefont{A.}~\bibnamefont{{Mesinger}}}, \bibinfo{journal}{\mnras} \textbf{\bibinfo{volume}{510}}, \bibinfo{pages}{3838} (\bibinfo{year}{2022}), \eprint{2102.02310}.

\bibitem[{\citenamefont{{Belladitta} et~al.}(2020)\citenamefont{{Belladitta}, {Moretti}, {Caccianiga}, {Spingola}, {Severgnini}, {Della Ceca}, {Ghisellini}, {Dallacasa}, {Sbarrato}, {Cicone} et~al.}}]{2020A&A...635L...7B}
\bibinfo{author}{\bibfnamefont{S.}~\bibnamefont{{Belladitta}}}, \bibinfo{author}{\bibfnamefont{A.}~\bibnamefont{{Moretti}}}, \bibinfo{author}{\bibfnamefont{A.}~\bibnamefont{{Caccianiga}}}, \bibinfo{author}{\bibfnamefont{C.}~\bibnamefont{{Spingola}}}, \bibinfo{author}{\bibfnamefont{P.}~\bibnamefont{{Severgnini}}}, \bibinfo{author}{\bibfnamefont{R.}~\bibnamefont{{Della Ceca}}}, \bibinfo{author}{\bibfnamefont{G.}~\bibnamefont{{Ghisellini}}}, \bibinfo{author}{\bibfnamefont{D.}~\bibnamefont{{Dallacasa}}}, \bibinfo{author}{\bibfnamefont{T.}~\bibnamefont{{Sbarrato}}}, \bibinfo{author}{\bibfnamefont{C.}~\bibnamefont{{Cicone}}}, \bibnamefont{et~al.}, \bibinfo{journal}{\aap} \textbf{\bibinfo{volume}{635}}, \bibinfo{eid}{L7} (\bibinfo{year}{2020}), \eprint{2002.05178}.

\bibitem[{\citenamefont{{Ba{\~n}ados} et~al.}(2021)\citenamefont{{Ba{\~n}ados}, {Mazzucchelli}, {Momjian}, {Eilers}, {Wang}, {Schindler}, {Connor}, {Andika}, {Barth}, {Carilli} et~al.}}]{2021ApJ...909...80B}
\bibinfo{author}{\bibfnamefont{E.}~\bibnamefont{{Ba{\~n}ados}}}, \bibinfo{author}{\bibfnamefont{C.}~\bibnamefont{{Mazzucchelli}}}, \bibinfo{author}{\bibfnamefont{E.}~\bibnamefont{{Momjian}}}, \bibinfo{author}{\bibfnamefont{A.-C.} \bibnamefont{{Eilers}}}, \bibinfo{author}{\bibfnamefont{F.}~\bibnamefont{{Wang}}}, \bibinfo{author}{\bibfnamefont{J.-T.} \bibnamefont{{Schindler}}}, \bibinfo{author}{\bibfnamefont{T.}~\bibnamefont{{Connor}}}, \bibinfo{author}{\bibfnamefont{I.~T.} \bibnamefont{{Andika}}}, \bibinfo{author}{\bibfnamefont{A.~J.} \bibnamefont{{Barth}}}, \bibinfo{author}{\bibfnamefont{C.}~\bibnamefont{{Carilli}}}, \bibnamefont{et~al.}, \bibinfo{journal}{\apj} \textbf{\bibinfo{volume}{909}}, \bibinfo{eid}{80} (\bibinfo{year}{2021}), \eprint{2103.03295}.

\bibitem[{\citenamefont{{Khusanova} et~al.}(2022)\citenamefont{{Khusanova}, {Ba{\~n}ados}, {Mazzucchelli}, {Rojas-Ruiz}, {Momjian}, {Walter}, {Decarli}, {Venemans}, {Farina}, {Meyer} et~al.}}]{2022A&A...664A..39K}
\bibinfo{author}{\bibfnamefont{Y.}~\bibnamefont{{Khusanova}}}, \bibinfo{author}{\bibfnamefont{E.}~\bibnamefont{{Ba{\~n}ados}}}, \bibinfo{author}{\bibfnamefont{C.}~\bibnamefont{{Mazzucchelli}}}, \bibinfo{author}{\bibfnamefont{S.}~\bibnamefont{{Rojas-Ruiz}}}, \bibinfo{author}{\bibfnamefont{E.}~\bibnamefont{{Momjian}}}, \bibinfo{author}{\bibfnamefont{F.}~\bibnamefont{{Walter}}}, \bibinfo{author}{\bibfnamefont{R.}~\bibnamefont{{Decarli}}}, \bibinfo{author}{\bibfnamefont{B.}~\bibnamefont{{Venemans}}}, \bibinfo{author}{\bibfnamefont{E.~P.} \bibnamefont{{Farina}}}, \bibinfo{author}{\bibfnamefont{R.}~\bibnamefont{{Meyer}}}, \bibnamefont{et~al.}, \bibinfo{journal}{\aap} \textbf{\bibinfo{volume}{664}}, \bibinfo{eid}{A39} (\bibinfo{year}{2022}), \eprint{2204.08973}.

\bibitem[{\citenamefont{{Gloudemans} et~al.}(2022)\citenamefont{{Gloudemans}, {Duncan}, {Saxena}, {Harikane}, {Hill}, {Zeimann}, {R{\"o}ttgering}, {Yang}, {Best}, {Ba{\~n}ados} et~al.}}]{2022A&A...668A..27G}
\bibinfo{author}{\bibfnamefont{A.~J.} \bibnamefont{{Gloudemans}}}, \bibinfo{author}{\bibfnamefont{K.~J.} \bibnamefont{{Duncan}}}, \bibinfo{author}{\bibfnamefont{A.}~\bibnamefont{{Saxena}}}, \bibinfo{author}{\bibfnamefont{Y.}~\bibnamefont{{Harikane}}}, \bibinfo{author}{\bibfnamefont{G.~J.} \bibnamefont{{Hill}}}, \bibinfo{author}{\bibfnamefont{G.~R.} \bibnamefont{{Zeimann}}}, \bibinfo{author}{\bibfnamefont{H.~J.~A.} \bibnamefont{{R{\"o}ttgering}}}, \bibinfo{author}{\bibfnamefont{D.}~\bibnamefont{{Yang}}}, \bibinfo{author}{\bibfnamefont{P.~N.} \bibnamefont{{Best}}}, \bibinfo{author}{\bibfnamefont{E.}~\bibnamefont{{Ba{\~n}ados}}}, \bibnamefont{et~al.}, \bibinfo{journal}{\aap} \textbf{\bibinfo{volume}{668}}, \bibinfo{eid}{A27} (\bibinfo{year}{2022}), \eprint{2210.01811}.

\bibitem[{\citenamefont{{Ba{\~n}ados} et~al.}(2024{\natexlab{a}})\citenamefont{{Ba{\~n}ados}, {Khusanova}, {Decarli}, {Momjian}, {Walter}, {Connor}, {Carilli}, {Mazzucchelli}, {Rojas-Ruiz}, and {Venemans}}}]{2024ApJ...977L..46B}
\bibinfo{author}{\bibfnamefont{E.}~\bibnamefont{{Ba{\~n}ados}}}, \bibinfo{author}{\bibfnamefont{Y.}~\bibnamefont{{Khusanova}}}, \bibinfo{author}{\bibfnamefont{R.}~\bibnamefont{{Decarli}}}, \bibinfo{author}{\bibfnamefont{E.}~\bibnamefont{{Momjian}}}, \bibinfo{author}{\bibfnamefont{F.}~\bibnamefont{{Walter}}}, \bibinfo{author}{\bibfnamefont{T.}~\bibnamefont{{Connor}}}, \bibinfo{author}{\bibfnamefont{C.~L.} \bibnamefont{{Carilli}}}, \bibinfo{author}{\bibfnamefont{C.}~\bibnamefont{{Mazzucchelli}}}, \bibinfo{author}{\bibfnamefont{S.}~\bibnamefont{{Rojas-Ruiz}}}, \bibnamefont{and} \bibinfo{author}{\bibfnamefont{B.~P.} \bibnamefont{{Venemans}}}, \bibinfo{journal}{\apjl} \textbf{\bibinfo{volume}{977}}, \bibinfo{eid}{L46} (\bibinfo{year}{2024}{\natexlab{a}}), \eprint{2407.07236}.

\bibitem[{\citenamefont{{Ba{\~n}ados} et~al.}(2024{\natexlab{b}})\citenamefont{{Ba{\~n}ados}, {Momjian}, {Connor}, {Belladitta}, {Decarli}, {Mazzucchelli}, {Venemans}, {Walter}, {Wang}, {Xie} et~al.}}]{2024NatAs.tmp..293B}
\bibinfo{author}{\bibfnamefont{E.}~\bibnamefont{{Ba{\~n}ados}}}, \bibinfo{author}{\bibfnamefont{E.}~\bibnamefont{{Momjian}}}, \bibinfo{author}{\bibfnamefont{T.}~\bibnamefont{{Connor}}}, \bibinfo{author}{\bibfnamefont{S.}~\bibnamefont{{Belladitta}}}, \bibinfo{author}{\bibfnamefont{R.}~\bibnamefont{{Decarli}}}, \bibinfo{author}{\bibfnamefont{C.}~\bibnamefont{{Mazzucchelli}}}, \bibinfo{author}{\bibfnamefont{B.~P.} \bibnamefont{{Venemans}}}, \bibinfo{author}{\bibfnamefont{F.}~\bibnamefont{{Walter}}}, \bibinfo{author}{\bibfnamefont{F.}~\bibnamefont{{Wang}}}, \bibinfo{author}{\bibfnamefont{Z.-L.} \bibnamefont{{Xie}}}, \bibnamefont{et~al.}, \bibinfo{journal}{Nature Astronomy}  (\bibinfo{year}{2024}{\natexlab{b}}).

\bibitem[{\citenamefont{{{\v{S}}oltinsk{\'y}} et~al.}(2023)\citenamefont{{{\v{S}}oltinsk{\'y}}, {Bolton}, {Molaro}, {Hatch}, {Haehnelt}, {Keating}, {Kulkarni}, and {Puchwein}}}]{2023MNRAS.519.3027S}
\bibinfo{author}{\bibfnamefont{T.}~\bibnamefont{{{\v{S}}oltinsk{\'y}}}}, \bibinfo{author}{\bibfnamefont{J.~S.} \bibnamefont{{Bolton}}}, \bibinfo{author}{\bibfnamefont{M.}~\bibnamefont{{Molaro}}}, \bibinfo{author}{\bibfnamefont{N.}~\bibnamefont{{Hatch}}}, \bibinfo{author}{\bibfnamefont{M.~G.} \bibnamefont{{Haehnelt}}}, \bibinfo{author}{\bibfnamefont{L.~C.} \bibnamefont{{Keating}}}, \bibinfo{author}{\bibfnamefont{G.}~\bibnamefont{{Kulkarni}}}, \bibnamefont{and} \bibinfo{author}{\bibfnamefont{E.}~\bibnamefont{{Puchwein}}}, \bibinfo{journal}{\mnras} \textbf{\bibinfo{volume}{519}}, \bibinfo{pages}{3027} (\bibinfo{year}{2023}), \eprint{2211.07655}.

\bibitem[{\citenamefont{{Niu} et~al.}(2025)\citenamefont{{Niu}, {Li}, {Xu}, {Guo}, and {Zhang}}}]{2025ApJ...978..145N}
\bibinfo{author}{\bibfnamefont{Q.}~\bibnamefont{{Niu}}}, \bibinfo{author}{\bibfnamefont{Y.}~\bibnamefont{{Li}}}, \bibinfo{author}{\bibfnamefont{Y.}~\bibnamefont{{Xu}}}, \bibinfo{author}{\bibfnamefont{H.}~\bibnamefont{{Guo}}}, \bibnamefont{and} \bibinfo{author}{\bibfnamefont{X.}~\bibnamefont{{Zhang}}}, \bibinfo{journal}{\apj} \textbf{\bibinfo{volume}{978}}, \bibinfo{eid}{145} (\bibinfo{year}{2025}), \eprint{2407.18136}.

\end{thebibliography}

\end{document}